\documentclass[11pt]{article}
\usepackage{graphicx}
\usepackage{amsmath}
\usepackage{amscd}

\allowdisplaybreaks
 \numberwithin{equation}{section}

\begin{document}


\begin{titlepage}
\thispagestyle{empty}
\begin{flushleft}
USCHEP/0304ib2
\hfill hep-th/0304005 \\
UT-03-08\hfill April, 2003 \\
\end{flushleft}

\vskip 1.5 cm
\bigskip

\begin{center}
\noindent{\Large \textbf{Fermionic Ghosts in Moyal String Field Theory}}\\
\noindent{
 }\\
\renewcommand{\thefootnote}{\fnsymbol{footnote}}

\vskip 2cm
{\large
I. Bars$^{a,}$\footnote{e-mail address: bars@usc.edu},
I. Kishimoto$^{b,}$\footnote{e-mail address:
 ikishimo@hep-th.phys.s.u-tokyo.ac.jp} and
Y. Matsuo$^{b,}$\footnote{e-mail address:
 matsuo@phys.s.u-tokyo.ac.jp} } \\
{\it
\noindent{ \bigskip }\\
$^{a)}$ Department of Physics and Astronomy,\\
University of Southern California, Los Angeles, CA 90089-0484, USA \\
\noindent{\smallskip  }\\
$^{b)}$ Department of Physics, Faculty of Science, University of Tokyo \\
Hongo 7-3-1, Bunkyo-ku, Tokyo 113-0033, Japan\\
\noindent{ \smallskip }\\
}
\bigskip
\end{center}
\begin{abstract}
We complete the construction of the
Moyal star formulation of bosonic
open string field theory (MSFT) by providing
a detailed study  of the fermionic ghost sector.
In particular, as in the case of the matter sector,
(1) we construct a map
from Witten's star product to the Moyal product,
(2) we propose a regularization scheme
which is consistent with the matter sector
and (3) as a check of the formalism,
we derive the ghost Neumann coefficients algebraically
directly from the Moyal product. The latter
satisfy the Gross-Jevicki nonlinear relations
even in the presence of the regulator, and
when the regulator is removed they coincide numerically
with the expression derived from conformal field theory.
After this basic construction, we derive a regularized action
of string field theory in the Siegel gauge and
define the Feynman rules. We give explicitly the
analytic expression of the off-shell four point function
for tachyons, including the ghost contribution.
Some of the results in this paper have already  been used in
our previous publications.
This paper provides the
technical details of the computations which were omitted there.
\end{abstract}
\end{titlepage}\vfill\setcounter{footnote}{0} \renewcommand{\thefootnote}{%
\arabic{footnote}} \newpage

\tableofcontents


\section{Introduction}

Witten's open bosonic string field theory \cite{Witten}, its operator
version \cite{GJ}, and its split string reformulations \cite
{split,rszSplit,GT,KO,B1}, led to the development of the
Moyal star formulation of string field theory (MSFT) during the past two
years \cite{B1}-\cite{BKM2}%
. This formulation has the following features:

\begin{enumerate}
\item  The star product in Witten's string field theory is mapped to the
Moyal star product \cite{B1} after an appropriate change of variables. The
string field $A\left( \bar{x},x_{e}.p_{e}\right) $ in the new basis is a
function of the string midpoint $\bar{x}$ and the phase space $\left(
x_{e},p_{e}\right) $ of even string modes $e=2,4,6,\cdots $. While Witten's
star product, written in terms of Neumann coefficients \cite{GJ}, is very
complicated to manipulate in computations, the Moyal star product in MSFT,
which is diagonal in mode space labelled by $e$, is the simplest form of the
star product that occurs in standard noncommutative geometry. This feature
vastly simplifies the structure of string field theory and is very helpful
in explicit computations.

\item  The change of variables introduces a set of simple but infinite
matrices $T_{eo},R_{oe},v_{o},w_{e},$ labelled by $e=2,4,6,\cdots ,$ and $%
o=1,3,5,\cdots ,$ which contain basic information about even $\left(
e\right) $ and odd $\left( o\right) $ string modes. These matrices obey a
matrix algebra that has an associativity anomaly, which in turn feeds into
an associativity anomaly among string fields \cite{BM1}. The origin of the
anomaly is the infinite number of string modes that cause the appearance of
ambiguous terms of the form $\infty /\infty .$ In order to resolve this
problem, we proposed \cite{BM1, BM2} a regularization by truncating the
number of the oscillators to finite $2N$, and defined a deformed set of
finite matrices $T,R,v,w$ as functions of the oscillator frequencies $\kappa
_{e},\kappa _{o}$ of the $2N$ modes. After such regulation, the
associativity is restored and all manipulations become well-defined. The
original open string field theory is restored by taking the original
frequencies, $\kappa _{e}=e,$ $\kappa _{o}=o,$ and the large $N$ limit at
the end of the computation. Through explicit computation of specific
examples, it has been shown that this regulation procedure correctly
reproduces results computed independently in conformal string theory.

\item  In the regularized basis, we computed the Neumann coefficients
analytically by using only the Moyal product. These coefficients are not
needed for computations in MSFT, but they provide a check of MSFT relative
to the operator formulation given in \cite{GJ}. We have shown that the
Neumann coefficients derived in the regularized MSFT framework satisfy the
Gross-Jevicki nonlinear relations, and thus provide a generalization of
Neumann coefficients for any set of frequencies $\kappa _{e},\kappa _{o}$
and any $N.$ This provided the first consistency check of MSFT \cite{BM2}.
Furthermore, we found that the Neumann coefficients for any $n$-point vertex
are all simple functions of a single matrix $t_{eo}=\kappa
_{e}^{1/2}T_{eo}\kappa _{o}^{-1/2}.$ Diagonalizing the matrix $t$
diagonalizes all the Neumann coefficients simultaneously for all $n$-point
vertices \cite{BM2}. At large $N$ our diagonal form agrees with the one
given in \cite{spectro}\cite{DLMZ}, and explains in particular why there is
Neumann spectroscopy for the 3-point vertex \cite{spectro}, and generalizes
it to any $n$-point vertex \cite{BM2}.

\item  One of the nice features of MSFT is that the star product is diagonal
in mode space (i.e. independent and same for each mode). The only cost for
this simplification is that the kinetic term given by the Virasoro operator $%
L_{0}$ becomes off-diagonal in mode space $\left( x_{e},p_{e}\right) $ \cite
{BM2}. However, this has not hindered explicit computations. In particular,
we have derived the Feynman rules, including the propagator $\left(
L_{0}-1\right) ^{-1},$ and shown that we can evaluate efficiently and
explicitly the Feynman graphs in open string field theory \cite{BKM1}.

\item  The off-diagonal part of the kinetic term depends on a specific
combination of momentum modes, namely $\hat{p}=\left( 1+\bar{w}w\right)
^{-1/2}\sum_{e}w_{e}p_{e}$ which we refer to as the \textquotedblleft
anomalous midpoint mode'' \cite{BM1}\cite{B2}. This mode appears in the
kinetic operator in the form $L_{0}=\gamma +\cdots ,$ with $\gamma \sim \hat{%
p}^{2}$, while the remaining part of the kinetic term is diagonal in mode
space. We named the term $\gamma $ the \textquotedblleft midpoint
correction'' . We found that if it were not for this midpoint correction, the
rest of the kinetic plus interaction terms would define a theory, equivalent
to an infinite matrix theory, that is vastly simpler and completely solvable 
\cite{BKM2}. However we determined that the $\gamma $ term is essential for
the correct definition of string field theory. Thus we have isolated the
hard part of string field theory in the form of the quadratic
\textquotedblleft midpoint correction'' term $\gamma .$ We have shown that
all classical solutions, including the true vacuum, of the interacting
string theory, are obtained analytically by first solving the nonlinear
equation explicitly by ignoring $\gamma ,$ and then including the effect of $%
\gamma $ in a closed formal expression that can be evaluated to any order in
a perturbative expansion in powers of $\gamma $ \cite{BKM2}.
\end{enumerate}

In our published work so far, we have demonstrated all of the above results
explicitly mainly in the matter sector. We have implied, and sometimes
explicitly included, the corresponding contribution of ghosts either in the
bosonic \cite{BM2} or fermionic \cite{BKM1,BKM2} version but the details
were not given explicitly. The purpose of the present paper is to provide
all the relevant material on fermionic ghosts which we used previously, in
an organized and comprehensive manner. In this sense, this paper completes
the basic formulation of MSFT.

Because we will try to be quite explicit, the content of this paper will be
rather technical. The construction of the Moyal product for fermionic ghosts
is basically parallel to the bosonic case \cite{B1,BM1,BM2}, except that we
need to be careful in some minor differences, including midpoint issues,
which appear in the ghost fields $b,c$.

The first point relates to the boundary conditions. While we needed to
consider Neumann type boundary conditions for the matter fields (open
strings on the D25 brane), Dirichlet type boundary conditions appear
for the ghost field. Therefore, the Fourier basis is different. The
regularization method developed in \cite{BM1, BM2} was based on the Fourier
modes, hence some care is needed to make the regularization compatible in
the matter and ghost sectors. The regularization developed in this paper can
be applied also to the treatment of the matter sector for lower D$p$ branes (%
$p<25$).

The second point relates to the overlapping conditions of split strings. For
the matter fields, we have to treat only overlapping conditions for the
split string degrees of freedom. For the ghost fields, on the other hand, we
should also consider the anti-overlapping conditions. This induces some
changes in the mapping of Witten's star to Moyal's star.

The third point is the fermionic nature of Moyal variables. The usual
bosonic derivatives which appear in the definition of the Moyal product get
replaced by derivatives of fermionic variables, and care is needed in the
ordering and signs.

The fourth point is the treatment of the midpoint mode. This is a rather
delicate issue since, as in the matter sector, it cannot be determined from
the split string formulation. The fermionic midpoint mode is not part of the
Moyal $\star $ product, and it is integrated in the definition of the
action. In the Siegel gauge, the dependence on this extra fermionic variable
becomes trivial, and it drops out in actual computations.

We mention the work of Erler \cite{Erler} where he defined the Moyal star
formulation for the ghost system mainly in the continuous basis
\cite{DLMZ}.\footnote{
There are some works on Moyal structure of Witten's
string field theory.\cite{conti}
}
There are overlaps of the current paper with his work especially in the
second and third points mentioned above. The correct treatment of the
midpoint appeared first in our work \cite{BKM1}, and his paper was modified
subsequently. Beyond the preliminary level, for correct computation with
ghosts, the results of the present paper are needed.

The definition of the fermionic Moyal product outlined above is given in
section \ref{sec:partI}. Because of its technical nature, we first summarize
the fundamental formulae at the beginning of the section, and explain the
full detail in the subsections. Readers who wish to skip the details of the
derivation can proceed to later sections by skipping the latter part. In
subsection \ref{sec:sine}, we first discuss the issue of boundary conditions
in general for fermions and the corresponding regularization scheme. We also
give a brief review of \cite{BM2} in appendix \ref{sec:review_app}. Together
with it, this paper provides the basic formulae for both the ghost and
matter sectors in a compact form. We then discuss the mapping from Witten's
star to Moyal's star in subsection \ref{sec:odd_Moyal}. Finally we apply
these techniques to the actual $bc$ ghost system in subsection \ref
{sec:Moyalbc}, and derive the correspondence between the conventional
oscillators and Moyal variables in subsection \ref{oscillators}.

After this preparation, in section \ref{sec:partII}, we define the monoid
algebra among gaussian string fields, which provides a useful tool for
computations (this is almost a group, except for inverse). We compute the
star product of $n$ monoid elements, which as a by-product give the Neumann
coefficients in the ghost sector. These were conjectured in \cite{BM2} by
using a nontrivial relation with the Neumann coefficients in the matter
sector. In this paper, we derive them directly from the fermionic Moyal
product. They satisfy the Gross-Jevicki- nonlinear relations exactly for any
frequencies $\kappa _{e},\kappa _{o}$ and any $N$. This fact confirms the
consistency of our construction including the midpoint prescription. As in
the matter sector, they are simple functions of the matrix $t_{eo}$ for
any $n$-point vertex, and they can be related to the corresponding
matter Neumann coefficients by the simple procedure of replacing the
matrix $t$ by its inverse.

We also give the direct numerical comparison between the analytic form of
the Neumann coefficients $\mathcal{M}$ obtained from conformal field theory
and our algebraic expression. We confirmed that the approximate value for
finite $N$ converges to its exact value as $N\rightarrow \infty $ with the
following universal behavior, ${\mathcal{M}_{nm}(N)}/{\mathcal{M}
_{nm}(cft)}\sim 1+a_{nm}N^{-\alpha }$ where the exponent is approximately 
$\alpha \sim 1.33$ for matter sector and $\alpha \sim 0.67$ for the ghost
sector for any components of the Neumann coefficients. While this analysis
is of different nature from the other parts in this paper, it is included
here since it gives strong support on the consistency of MSFT in \cite{BM2}
and this paper. It also provides the basis for numerical computation of MSFT
in our future study.

In section {\ref{sec:partIII} we apply the formalism. First}, we present the
derivation of the open string field action in the Siegel gauge by including
both matter and ghost fields. This action was the starting point in our
recent work \cite{BKM1,BKM2} where we used the results of the present paper
without providing the details. Finally, we also compute the ghost
contribution to the Feynman graphs for the four-point scattering amplitude
for off-shell tachyons, whose matter sector was discussed in \cite{BKM1}.

\section{Moyal's star from Witten's star in fermionic ghost sector \label
{sec:partI}}

In this section we construct the map from Witten's star product to the Moyal
star product for fermionic variables. The fermionic version of the Moyal $%
\star $ product was called ``anti-Moyal star product''\ in the literature 
\cite{BFFLS}. More general star products have also been considered in the
context of deformation quantization of super-Poisson brackets \cite
{Bordemann}. The anti-Moyal star product will simply be referred to as the
Moyal star product in the following. A first basic construction for the
ghost $bc$ system, following the one in the matter sector \cite{B1}, was
previously discussed in \cite{Erler}, while the correct treatment of the
midpoint was first given in \cite{BKM1}. In this section we give the
complete treatment, including the consistent regularization with the matter
sector.

We also define the even basis of ghost modes that is most transparent for
our computations, after defining some other bases as well. We first
summarize the main results of this rather lengthy and technical section:

\begin{itemize}
\item  The Moyal $\star $ acts on fermionic ghost modes $\xi \equiv
(x_{o},p_{o},y_{o},q_{o})$ ($o=1,3,5,\cdots ,2N-1$) as 
\begin{equation}
(A\star B)(x_{o},p_{o},y_{o},q_{o})=A\exp \left( {\frac{\theta ^{\prime }}{2}%
}\sum_{o>0}\left( {\frac{\overleftarrow{\partial }}{\partial x_{o}}}{\frac{%
\overrightarrow{\partial }}{\partial p_{o}}}+{\frac{\overleftarrow{\partial }%
}{\partial y_{o}}}{\frac{\overrightarrow{\partial }}{\partial q_{o}}}+{\frac{%
\overleftarrow{\partial }}{\partial p_{o}}}{\frac{\overrightarrow{\partial }%
}{\partial x_{o}}}+{\frac{\overleftarrow{\partial }}{\partial q_{o}}}{\frac{%
\overrightarrow{\partial }}{\partial y_{o}}}\right) \right) B\,,
\label{e_M1}
\end{equation}
where $\theta ^{\prime }$ is a parameter which absorbs units, and if
desired, could be absorbed away by a rescaling of the variables which
amounts to a choice of units. We note the canonical structure $\left\{
x_{o},p_{o^{\prime }}\right\} _{\star }=\theta ^{\prime }\delta _{oo^{\prime
}}=\left\{ y_{o},q_{o^{\prime }}\right\} .$ This odd basis $\left( o\right) $
of ghost modes, which was used in \cite{BKM1}, is naturally defined in the
process of mapping the Witten star to the Moyal star, including the
treatment of the midpoint. However, a more transparent basis that is more
parallel to the matter sector, which simplifies the overall formalism, is
obtained by rewriting the odd basis, through the following linear \textit{%
canonical} transformation, in terms of an even basis $%
x_{e}^{b},p_{e}^{b},x_{e}^{c},p_{e}^{c}$ ($e=2,4,\cdots ,2N$), where the
labels $b,c$ refer to the modes of the usual $b,c$ ghosts 
\begin{equation}
x_{e}^{b}:=\kappa _{e}^{-1}\sum_{o>0}S_{eo}x_{o}\,,\quad p_{e}^{b}:=\kappa
_{e}\sum_{o>0}S_{eo}p_{o}\,,\quad x_{e}^{c}:=\sum_{o>0}T_{eo}y_{o}\,,\quad
p_{e}^{c}:=\sum_{o>0}q_{o}R_{oe}\,.
\end{equation}
This even basis is different than the one defined in \cite{Erler}. The Moyal
product is rewritten as 
\begin{equation}
(A\star B)(x_{e}^{b},p_{e}^{b},x_{e}^{c},p_{e}^{c})=A\exp \left( {\frac{%
\theta ^{\prime }}{2}}\sum_{e>0}\left( {\frac{\overleftarrow{\partial }}{%
\partial x_{e}^{b}}}{\frac{\overrightarrow{\partial }}{\partial p_{e}^{b}}}+{%
\frac{\overleftarrow{\partial }}{\partial x_{e}^{c}}}{\frac{\overrightarrow{%
\partial }}{\partial p_{e}^{c}}}+{\frac{\overleftarrow{\partial }}{\partial
p_{e}^{b}}}{\frac{\overrightarrow{\partial }}{\partial x_{e}^{b}}}+{\frac{%
\overleftarrow{\partial }}{\partial p_{e}^{c}}}{\frac{\overrightarrow{%
\partial }}{\partial x_{e}^{c}}}\right) \right) B\,.  \label{e_M2}
\end{equation}
We note the canonical structure $\left\{ x_{e}^{b},p_{e^{\prime
}}^{b}\right\} _{\star }=\theta ^{\prime }\delta _{ee^{\prime }}=\left\{
x_{e}^{c},p_{e^{\prime }}^{c}\right\} _{\star }.$ The linear transformation
matrices, $T,R$ (and matrices $U$ and vectors $v,w$ which appear in the
following) were defined in \cite{BM2}, while the matrix $S$ appears for the
first time in this paper. Their properties are derived explicitly in
subsection \ref{sec:sine}. They play a central role in MSFT since they
define a Bogoliubov transformation from the oscillators in the operator
formalism to the Moyal coordinates, and thus carry essential information
about string theory. For instance, see Eqs.(\ref{e_bra},\ref{e_def_M_lambda}%
) in the next paragraph. For the moment, we just mention that they satisfy
the inverse properties $TR=RT=S\bar{S}=\bar{S}S=1$ which prove that the two
versions of the star product (\ref{e_M1}) and (\ref{e_M2}) are canonically
equivalent. Throughout this paper, the bar ($\bar{\,\,}$) means the
transpose of matrices or vectors. The relation of split strings to the Moyal
star is discussed in subsection \ref{sec:odd_Moyal}.

\item  In the operator formulation, the ghost sector of the string field $%
|\Psi \rangle $ is represented in the Fock space of the $bc$ ghost
oscillators. The transformation to the Moyal field $A\left( \xi \right) $ as
a function of noncommutative coordinates $\xi =\left( x,p\right) $ is
obtained through the Fourier transformation \cite{B1} of the coordinate
representation of $|\Psi \rangle $. The whole procedure is more neatly
expressed, as in the matter sector \cite{BM2}, by the inner product with a
particular bra state $\langle \xi _{0},\xi _{1},\xi _{2}|$ in the Fock
space, where $\bar{\xi}_{1}=\left( x_{o},p_{o}\right) ,\bar{\xi}_{2}=\left(
y_{o},q_{o}\right) $ are the noncommutative fermionic coordinates, and $\xi
_{0}$ is a fermionic variable related to the zero mode dependence of $|\Psi
\rangle $ 
\begin{eqnarray}
&&\langle \xi _{0},\xi _{1},\xi _{2}| =-2^{-2N}(1+\bar{w}w)^{-{\frac{1}{4}}%
}\langle \Omega |\hat{c}_{-1}e^{-\xi _{0}(\hat{c}_{0}-\sqrt{2}\bar{w}\hat{c}%
_{e})}e^{\hat{c}_{e}\hat{b}_{e}-\hat{c}_{o}\hat{b}_{o}-2i\bar{\xi}
_{1}M_{0}^{(o)}\xi_{2}-\bar{\xi}_{1}{\lambda}_{1}-\bar{\xi}_{2}{\lambda}_{2}}
,~~~~
\label{e_bra} \\
&&\hat{A}(\xi _{0},\xi _{1},\xi _{2}) =\langle \xi _{0},\xi _{1},\xi
_{2}|\Psi \rangle \,.
\end{eqnarray}
Here $\langle \Omega |$ is the SL$(2,R)$ invariant bra Fock vacuum, $\hat{b}%
_{n},\hat{c}_{n}$ are the conventional ghost oscillators, and the matrices $%
M_{0}^{\left( o\right) }$ and $\lambda $ are defined as 
\begin{equation}
M_{0}^{\left( o\right) }=\left( 
\begin{array}{cc}
{\frac{1}{2}}\mathbf{1}_{o} & 0 \\ 
0 & {\frac{2}{{\theta ^{\prime }}^{2}}}\left( \bar{S}\bar{R}\right)
_{oo^{\prime }}
\end{array}
\right) \,,~\lambda _{1}=\left( 
\begin{array}{c}
-i\sqrt{2}\hat{c}_{o} \\ 
\frac{2i}{\theta ^{\prime }}\bar{S}_{oe}\left( -\sqrt{2}\hat{b}_{e}+w_{e}\xi
_{0}\right) 
\end{array}
\right) \,,~\lambda _{2}=\left( 
\begin{array}{c}
-\sqrt{2}\hat{b}_{o} \\ 
-{\frac{2\sqrt{2}}{\theta ^{\prime }}}R_{oe}\hat{c}_{e}
\end{array}
\right)\,.   \label{e_def_M_lambda}
\end{equation}
The matrices $w_{e},S_{eo},R_{oe}$ are functions of the oscillator
frequencies $\kappa _{e},\kappa _{o}$ as given below. Through Eqs.(\ref{e_M1}%
,\ref{e_bra}) we map Witten's star ($\star ^{W}$) into the Moyal's star $%
\star $ as follows 
\begin{eqnarray}
\langle \xi _{0},\xi |\Psi _{1}\star ^{W}\Psi _{2}\rangle  &\sim &\langle
\xi _{0},\xi |\Psi _{1}\rangle \star \langle \xi _{0},\xi |\Psi _{2}\rangle
\,\,, \\
|\Psi _{1}\star ^{W}\Psi _{2}\rangle _{3} &=&{}_{1}\langle \Psi
_{1}|{}_{2}\langle \Psi _{2}|V_{3}\rangle _{123},\quad {}_{1}\langle \Psi
_{1}|={}_{14}\langle V_{2}|\Psi _{1}\rangle _{4},\,{}_{2}\langle \Psi
_{2}|={}_{25}\langle V_{2}|\Psi _{2}\rangle _{5}\,.  \label{eq:Witten's*}
\end{eqnarray}
We note that the product is local in $\xi _{0},~$while $\xi _{0}$ plays a
similar role to the midpoint coordinate $\bar{x}^{\mu }$ in the matter
sector. The Moyal star reproduces correctly the three string vertex 
$|V_{3}\rangle$ and the reflector $|V_{2}\rangle $ of the operator
formalism \cite{GJ}. The details are given in subsection
\ref{sec:Moyalbc}. The precise correspondence including the zero
mode is in section {\ref{sec:partII}}.

\item  Eq.(\ref{e_bra}) is enough to derive the connection between the
conventional operator formalism and the Moyal star formalism. For example,
the action of the standard oscillators on the Fock space field $|\Psi
\rangle $ can be rewritten in terms of their Moyal images acting on the
field $\hat{A}=\langle \xi _{0},\xi _{1},\xi _{2}|\Psi \rangle $ through the
star product, as follows 
\begin{eqnarray}
\hat{b}_{0}|\Psi \rangle  &\leftrightarrow &-\xi _{0}\hat{A}\,,  \notag \\
\hat{c}_{0}|\Psi \rangle  &\leftrightarrow &\left( -{\frac{\partial }{%
\partial \xi _{0}}}+{\frac{\theta ^{\prime }}{2}}\bar{v}{\frac{\partial }{%
\partial q_{o}}}\right) \hat{A}\,,  \notag \\
\hat{b}_{o}|\Psi \rangle  &\leftrightarrow &{\frac{1}{\sqrt{2}}}\left( \beta
_{o}^{b}\star \hat{A}-(-1)^{|A|}\hat{A}\star \beta _{-o}^{b}\right) \,, 
\notag \\
\hat{c}_{o}|\Psi \rangle  &\leftrightarrow &{\frac{1}{\sqrt{2}}}\left( \beta
_{o}^{c}\star \hat{A}+(-1)^{|A|}\hat{A}\star \beta _{-o}^{c}\right) \,, \\
\hat{b}_{e}|\Psi \rangle  &\leftrightarrow &{\frac{1}{\sqrt{2}}}\left( \beta
_{e}^{b}\star \hat{A}+(-1)^{|A|}\hat{A}\star \beta _{-e}^{b}\right)
+w_{e}^{\prime }\,\xi _{0}\hat{A}\,,  \notag \\
\hat{c}_{e}|\Psi \rangle  &\leftrightarrow &{\frac{1}{\sqrt{2}}}\left( \beta
_{e}^{c}\star \hat{A}-(-1)^{|A|}\hat{A}\star \beta _{-e}^{c}\right) \,, 
\notag
\end{eqnarray}
where $\beta _{o}^{b},\beta _{o}^{c}$ are fields in Moyal space which obey
oscillator relations under the $\star $ product 
\begin{equation}
\beta _{o}^{b}:={\frac{1}{2}}\left( {\frac{2}{{\theta ^{\prime }}}}%
q_{|o|}-i\epsilon (o)x_{|o|}\right) ,\;\beta _{o}^{c}:={\frac{1}{2}}\left(
y_{|o|}-i{\frac{2}{{\theta ^{\prime }}}}\epsilon (o)p_{|o|}\right)
\,,\;\{\beta _{o}^{b},\beta _{o^{\prime }}^{c}\}_{\star }=\delta
_{o+o^{\prime }}\,.~~~~  \label{oddmode_fields}
\end{equation}
On the other hand $\beta _{e}^{b},\beta _{e}^{c}$ are not independent from
the $\beta _{o}^{b},\beta _{o}^{c}.$ Rather, they are their Bogoliubov
transforms which obey the following relations 
\begin{equation}
\beta _{e}^{b}=\sum_{o}\beta _{o}^{b}\,U_{-o,e}^{-1}\,,\;\beta
_{e}^{c}=\sum_{o}U_{e,-o}\beta _{o}^{c}\,,\;\{\beta _{e}^{b},\beta
_{e^{\prime }}^{c}\}_{\star }=\delta _{e+e^{\prime }}\,,  \label{even_fields}
\end{equation}
and $\left\{ \beta _{-o}^{b},\beta _{e}^{c}\right\}_{\star} =U_{e,-o}$ 
where the
matrix $U_{e,-o}$ will be given below. With these formulas, one can directly
translate operators in Fock space into their images which act in Moyal
space. The proof of these formulas and some variants are discussed in
subsection \ref{oscillators}. In this way we derive the explicit form of the
Virasoro operator $L_{0}$ which acts in Moyal space 
\begin{eqnarray}
L_{0} &=&\sum_{k=1}^{2N}\kappa _{k}(\hat{\beta}_{-k}^{b}\hat{\beta}_{k}^{c}+%
\hat{\beta}_{-k}^{c}\hat{\beta}_{k}^{b}) \\
&=&\sum_{k=1}^{2N}\kappa _{k}+i\sum_{o>0}\kappa _{o}\left( x_{o}y_{o}+{\frac{%
\partial }{\partial x_{o}}}{\frac{\partial }{\partial y_{o}}}\right)
+i\sum_{o>0}\kappa _{o}\left( {\frac{4}{{\theta ^{\prime }}^{2}}}p_{o}q_{o}+{%
\frac{{\theta ^{\prime }}^{2}}{4}}{\frac{\partial }{\partial p_{o}}}{\frac{%
\partial }{\partial q_{o}}}\right)   \notag \\
&&+{\frac{4i}{{\theta ^{\prime }}^{2}}}(1+\bar{w}w)\left( \sum_{o>0}\kappa
_{o}v_{o}p_{o}\right) \left( \sum_{o^{\prime }>0}v_{o^{\prime }}q_{o^{\prime
}}\right) +{\frac{2i}{{\theta ^{\prime }}}}(1+\bar{w}w)\left(
\sum_{o>0}v_{o}\kappa _{o}p_{o}\right) \xi _{0}\, . \notag
\end{eqnarray}
This was used in \cite{BKM1,BKM2}. Here $\hat{\beta}_{k}^{b},\hat{\beta}%
_{k}^{c}$ are not the Moyal fields $\beta _{o}^{b},\beta _{o}^{c}$ or $\beta
_{e}^{b},\beta _{e}^{c}$ given in Eqs.(\ref{oddmode_fields},\ref{even_fields}%
); rather, they are differential operators that obey the standard oscillator
relations, and which are derived from the star products of the fields $\beta
_{o}^{b},\beta _{o}^{c}$ or $\beta _{e}^{b},\beta _{e}^{c}$ as will be shown
below.
\end{itemize}

\subsection{Half string formalism and regularization \label{sec:sine}}

We start from full string functions $\psi (\sigma ),$ $0\leq \sigma \leq \pi 
$ with Dirichlet boundary conditions at $\sigma =0,\pi ,$ and discuss their
split string formulation in the interval $0\leq \sigma \leq \pi /2$. Such
functions have a Fourier expansion with only \textit{sine} modes in the full
string formalism. By contrast, the corresponding problem in the matter
sector involved only \textit{cosine} modes because of Neumann boundary
conditions at $\sigma =0,\pi $. We collect the basic formulae in the
appendix \ref{sec:cosine}. The essential step was the construction of the
regularization \cite{BM1,BM2} which is needed to avoid the associativity
anomaly. We give a regularization of the ghost sector for the sine mode
expansion which is compatible with the previous results.

A full string function $\psi (\sigma )$ which satisfies Dirichlet boundary
conditions 
$\psi (0)=\psi (\pi )=0$ is expanded as 
\begin{equation}
\psi (\sigma )=\sqrt{2}\sum_{n=1}^{\infty }\psi _{n}\sin n\sigma \,,\;\psi
_{n}={\frac{\sqrt{2}}{\pi }}\int_{0}^{\pi }d\sigma \psi (\sigma )\sin
n\sigma \,.  \label{eq_psi_exp}
\end{equation}
We decompose such a field $\psi (\sigma )$ into left half $l(\sigma )$ and
right half $r(\sigma )$ as follows 
\begin{equation}
\psi (\sigma )=\left\{ 
\begin{array}{lc}
l(\sigma ) & \quad 0\leq \sigma \leq {\frac{\pi }{2}} \\ 
r(\pi -\sigma ) & \quad {\frac{\pi }{2}}\leq \sigma \leq \pi
\end{array}
\right. \,,\;\psi _{n}={\frac{\sqrt{2}}{\pi }}\int_{0}^{\frac{\pi }{2}%
}d\sigma (l(\sigma )-(-1)^{n}r(\sigma ))\sin n\sigma \,.  \label{eq:psilrExp}
\end{equation}
As we have seen in \cite{B1,BM1}, we have some arbitrariness in the choice
of the boundary condition for the split string functions $l(\sigma
),r(\sigma )$ at the midpoint $\sigma =\pi /2$. They can be expanded using
either odd or even modes according to the two possible choices of the
boundary condition at the midpoint $\sigma ={\frac{\pi }{2}}$ as discussed
below.

\subsubsection{Dirichlet at end point, Neumann at midpoint (DN) \label%
{sec:sinDN}}

First we consider 
Neumann boundary conditions at the midpoint $\sigma =\pi /2,$ while we have
Dirichlet boundary conditions at the end point for the split string
functions $l(\sigma ),r(\sigma )$: 
\begin{equation}
l(0)=r(0)=0\,,\qquad l^{\prime }(\pi /2)=r^{\prime }(\pi /2)=0\,.
\end{equation}
We can expand them using odd sine modes $o=1,3,5,\cdots $ 
\begin{eqnarray}
&&l(\sigma )=\sqrt{2}\sum_{o=1}^{\infty }l_{o}\sin o\sigma \,,\qquad l_{o}={%
\frac{2\sqrt{2}}{\pi }}\int_{0}^{\frac{\pi }{2}}d\sigma \,l(\sigma )\sin
o\sigma \,,  \label{eq:lexpDN} \\
&&r(\sigma )=\sqrt{2}\sum_{o=1}^{\infty }r_{o}\sin o\sigma \,,\qquad r_{o}={%
\frac{2\sqrt{2}}{\pi }}\int_{0}^{\frac{\pi }{2}}d\sigma \,r(\sigma )\sin
o\sigma \,.  \label{eq:rexpDN}
\end{eqnarray}
By comparing the mode expansions
Eqs.(\ref{eq_psi_exp},\ref{eq:lexpDN},\ref{eq:rexpDN}) with
(\ref{eq:psilrExp}), we obtain the correspondence between
the full and split string variables, 
\begin{equation}
l_{o}=\psi _{o}+\bar{S}_{oe}\psi _{e}\,,~~~r_{o}=\psi _{o}-\bar{S}_{oe}\psi
_{e}\,,  \label{eq:sinDN_psi2lr}
\end{equation}
or the inverse 
\begin{equation}
\psi _{e}={\frac{1}{2}}S_{eo}(l_{o}-r_{o})\,,~~~\psi _{o}={\frac{1}{2}}%
(l_{o}+r_{o}),  \label{eq:sinDN_lr2psi}
\end{equation}
where $e=2,4,6,\cdots ,$ and the matrix $S_{eo}$ is given by 
\begin{equation}
S_{eo}={\frac{4}{\pi }}\int_{0}^{\frac{\pi }{2}}d\sigma \sin e\sigma \sin
o\sigma ={\frac{4i^{o-e+1}e}{\pi (e^{2}-o^{2})}}\,.  \label{S}
\end{equation}
The above mappings are consistent because $S_{eo}$ is an orthogonal matrix: 
\begin{equation}
\bar{S}S=S\bar{S}=1\,.  \label{eq:tilRconsistency}
\end{equation}
The continuity condition at the midpoint $\psi (\pi /2)=l(\pi /2)=r(\pi /2)$
is 
\begin{equation}
\sum_{o=1}^{\infty }{\tilde{w}}_{o}\psi _{o}=\sum_{o=1}^{\infty }{\tilde{w}}%
_{o}l_{o}=\sum_{n=1}^{\infty }{\tilde{w}}_{o}r_{o}  \label{CONTI}
\end{equation}
where we defined the odd vector $\tilde{w}$ associated with the midpoint 
\begin{equation}
\tilde{w}_{o}=\sqrt{2}\sin {\frac{o\pi }{2}}=\sqrt{2}\,i^{o-1}\,.
\label{eq:tilwinfty}
\end{equation}
Eq.(\ref{CONTI}) holds thanks to the identity $(S{\tilde{w}}
)_{e}=\sum_{o=1}^{\infty }S_{eo}{\tilde{w}}_{o}=0.$ This equation implies
that $S$ has a singular eigenvector even though it has an inverse, which is
just its transpose $\bar{S},$ as stated in (\ref{eq:tilRconsistency}). This
esoteric relation is possible because $S$ is an infinite matrix. However it
causes an associativity anomaly with respect to matrix products, which in
turn feeds into associativity anomaly of string field star products, as
discussed in \cite{BM1}.

An example of the associativity anomaly is $\bar{S}(S\tilde{w})=0,$
but $(\bar{S}S)\tilde{w}=\tilde{w}.$ 
Each single sum indicated by the parentheses
has a unique answer, but the double sums are ambiguous. The reason is that,
due to infinite sums, in the first expression there are terms of the form $%
\infty /\infty $ which are ambiguous. Since these matrices appear in many
physical computations we must give an unambiguous definition of the matrix
product. We will resolve this ambiguity in computations successfully by a
regularization procedure.

\subsubsection{Dirichlet at end point, Dirichlet at midpoint (DD) \label%
{sec:sinDD}}

We consider another possibility: we define the midpoint of the string ${\bar{%
\psi}}:=\psi (\pi /2),$ and impose Dirichlet boundary conditions at both $%
\sigma =0,\pi /2$ on the split string functions $l(\sigma ),r(\sigma )$, 
\begin{equation}
l(0)=r(0)=0\,,\quad l(\pi /2)=r(\pi /2)={\bar{\psi}}.
\end{equation}
We note that the midpoint value ${\bar{\psi}}$ is an additional degree of
freedom in the split string basis and cannot be chosen arbitrarily. We
expand $l(\sigma ),r(\sigma )$ using even sine modes, $e=2,4,6,\cdots $ 
\begin{eqnarray}
&&l(\sigma )={\frac{2}{\pi }}\sigma {\bar{\psi}}+\sqrt{2}\sum_{e=2}^{\infty
}l_{e}\sin e\sigma \,,~~~~l_{e}={\frac{2\sqrt{2}}{\pi }}\int_{0}^{\frac{\pi 
}{2}}d\sigma \left( l(\sigma )-{\frac{2}{\pi }}\sigma \bar{\psi}\right) \sin
e\sigma \,,  \label{eq:lexpDD} \\
&&r(\sigma )={\frac{2}{\pi }}\sigma {\bar{\psi}}+\sqrt{2}\sum_{e=2}^{\infty
}r_{e}\sin e\sigma \,,~~~~r_{e}={\frac{2\sqrt{2}}{\pi }}\int_{0}^{\frac{\pi 
}{2}}d\sigma \left( r(\sigma )-{\frac{2}{\pi }}\sigma \bar{\psi}\right) \sin
e\sigma \,.  \label{eq:rexpDD}
\end{eqnarray}
Again, by comparing the mode expansions Eqs.(\ref{eq_psi_exp},\ref{eq:lexpDD}%
,\ref{eq:rexpDD}) with (\ref{eq:psilrExp}), the correspondence between
split and full string variables is obtained 
\begin{equation}
\bar{\psi}={\tilde{w}}_{o}\psi _{o}\,,~~~l_{e}=\psi _{e}+\tilde{T}_{eo}\psi
_{o}\,,~~~r_{e}=-\psi _{e}+\tilde{T}_{eo}\psi _{o}\,,
\label{eq:sinDD_psi2lr}
\end{equation}
or the inverse 
\begin{equation}
\psi _{e}={\frac{1}{2}}(l_{e}-r_{e})\,,~~~\psi _{o}={\tilde{u}}_{o}\bar{\psi}%
+{\frac{1}{2}}\bar{S}_{oe}(l_{e}+r_{e})\,,  \label{eq:sinDD_lr2psi}
\end{equation}
where 
\begin{eqnarray}
{\tilde{T}}_{eo} &=&{\frac{4o^{2}~i^{o-e+1}}{\pi e(e^{2}-o^{2})}}\,=S_{eo}+{%
\tilde{v}}_{e}{\tilde{w}}_{o}=\frac{1}{e^{2}}S_{eo}o^{2}, \\
{\tilde{u}}_{o} &=&{\frac{4\sqrt{2}}{\pi ^{2}}}\int_{0}^{\frac{\pi }{2}%
}d\sigma \sigma \sin o\sigma ={\frac{4\sqrt{2}\,i^{o-1}}{\pi ^{2}o^{2}}}\,, \\
{\tilde{v}}_{e} &=&-{\frac{4\sqrt{2}}{\pi ^{2}}}\int_{0}^{\frac{\pi }{2}%
}d\sigma \sigma \sin e\sigma ={\frac{2\sqrt{2}\,i^{e}}{e\pi }}\,.
\label{eq:tilTuvinfty}
\end{eqnarray}
The maps $(\bar{\psi},l_{e},r_{e})\leftrightarrow (\psi _{n})$ are
consistent by the relation 
\begin{equation}
{\tilde{u}}_{o}{\tilde{w}}_{o}=1,\quad {\tilde{T}}\bar{S}=1\,,\quad {\tilde{T%
}}{\tilde{u}}=0,\quad S\tilde{w}=0\,.
\end{equation}

We can prove the following relations among infinite matrices $S,{\tilde{T}}$
and vectors ${\tilde{u}},{\tilde{v}},{\tilde{w}}$ by straightforward
computation: 
\begin{eqnarray}
&&S=\kappa _{e}^{2}\,{\tilde{T}}\,\kappa _{o}^{-2}\,,\quad S={\tilde{T}}-{%
\tilde{v}}{\bar{\tilde{w}}}\,,\quad \bar{S}{\tilde{v}}=-{\tilde{u}}\,,\quad S%
{\tilde{u}}=-{\tilde{v}}\,,\quad \bar{{\tilde{T}}}\tilde{v}=-{\tilde{u}}+{%
\frac{1}{3}}{\tilde{w}}\,,  \label{ss} \\
&&\bar{S}S=S\bar{S}={\tilde{T}}\bar{S}=1,\quad \bar{S}{\tilde{T}}=1-{\tilde{u%
}}{\bar{\tilde{w}}}\,,\quad S\tilde{w}={\tilde{T}}{\tilde{u}}=0\,,\quad {%
\bar{\tilde{w}}}{\tilde{u}}=1\,,\quad {\bar{\tilde{v}}}{\tilde{v}}={\bar{%
\tilde{u}}}{\tilde{u}}={\frac{1}{3}}\,,~~~  \label{eq:til_mat_rel_infty}
\end{eqnarray}
where $\kappa _{e},\kappa _{o}$ are the diagonal matrices $\kappa
_{e}=diag(2,4,6\cdots )$ and $\kappa _{o}=diag(1,3,5,\cdots )$. This algebra
is similar to the one among the infinite matrices $T,R,v,w$ which appeared
in the matter sector \cite{BM1}, where a full string function $\psi (\sigma
) $ is expanded in terms of cosine modes (see \S \ref{sec:cosine}).

\subsubsection{Regularization \label{sec:regularization_ghost}}

In the split string formulation given in \S \ref{sec:cosND} \S \ref
{sec:cosNN} \S \ref{sec:sinDN} \S \ref{sec:sinDD}, we encountered a set of
infinite dimensional matrices $T,R,S,{\tilde{T}}$ and vectors $w,v,{\tilde{w}%
},{\tilde{v}},{\tilde{u}}$. These represent Bogoliubov transformations
between odd and even modes, with $(T,R,w,v)$ appearing when the full string
is expanded in terms of cosine modes, and $(S,\tilde{T},\tilde{w},\tilde{v},%
\tilde{u})$ appearing when the full string is expanded in terms of sine
modes. Such transformations are essential in the Moyal formulation since
they carry basic information about string theory. We note that, the Moyal
star product itself, which is applied independently for each mode, has no
specific information about string theory, and as such is a more general
structure.

In the analysis of the matter sector \cite{BM1} as well as the sine mode
expansion given so far, there appears an associativity anomaly in the matrix
algebra of these matrices. This originates from the infinite dimensionality
of these matrices. It produces ambiguities in computations in string field
theory. To have a well defined theory, it is mandatory to define a deformed,
unambiguous, associative algebra that preserves the basic matrix algebraic
structure of these matrices \cite{BM1,BM2}. Such a deformation contains a
parameter $N,$ that corresponds to the rank of the matrices. The original
definition of these matrices given above is reproduced by taking the limit $%
N\rightarrow \infty $ of this parameter. All computations in the open string
field theory are performed unambiguously with finite $N,$ and the correct
value in string theory is obtained at the end of the computation by taking
the limit $N\rightarrow \infty $. This is the basic strategy for practical
computations in the MSFT proposal.

The deformed set of matrices for $T,R,w,v$ that have correctly reproduced
string theory was proposed in \cite{BM1,BM2}. Since an additional set of
matrices $S,\tilde{T},\tilde{w},\tilde{v},\tilde{u}$ have appeared in the
ghost sector we need to obtain their deformation consistently with the
matter sector. For that purpose, we start from the infinite dimensional
matrix $U$ and vectors $w^{\prime },v^{\prime }$ defined in \cite{BM2}, 
\begin{equation}
U_{-e,o}={\frac{2}{\pi }}{\frac{i^{o-e-1}}{o-e}}\,,\quad U_{-o,e}^{-1}={%
\frac{2}{\pi }}{\frac{e}{o}}{\frac{i^{o-e-1}}{o-e}}\,,\quad w_{e}^{\prime
}=i^{-e+2}\,,\quad v_{o}^{\prime }={\frac{2}{\pi }}{\frac{i^{o-1}}{o}}\,,
\label{eq:UU^-1limit}
\end{equation}
where $e$ ($o$) now run over both positive and negative integers $\pm 2,\pm
4,\cdots $ (resp. $\pm 1,\pm 3,\cdots $). In single sums these matrices
satisfy 
\begin{equation}
\sum_{o}U_{-e,o}U_{-o,e^{\prime }}^{-1}=\delta _{e,e^{\prime }}\,,\quad
\sum_{e\neq 0}U_{-o,e}^{-1}U_{-e,o^{\prime }}=\delta _{o,o^{\prime }}\,,
\label{eq_UUinv}
\end{equation}
which we denote $UU^{-1}=1_{e}$, $U^{-1}U=1_{o}$  for short in the
following. More importantly, there exists the following matrix relations
among them, 
\begin{equation}
U^{-1}={\kappa _{o}^{\prime }}^{-1}{\bar{U}}\kappa _{e}^{\prime }\,,\quad
U^{-1}=\bar{U}+v^{\prime }\bar{w}^{\prime }\,,\quad v^{\prime }=\bar{U}%
w^{\prime }\,,\quad w^{\prime }=\bar{U}^{-1}v^{\prime }\,,  \label{eq:Udef}
\end{equation}
where $\kappa _{e}^{\prime }=diag(\cdots ,-4,-2,2,4,\cdots )$, and $\kappa
_{o}^{\prime }=diag(\cdots ,-3,-1,1,3,\cdots )$ are the diagonal matrices
which specify the spectrum$.$ These relations will be used as the defining
relations. From them it is possible to derive the matrices themselves, as
given in Eq.(\ref{eq:UU^-1limit}). Therefore, they will be used as the basic
relations that are also satisfied by the deformed matrices, as given below.
The first relation implies that $U$ defines an invertible Bogoliubov
transformation between even and odd spectra (see also Eq.(\ref{even_fields}%
)). The second relation shows that the transformation $U$ is almost
orthogonal except for the vectors $v^{\prime },w^{\prime }$ which are
associated with the midpoint mode. Finally the last two define the relation
between the vectors.

On the other hand, the matrices (\ref{eq:UU^-1limit}) also satisfy 
\begin{equation}
U\bar{U}=1_{e},\quad Uv^{\prime }=0,\quad \bar{v}^{\prime }v^{\prime }=1\,,
\label{e_UUanomaly}
\end{equation}
which break the associativity \cite{BM1,BM2}. Therefore, these relations
will be deformed in the regulated theory, as seen below. Of course these
equations will hold when the regulator is removed.

All the other matrices are written in terms of $U,w',v'$, (for $e,o>0$), 
\begin{eqnarray}
&&T_{eo}=U_{-e,o}+U_{e,o}\,,\quad R_{oe}=U_{-o,e}^{-1}+U_{o,e}^{-1}\,,\quad
w_{e}=\sqrt{2}w_{e}^{\prime },\quad v_{o}=\sqrt{2}v_{o}^{\prime }\,,
\label{eqs_defns1} \\
&&S_{eo}=U_{-e,o}-U_{e,o}=U_{-o,e}^{-1}-U_{o,e}^{-1}\,,
\quad \tilde{T}_{eo}=\kappa _{e}^{-1}T\kappa _{o}\,,  \label{eqs_defns2} \\
&&\tilde{u}={\frac{2}{\pi }}\kappa _{o}^{-1}v\,,
\quad \tilde{w}={\frac{\pi }{2}%
}\kappa _{o}v\,,\quad \tilde{v}=-{\frac{2}{\pi }}\kappa _{e}^{-1}w\,,
\label{eqs_defns3}
\end{eqnarray}
where $\kappa _{o}$ and $\kappa _{e}$ are restrictions of $\kappa
_{o}^{\prime }$ and $\kappa _{e}^{\prime }$ to the positive sector. These
definitions in terms of $U,w^{\prime },v^{\prime }$ together with the
relations (\ref{eq:Udef}) among $U,w^{\prime },v^{\prime }$ are
sufficient to derive all the relations among $T,R,w,v,S,\tilde{T},\tilde{u},%
\tilde{v},\tilde{w}$. Therefore, we may use these relations as the \emph{%
definitions} of these matrices and vectors even when we use the
regularization of $U,w^{\prime },v^{\prime }$.

In \cite{BM1,BM2,BKM1}, the regularization of $U,v^{\prime },w^{\prime }$ is
given explicitly. We truncate the size of $U,v^{\prime },w^{\prime }$ to $2N$
while keeping their property of Bogoliubov transformation between even and
odd spectrum. It turns out that one may take the even and odd frequencies $%
\kappa _{e},\kappa _{o}$ as arbitrary functions of the positive integers $%
\left( e,o\right) ,$ while keeping the reflection property of $\kappa
_{e,o}^{\prime }$ to extend the definition to negative integers $\kappa
_{-e}^{\prime }=-\kappa _{e}^{\prime }$, $\kappa _{-o}^{\prime }=-\kappa
_{o}^{\prime }$. Therefore we put 
\begin{equation}
\kappa _{e}^{\prime }=\epsilon (e)\kappa _{|e|}\,,\quad \kappa _{o}^{\prime
}=\epsilon (o)\kappa _{|o|}\,.
\label{eq:kekoprime}
\end{equation}
We suppose implicitly that $\kappa _{e}^{\prime }\neq 0,\kappa _{o}^{\prime
}\neq 0$ and that these are not degenerate.

The matrices $U,U^{-1}$ and vectors $w^{\prime },v^{\prime }$ can then be 
\textit{derived} from the defining relations (\ref{eq:Udef}) as functions of
the arbitrary spectral parameters $\kappa _{e},\kappa _{o}$ as (see appendix 
\ref{sec:are} for details of the derivation) 
\begin{eqnarray}
&&U_{-e,o}={\frac{w_{e}^{\prime }v_{o}^{\prime }\kappa _{o}^{\prime }}{%
\kappa _{e}^{\prime }-\kappa _{o}^{\prime }}}\,,\quad U_{-o,e}^{-1}={\frac{%
w_{e}^{\prime }v_{o}^{\prime }\kappa _{e}^{\prime }}{\kappa _{e}^{\prime
}-\kappa _{o}^{\prime }}}\,,~~
U_{-e,o}=U_{e,-o}\,,~~U_{-o,e}^{-1}=U_{o,-e}^{-1}\,,
\label{eq:UU^-1} \\
&&\sqrt{2}\,w_{e}^{\prime }=w_{\left\vert e\right\vert }={i^{2-e}}\frac{
\prod_{o^{\prime }>0}\left\vert \kappa _{|e|}^{2}/\kappa _{o^{\prime
}}^{2}-1\right\vert ^{\frac{1}{2}}}{\prod_{e'>0, e^{\prime }\neq |e|}\left\vert
\kappa _{|e|}^{2}/\kappa _{e^{\prime }}^{2}-1\right\vert ^{\frac{1}{2}}}\,,
~~w_{e}^{\prime }=w_{-e}^{\prime }\,,\label{we}\\
&&\sqrt{2}\,v_{o}^{\prime }=v_{|o|}={i^{|o|-1}}\frac{\prod_{e^{\prime}>0}
\left\vert 1-\kappa _{|o|}^{2}/\kappa _{e^{\prime }}^{2}
\right\vert ^{\frac{1}{2}}}{\prod_{o'>0, o^{\prime }\neq |o|}
\left\vert 1-\kappa _{|o|}^{2}/\kappa_{o^{\prime }}^{2}\right\vert ^{\frac{1}{2}}}\,,~~
v_{o}^{\prime}=v_{-o}^{\prime }
\label{prime}
\end{eqnarray}
where now the indices $\left( e,o\right) $ run over the finite set $e=\pm
2,\pm 4,\dots ,\pm 2N$ and $o=\pm 1,\pm 3,\dots ,\pm (2N-1)$.
It is easy to check explicitly that in the limit $N\rightarrow \infty $ 
and $\kappa_{e}=e, $ $\kappa _{o}=o,$ these expressions reduce to 
Eq.(\ref{eq:UU^-1limit}).

The regulated expressions for these matrices look considerably more
complicated than their large $N$ limit. Therefore, it may appear that this
would create a problem in analytic computations. Actually this is not the
case at all, because in analytic computations one uses the matrix relations
satisfied by these matrices rather than the explicit matrices themselves.
The relations are preserved in the regularized version for any $N,$ and they
look the same as their $N=\infty $ counterpart. Therefore analytically the
expressions in any computation look the same in the regulated or infinite
versions as long as they are written in terms of these matrices without
using their explicit form. The explicit construction of the regulated
version insures associativity and eliminates the ambiguity of the
associativity anomaly as explained below. Thus, in addition to the basic
defining relations (\ref{eq:Udef}) which are the same for any $N,$ including 
$N=\infty $, there are more relations among $U,U^{-1},v^{\prime },w^{\prime }
$ that can now be derived from the defining relations alone for any $%
N,\kappa _{e}^{\prime },\kappa _{o}^{\prime }$ : 
\begin{eqnarray}
&&UU^{-1}=1\,,\quad U^{-1}U=1\,,\quad {\bar{U}}^{-1}U^{-1}=1+w^{\prime }\bar{%
w}^{\prime }\,,\quad \bar{U}U=1-v^{\prime }\bar{v}^{\prime }\,,
\label{eq:Umama} \\
&&U\bar{U}=1-{\frac{w^{\prime }\bar{w}^{\prime }}{1+\bar{w}^{\prime
}w^{\prime }}}\,,\quad Uv^{\prime }={\frac{w^{\prime }}{1+\bar{w}^{\prime
}w^{\prime }}}\,,\quad \bar{v}^{\prime }v^{\prime }={\frac{\bar{w}^{\prime
}w^{\prime }}{1+\bar{w}^{\prime }w^{\prime }}}\,,  \label{eq:Umodify} \\
&&U^{-1}w^{\prime }=v^{\prime }(1+{\bar{w}}^{\prime }w^{\prime })\,,\quad
U^{-1}\bar{U}^{-1}=1+v^{\prime }\bar{v}^{\prime }(1+\bar{w}^{\prime
}w^{\prime })\,,  \label{eq:Umodify2} \\
&&1+\bar{w}^{\prime }w^{\prime }={\frac{\prod_{e}\kappa _{e}^{\prime }}{%
\prod_{o}\kappa _{o}^{\prime }}}\,={\frac{\prod_{e>0}\kappa _{e}^{2}}{%
\prod_{o>0}\kappa _{o}^{2}}}\,.  \label{w2+1}
\end{eqnarray}
In particular note that Eqs.(\ref{e_UUanomaly}) are now deformed into Eqs.(%
\ref{eq:Umodify}), and that $\bar{w}^{\prime }w^{\prime }\rightarrow
2N\rightarrow \infty $ in the large $N$ limit. Thus, the deformed algebra
actually holds also at $N=\infty ;$ it simply makes explicit the behavior as
a function of $N.$ With this, the associativity anomaly hidden in the
original algebra is now resolved and the matrix algebra for all the matrices 
$U,U^{-1},T,R,S,\tilde{T},v,w,\tilde{v},\tilde{w},\tilde{u}$ defined through
Eqs.(\ref{eqs_defns1}--\ref{prime}) becomes associative.

In particular, from Eq.(\ref{eqs_defns1}--\ref{eqs_defns3}) we obtain the
regularized matrices, such as $T,R,S$, 
\begin{equation}
T_{eo}={\frac{w_{e}v_{o}\kappa _{o}^{2}}{\kappa _{e}^{2}-\kappa _{o}^{2}}}%
\,,\quad R_{oe}={\frac{w_{e}v_{o}\kappa _{e}^{2}}{\kappa _{e}^{2}-\kappa
_{o}^{2}}}\,,\quad S_{eo}={\frac{w_{e}v_{o}\kappa _{e}\kappa _{o}}{\kappa
_{e}^{2}-\kappa _{o}^{2}}}\,.  \label{eqs:defns}
\end{equation}
From the relations among $U,w^{\prime },v^{\prime }$ we can derive the
relations among $T,R,v,w,S,\tilde{T},\tilde{u},\tilde{v},\tilde{w}$. The
deformed algebra among $T,R,w,v$ (\ref{relations}) is already given in \cite
{BM1}, while the deformed algebra among $S,\tilde{T},\tilde{u},\tilde{v},
\tilde{w}$ which replaces (\ref{ss},\ref{eq:til_mat_rel_infty}) is\footnote{
Note that for finite $N$, the continuity condition at the midpoint (\ref
{CONTI}) is not satisfied because $S\tilde{w}\neq 0$. However, we recover
it, as well as all other infinite matrix relations by taking the open string
limit $\kappa _{e}=e,\kappa _{o}=o,N=\infty $.}, 
\begin{eqnarray}
&&S=\kappa _{e}^{2}\tilde{T}\kappa _{o}^{-2}\,,\quad S=\tilde{T}-\tilde{v}%
\bar{\tilde{w}}\,,\quad \bar{S}\tilde{v}=-\tilde{u}\,,\quad S\tilde{u}=-%
\tilde{v}\,,   \\
&&\bar{S}S=1\,,\quad S\bar{S}=1\,,\quad \bar{S}\tilde{T}=1-\tilde{u}\bar{%
\tilde{w}}\,,\quad \tilde{T}\bar{S}=1-{\frac{\kappa _{e}^{-1}w\bar{w}\kappa
_{e}}{1+\bar{w}w}}\,,   \\
&&S\tilde{w}={\frac{2}{\pi }}{\frac{\kappa _{e}w}{1+\bar{w}w}}\,,\quad 
\tilde{T}\tilde{u}={\frac{2}{\pi }}{\frac{\kappa _{e}^{-1}w}{1+\bar{w}w}}%
\,,\quad \bar{\tilde{T}}\tilde{v}=-\tilde{u}+\left( {\frac{2}{\pi }}\right)
^{2}\bar{w}\kappa _{e}^{-2}w\tilde{w}\,,   \\
&&\bar{\tilde{w}}\tilde{u}=\bar{v}v={\frac{\bar{w}w}{1+\bar{w}w}}\,,\quad 
\bar{\tilde{v}}\tilde{v}=\left( {\frac{2}{\pi }}\right) ^{2}\bar{w}\kappa
_{e}^{-2}w\,,\quad \bar{\tilde{u}}\tilde{u}=\left( {\frac{2}{\pi }}\right)
^{2}\bar{v}\kappa _{o}^{-2}v\,.
\end{eqnarray}
Furthermore, note that for any $N,\kappa _{e,}\kappa _{o}$ we have 
\begin{equation}
\bar{w}\kappa _{e}^{-2}w=\bar{v}\kappa _{o}^{-2}v=\sum_{o>0}\kappa
_{o}^{-2}-\sum_{e>0}\kappa _{e}^{-2}.  \label{w2w}
\end{equation}
The right hand side converges to ${\frac{\pi ^{2}}{12}}$ if we take the open
string limit $\kappa _{e}=e,~\kappa _{o}=o,~N=\infty $. More relations of
this type can be found in the next subsection.

\subsubsection{GL$\left( N|N\right) $ supergroup property of the regulator 
\label{w2relations}}

In this subsection we take a small detour to make an observation on the
regulator whose significance for computations in MSFT is not yet fully
apparent, but which is mathematically interesting, and could be useful in
future applications. Many computations in MSFT boil down to expressions of
the form $\bar{w}f\left( \kappa \right) w$ where $f\left( \kappa \right) $
is a matrix constructed from the frequencies $\kappa _{e},\kappa _{o}$
through the regulated matrices we discussed above. Therefore, we are
interested in developing analytic methods of computation involving such
expressions, in particular for arbitrary frequencies $\kappa _{e},\kappa
_{o} $. In such computations the properties of the supergroup GL$\left(
N|N\right) $ mysteriously makes an appearance, as follows.

By using explicitly the expression for $w_{e}$ given in Eq.(\ref{we}), we
have 
\begin{equation}
\bar{w}f\left( \kappa _{e}^{2}\right) w=\sum_{e}f\left( \kappa
_{e}^{2}\right) \frac{\det_{o}\left( \kappa _{e}^{2}\kappa
_{o}^{-2}-1\right) }{\det_{e^{\prime }\neq e}\left( \kappa _{e}^{2}\kappa
_{e^{\prime }}^{-2}-1\right) }=\oint \frac{dz}{2\pi i}\frac{f\left(
z\right) }{z}\frac{\det \left( 1-z\kappa _{o}^{-2}\right) }{\det \left(
1-z\kappa _{e}^{-2}\right) }\,,
\end{equation}
where the contour encircles only the poles at $z=\kappa _{e}^{2}.$ The
contour may then be deformed to evaluate the integral. When $f\left( \kappa
_{e}^{2}\right) =1$ or $\kappa _{e}^{-2}$ the results have already been
given in Eqs.(\ref{w2+1},\ref{w2w}). We note that these may be written in
the form 
\begin{equation}
\bar{w}w=-1+\text{Sdet}\left( \kappa ^{2}\right)\,,~~~\bar{w}\kappa
_{e}^{-2}w=-Str\left( \kappa ^{-2}\right)\,,  \label{sdet}
\end{equation}
where we used the superdeterminant (Sdet) and supertrace $\left( Str\right) $
by treating the matrix 
\begin{equation}
\kappa =\left( 
\begin{array}{cc}
\kappa _{e} & 0 \\ 
0 & \kappa _{o}
\end{array}
\right) 
\end{equation}
\textit{as if} it is a graded GL$\left( N|N\right) $ super matrix. When 
$f\left( \kappa _{e}^{2}\right) =\kappa _{e}^{-2n},$ with $n=1,2,\cdots
,$ we note that the contour integral is precisely the integral
representation of
the supercharacter of GL$\left( N|N\right) $ for the representation of GL$%
\left( N|N\right) $ described by a Young supertableau with a single row with 
$n$ superboxes \cite{supercharacter} 
\begin{equation}
\bar{w}\kappa _{e}^{-2n}w=-\chi _{n}\left( \kappa ^{-2}\right) .
\label{charn}
\end{equation}
The expression for the supercharacter $\chi _{n}\left( M\right) $ for any
supermatrix $M,$ can also be written in terms of supertraces of powers of $M$
in the fundamental representation, as given in \cite{supercharacter}. By
taking advantage of this observation we evaluate $\chi _{n}\left( \kappa
^{-2}\right) $ in terms of the supertraces of powers of $\kappa ^{-2}.$ For
example, $\chi _{2}\left( M\right) =\frac{1}{2}Str\left( M^{2}\right) +\frac{%
1}{2}\left( Str\left( M\right) \right) ^{2},$ which gives the following
interesting expression for the sum 
\begin{equation}
\bar{w}\kappa _{e}^{-4}w=-\chi _{2}\left( \kappa ^{-2}\right) =-\frac{1}{2}%
\left[ Tr\left( \kappa _{e}^{-4}\right) -Tr\left( \kappa _{o}^{-4}\right) 
\right] -\frac{1}{2}\left[ Tr\left( \kappa _{e}^{-2}\right) -Tr\left( \kappa
_{o}^{-2}\right) \right] ^{2}.  \label{w4w}
\end{equation}
We can check the correctness of Eq.(\ref{charn}) in the limit $\kappa
_{e}=e,~\kappa _{o}=o,~N=\infty $. In this limit, since $w_{e}\rightarrow 
\sqrt{2}\,i^{2-e},$ we can evaluate the sum on the left side directly in terms
of the zeta function, $\bar{w}\kappa _{e}^{-2n}w\rightarrow
2\sum_{k=1}^{\infty }\left( 2k\right) ^{-2n}=\frac{2}{2^{2n}}\zeta \left(
2n\right) ,$ and then compare it to the value generated by the
supercharacter $-\chi _{n}\left( \kappa ^{-2}\right) $ in the limit.

For example, for $n=1$ the right hand side of Eq.(\ref{charn}) is already
given following Eq.(\ref{w2w}), and this agrees with the left hand side
which is $\frac{2}{2^{2}}\zeta \left( 2\right) $. Similarly, for $n=2$ the
left hand side of Eq.(\ref{w4w}) gives $\frac{2}{2^{4}}\zeta \left( 4\right)
=\allowbreak \frac{1}{720}\pi ^{4}$ while the right hand side gives $\frac{7%
}{16}\zeta \left( 4\right) -\frac{1}{8}\left( \zeta \left( 2\right) \right)
^{2}=\allowbreak \frac{1}{720}\pi ^{4}$ which agree. The GL$\left(
N|N\right) $ supergroup property of these sums is intriguing. It may be the
signal of an underlying mathematical structure that could be helpful in
computations in MSFT.

\subsection{Moyal $\star $ from Witten's $\ast $ for fermionic modes \label%
{sec:odd_Moyal}}

The second step in constructing the map from Witten's star to Moyal's star
is to perform the Fourier transformation from position space to momentum
space for a subset of string modes \cite{B1}. We recall the definition of
Witten's $\ast $-product for functions of split strings in the ghost sector 
\begin{equation}
\Psi _{1}\ast \Psi _{2}[l(\sigma ),r(\sigma )]\sim \int \mathcal{D}z(\sigma
)\Psi _{1}[l(\sigma ),z(\sigma )]\Psi _{2}[\pm z(\sigma ),r(\sigma )]\,\,.
\end{equation}
For the $bc$ ghost sector, we consider two types of overlapping conditions 
\cite{GJ}. The anti-overlapping condition (resp. overlapping condition) is
defined by choosing the minus (resp. plus) sign in this formula. In both
cases the Witten $\ast $ product is mapped to the Moyal $\star $ product as
follows.

We denote the string field as $\Psi \lbrack l,r]$ in the split string
formulation and as ${\hat{\Psi}}[x,p]$ in the Moyal formulation. The
variables $l,r,x,p$ are all fermionic and we consider the simplified
situation where each of them represents a single degree of freedom. The
generalization to multiple variables is straightforward. For the simplified
setup we define Witten's star product in the split string formalism
(ignoring the midpoint for the time being) as 
\begin{equation}
\Psi _{1}\ast \Psi _{2}[l,r]=(-1)^{|\Psi _{1}|}\int dw\,\Psi _{1}[l,w]\,\Psi
_{2}[\pm w,r]\,.  \label{eq:anti-star}
\end{equation}
The sign factor $(-1)^{|\Psi _{1}|}$ (Grassmann parity of $\Psi _{1}$) is
needed to make the $\ast $ product associative. We define the mapping from a
string field in the split string picture $\Psi \lbrack l,r]$ to the Moyal
picture ${\hat{\Psi}}[x,p]$ by using the Fourier transform\footnote{%
If $l$ and $r$ consist of $N$ variables, the sign factor on the right hand
sides of Eqs.(\ref{eq:anti-star})(\ref{anti-from half})(\ref{anti-to half})
become $(-1)^{|\Psi _{1}|N},(\pm 1)^{N},(\pm 1)^{N}$ respectively. In
particular, they are trivial in the case that $N$ is even.} 
\begin{eqnarray}
{\hat{\Psi}}[x,p] &=&\pm \int dy\,e^{-py}\,\Psi \left[ \pm x+{\frac{y}{2}}%
,x\mp {\frac{y}{2}}\right] \,,  \label{anti-from half} \\
\Psi \lbrack l,r] &=&\pm \int dp\,e^{p(l\mp r)}\,{\hat{\Psi}}\left[ {\frac{%
r\pm l}{2}},p\right] \,.  \label{anti-to half}
\end{eqnarray}
Witten's star for $\Psi $ is then mapped to Moyal's star for $\hat{\Psi}$: 
\begin{equation}
{\hat{\Psi}}_{1}\star {\hat{\Psi}}_{2}[x,p]={\hat{\Psi}}_{1}[x,p]\exp \left(
\mp {\frac{1}{2}}\left( {\frac{\overleftarrow{\partial }}{\partial x}}{\frac{%
\overrightarrow{\partial }}{\partial p}}+{\frac{\overleftarrow{\partial }}{%
\partial p}}{\frac{\overrightarrow{\partial }}{\partial x}}\right) \right) {%
\hat{\Psi}}_{2}[x,p]\,.  \label{eq:anti-Moyal}
\end{equation}
The derivation of this correspondence is completely parallel to the bosonic
case \cite{B1}: 
\begin{eqnarray}
&&{\hat{\Psi}}_{1}[x,p]e^{\mp {\frac{1}{2}}\left( {\frac{\overleftarrow{%
\partial }}{\partial x}}{\frac{\overrightarrow{\partial }}{\partial p}}+{%
\frac{\overleftarrow{\partial }}{\partial p}}{\frac{\overrightarrow{\partial 
}}{\partial x}}\right) }{\hat{\Psi}}_{2}[x,p]  \notag \\
&=&\left( \pm \int dy_{1}\,e^{-py_{1}}\,\Psi _{1}\left[ \pm x+{\frac{y_{1}}{2%
}},x\mp {\frac{y_{1}}{2}}\right] \right) e^{\mp {\frac{1}{2}}\left( {\frac{%
\overleftarrow{\partial }}{\partial x}}{\frac{\overrightarrow{\partial }}{%
\partial p}}+{\frac{\overleftarrow{\partial }}{\partial p}}{\frac{%
\overrightarrow{\partial }}{\partial x}}\right) }\left( \pm \int
dy_{2}\,e^{-py_{2}}\,\Psi _{2}\left[ \pm x+{\frac{y_{2}}{2}},x\mp {\frac{%
y_{2}}{2}}\right] \right)  \notag \\
&=&(-1)^{|\Psi _{1}|}\int dy_{1}dy_{2}\left( \Psi _{1}\left[ \pm x+{\frac{%
y_{1}}{2}},x\mp {\frac{y_{1}}{2}}\right] e^{-py_{1}}\right) e^{\mp {\frac{1}{%
2}}\left( {\frac{\overleftarrow{\partial }}{\partial x}}{\frac{%
\overrightarrow{\partial }}{\partial p}}+{\frac{\overleftarrow{\partial }}{%
\partial p}}{\frac{\overrightarrow{\partial }}{\partial x}}\right) }\left(
e^{-py_{2}}\,\Psi _{2}\left[ \pm x+{\frac{y_{2}}{2}},x\mp {\frac{y_{2}}{2}}%
\right] \right)  \notag \\
&=&(-1)^{|\Psi _{1}|}\int dy_{1}dy_{2}\,\Psi _{1}\left[ \pm x+{\frac{y_{1}}{2%
}},x\mp {\frac{y_{1}}{2}}\right] e^{\pm {\frac{1}{2}}{\frac{\overleftarrow{%
\partial }}{\partial x}}y_{2}}e^{-p(y_{1}+y_{2})}e^{\mp {\frac{1}{2}}y_{1}{%
\frac{\overrightarrow{\partial }}{\partial x}}}\Psi _{2}\left[ \pm x+{\frac{%
y_{2}}{2}},x\mp {\frac{y_{2}}{2}}\right]  \notag \\
&=&(-1)^{|\Psi _{1}|}\int dy_{1}dy_{2}\,e^{-p(y_{1}+y_{2})}\Psi _{1}\left[
\pm x+{\frac{y_{1}+y_{2}}{2}},x\mp {\frac{y_{1}-y_{2}}{2}}\right] \Psi _{2}%
\left[ \pm x-{\frac{y_{1}-y_{2}}{2}},x\mp {\frac{y_{1}+y_{2}}{2}}\right] 
\notag \\
&=&\pm (-1)^{|\Psi _{1}|}\int dye^{-py}\int dz\,\Psi _{1}\left[ \pm x+{\frac{%
y}{2}},z\right] \Psi _{2}\left[ \pm z,x\mp {\frac{y}{2}}\right]  \notag \\
&=&\pm \int dye^{-py}\,\left( \Psi _{1}\ast \Psi _{2}\right) \left[ \pm x+{%
\frac{y}{2}},x\mp {\frac{y}{2}}\right] =:\hat{\Psi}_{1}\star \hat{\Psi}%
_{2}[x,p]\,.
\end{eqnarray}
The form of the product in Eq.(\ref{eq:anti-Moyal}) is similar to the
ordinary Moyal product although the derivatives in the exponential are for
fermionic variables. This Moyal $\star $ product is associative and
non-commutative.

From Eq.(\ref{anti-to half}), we also obtain the correspondence between the
definition of trace in the split-string formulation, which we take with the
anti-periodic condition, and that of Moyal one which is given by an
integration in \textquotedblleft phase space\textquotedblright\ 
\begin{equation}
\mathrm{Tr}\,\Psi :=\int dz\,\Psi \lbrack \pm z,z]=\pm \int dxdp\,\hat{\Psi}[
x,p]=:\pm \mathrm{Tr}\,\hat{\Psi}\,.
\end{equation}

\subsection{Moyal $\star $ product in the $bc$ ghost sector \label%
{sec:Moyalbc}}

In this section we define the Moyal $\star $ product which represents
Witten's star product using the results in \S \ref{sec:sine}, 
\S \ref{sec:odd_Moyal}, \S \ref{sec:cosine}. We first review the
conventional operator formalism to fix the
notation in the $bc$ ghost sector. We take the ghost coordinates $b(\sigma
),c(\sigma )$ and their conjugates $\pi _{b}(\sigma ),\pi _{c}(\sigma )$ 
\begin{equation}
b^{\pm }(\sigma )=\sum_{n=-\infty }^{\infty }\hat{b}_{n}e^{\pm in\sigma
}=\pi _{c}(\sigma )\mp ib(\sigma )\,,\qquad c^{\pm }(\sigma
)=\sum_{n=-\infty }^{\infty }\hat{c}_{n}e^{\pm in\sigma }=c(\sigma )\pm i\pi
_{b}(\sigma )\,.  \label{eq:ghosts_coord}
\end{equation}
We introduce the fermionic variables $\hat{x}_{n},\hat{y}_{n}$ and their
conjugates $\hat{p}_{n},\hat{q}_{n}$ as follows 
\begin{eqnarray}
&&b(\sigma )=-\sum_{n=1}^{\infty }(\hat{b}_{n}-\hat{b}_{-n})\sin n\sigma =i%
\sqrt{2}\sum_{n=1}^{\infty }\hat{x}_{n}\sin n\sigma \,,  \label{eq:bsigma} \\
&&c(\sigma )=\hat{c}_{0}+\sum_{n=1}^{\infty }(\hat{c}_{n}+\hat{c}_{-n})\cos
n\sigma =\hat{c}_{0}+\sqrt{2}\sum_{n=1}^{\infty }\hat{y}_{n}\cos n\sigma \,,
\label{eq:csigma} \\
&&\pi _{b}(\sigma )=\sum_{n=1}^{\infty }(\hat{c}_{n}-\hat{c}_{-n})\sin
n\sigma =-i\sqrt{2}\sum_{n=1}^{\infty }\hat{p}_{n}\sin n\sigma \,, \\
&&\pi _{c}(\sigma )=\hat{b}_{0}+\sum_{n=1}^{\infty }(\hat{b}_{n}+\hat{b}%
_{-n})\cos n\sigma =\hat{b}_{0}+\sqrt{2}\sum_{n=1}^{\infty }\hat{q}_{n}\cos
n\sigma \,.
\end{eqnarray}
The nonzero modes $\hat{x}_{n},\hat{y}_{n},\hat{p}_{n},\hat{q}_{n}$ are
related to $\hat{b}_{n},\hat{c}_{n}$: 
\begin{equation}
\hat{x}_{n}={\frac{i}{\sqrt{2}}}(\hat{b}_{n}-\hat{b}_{-n})\,,\quad \hat{y}%
_{n}={\frac{1}{\sqrt{2}}}(\hat{c}_{n}+\hat{c}_{-n})\,,\quad \hat{p}_{n}={%
\frac{i}{\sqrt{2}}}(\hat{c}_{n}-\hat{c}_{-n})\,,\quad \hat{q}_{n}={\frac{1}{%
\sqrt{2}}}(\hat{b}_{n}+\hat{b}_{-n})\,,
\end{equation}
and the canonical commutation relation $\{\hat{b}_{n},\hat{c}_{m}\}=\delta
_{m+n,0}$ can be rewritten as 
\begin{equation}
\{\hat{x}_{n},\hat{p}_{m}\}=\delta _{n,m}\,,\quad \{\hat{y}_{n},\hat{q}%
_{m}\}=\delta _{n,m}\,,\qquad n,m=1,2,\cdots \,.
\end{equation}
We represent the string field by treating $c_{0}$ and $x_{n},y_{n}$ as the
``position'' coordinates. The translation between Fock space representation
and position representation is made through 
\begin{equation}
\Psi (c_{0},x_{n},y_{n})=\langle c_{0},x_{n},y_{n}|\Psi \rangle \,.
\end{equation}
Here we introduced the bra state $\langle c_{0},x_{n},y_{n}|$ (and the
corresponding ket state) as states in Fock space which satisfy the
eigenvalue conditions for the operators $\hat{c}_{0},\hat{x}_{n},\hat{y}_{n}$%
, 
\begin{eqnarray}
&&\langle c_{0},x_{n},y_{n}|\hat{c}_{0}=\langle
c_{0},x_{n},y_{n}|c_{0}\,,~~\langle c_{0},x_{n},y_{n}|\hat{x}_{n}=\langle
c_{0},x_{n},y_{n}|x_{n}\,,~~\langle c_{0},x_{n},y_{n}|\hat{y}_{n}=\langle
c_{0},x_{n},y_{n}|y_{n}\,,\nonumber\\
&&\hat{c}_{0}|c_{0},x_{n},y_{n}\rangle =c_{0}|c_{0},x_{n},y_{n}\rangle \,,~~%
\hat{x}_{n}|c_{0},x_{n},y_{n}\rangle =x_{n}|c_{0},x_{n},y_{n}\rangle \,,~~%
\hat{y}_{n}|c_{0},x_{n},y_{n}\rangle =y_{n}|c_{0},x_{n},y_{n}\rangle\,.
\nonumber
\end{eqnarray}
Explicitly these are given by 
\begin{eqnarray}
&&\langle c_{0},x_{n},y_{n}|=\langle \Omega |{\hat{c}}_{-1}{\hat{c}}_{0}\exp
\left( c_{0}\hat{b}_{0}+\sum_{n=1}^{\infty }\left( -\hat{c}_{n}\hat{b}_{n}-i%
\sqrt{2}\hat{c}_{n}x_{n}+\sqrt{2}y_{n}\hat{b}_{n}+iy_{n}x_{n}\right) \right)
\,, \\
&&|c_{0},x_{n},y_{n}\rangle =\exp \left( \hat{b}_{0}c_{0}+\sum_{n=1}^{\infty
}\left( -\hat{b}_{-n}\hat{c}_{-n}+i\sqrt{2}x_{n}\hat{c}_{-n}+\sqrt{2}\hat{b}%
_{-n}y_{n}-ix_{n}y_{n}\right) \right) \hat{c}_{0}\hat{c}_{1}|\Omega \rangle
~~~  \label{e_bra_xy}
\end{eqnarray}
where $\langle \Omega |,|\Omega \rangle $ represents the conformal vacuum%
\footnote{%
We take the convention that $|\Omega \rangle $ is Grassmann even and $%
\langle \Omega |$ is odd.} normalized as $\langle \Omega |\hat{c}_{-1}\hat{c}%
_{0}\hat{c}_{1}|\Omega \rangle =1$. These bras and kets satisfy the
normalization and completeness relations 
\begin{eqnarray}
&&\langle c_{0},x_{n},y_{n}|c_{0}^{\prime },x_{n}^{\prime },y_{n}^{\prime
}\rangle =-(c_{0}-c_{0}^{\prime })\prod_{n=1}^{\infty }\left(
-2i(x_{n}-x_{n}^{\prime })(y_{n}-y_{n}^{\prime })\right) \,, \\
&&-\int dc_{0}\int \prod_{n=1}^{\infty }{\frac{dx_{n}dy_{n}}{2i}}%
|c_{0},x_{n},y_{n}\rangle \langle c_{0},x_{n},y_{n}|=1\,.
\end{eqnarray}
Witten's star product for the ghost sector is defined by the
(anti-)overlapping conditions \cite{GJ}, 
\begin{equation}
b^{\pm (r)}(\sigma )-b^{\pm (r-1)}(\pi -\sigma )=0\,,\quad c^{\pm
(r)}(\sigma )+c^{\pm (r-1)}(\pi -\sigma )=0\,,
\end{equation}
for $\quad r=1,2,3\,\mathrm{mod}\,3\,,~~\sigma \in \lbrack 0,\pi /2],\,$\ or
equivalently 
\begin{eqnarray}
&&b^{(r)}(\sigma )-b^{(r-1)}(\pi -\sigma )=0\,,\quad c^{(r)}(\sigma
)+c^{(r-1)}(\pi -\sigma )=0\,,  \label{eq:bc_overlapping} \\
&&\pi _{b}^{(r)}(\sigma )+\pi _{b}^{(r-1)}(\pi -\sigma )=0\,,\quad \pi
_{c}^{(r)}(\sigma )-\pi _{c}^{(r-1)}(\pi -\sigma )=0\,\,.
\end{eqnarray}
These (anti-)overlapping conditions for $bc$ ghost will be used to define
the mapping from Witten's $\ast $ to Moyal's $\star $ by using Eq.(\ref
{anti-from half}) defined in the previous section.

To apply the formulation in \S \ref{sec:odd_Moyal}, we need to specify the
boundary conditions of the split string variables, since we have to use the
Bogoliubov transformation given in \S \ref{sec:sine}/\S \ref{sec:cosine}
accordingly. At $\sigma =0,\pi $, $b(\sigma )$ (resp. $c(\sigma )$)
satisfies the Dirichlet (resp. Neumann) boundary condition. On the other
hand, at the midpoint, there are two options, namely Neumann or Dirichlet
type boundary conditions. In the following we choose the Neumann condition
for $b(\sigma )$ and Dirichlet condition for $c(\sigma )$\footnote{%
The other choice (Neumann for $b$ and Dirichlet for $c$ at the midpoint) is
discussed in the appendix \ref{DDNN}. It gives equivalent but more
complicated expression for the Moyal formulation.}.

In the split string language, the left and right halves of $b(\sigma
)$,  $l^{b}(\sigma ),r^{b}(\sigma ),$ satisfy Dirichlet at $\sigma =0$
and Neumann at $\sigma =\pi /2$, while the left and right halves of
$c(\sigma )$, $l^{c}(\sigma ),r^{c}(\sigma ),$ 
satisfy Neumann at $\sigma =0$ and Dirichlet
at $\pi /2$. With this choice, $l^{b}(\sigma ),r^{b}(\sigma )$ are expanded
by using \textit{odd} sine modes: $\{\sin o\sigma ,~o=1,3,5,\cdots \}$,
and $l^{c}(\sigma ),r^{c}(\sigma )$ by using \textit{odd} cosine modes: 
$\{\cos o\sigma ,~o=1,3,5,\cdots \}$ : 
\begin{eqnarray}
l^{b}(\sigma ) &=&i\sqrt{2}\sum_{o=1}^{\infty }l_{o}^{b}\sin o\sigma
\,,\;\;r^{b}(\sigma )=i\sqrt{2}\sum_{o=1}^{\infty }r_{o}^{b}\sin o\sigma \,,
\\
l^{c}(\sigma ) &=&\bar{c}+\sqrt{2}\sum_{o=1}^{\infty }l_{o}^{c}\cos o\sigma
\,,\;\;r^{c}(\sigma )=\bar{c}+\sqrt{2}\sum_{o=1}^{\infty }r_{o}^{c}\cos
o\sigma \,.
\end{eqnarray}
From Eqs.(\ref{eq:bsigma})(\ref{eq:csigma})(\ref{eq:sinDN_psi2lr})(\ref
{eq:oddcos}), we have the relations between split- and full-string variables 
\begin{eqnarray}
&&l_{o}^{b}=\bar{S}x_{e}+x_{o},\quad r_{o}^{b}=-\bar{S}x_{e}+x_{o}\,,
\label{half_odd_b} \\
&&\bar{c}=c_{0}-\bar{w}y_{e},\quad l_{o}^{c}=Ry_{e}+y_{o},\quad
r_{o}^{c}=Ry_{e}-y_{o}\,  \label{half_odd_c}
\end{eqnarray}
where we used a matrix notation. Witten's $\ast $ product for the split
string formulation is written as\footnote{%
Here we consider \textit{naive} overlapping condition. To obtain the
conventional Witten's star product, as we will show, we should treat
midpoint variable $\bar{c}$ more carefully. The phase factor $i$ in the
measure is only convention so that it is ``real''\ $(idx_{e}dy_{e})^{\dagger
}=idx_{e}dy_{e}$. Here we define complex conjugate for fermionic variables $%
\xi ,\xi ^{\prime }$ as $(\xi \xi ^{\prime })^{\dagger }=(\xi ^{\prime
})^{\dagger }(\xi )^{\dagger }$\thinspace .} 
\begin{equation}
{\tilde{A}}\ast {\tilde{B}}(\bar{c}%
,l_{o}^{b},l_{o}^{c},r_{o}^{b},r_{o}^{c})=\int \prod_{o>0}\left( id\eta
_{o}^{b}d\eta _{o}^{c}\right) \,{\tilde{A}}(\bar{c},l_{o}^{b},l_{o}^{c},\eta
_{o}^{b},\eta _{o}^{c}){\tilde{B}}(\bar{c},\eta _{o}^{b},-\eta
_{o}^{c},r_{o}^{b},r_{o}^{c})\,.
\end{equation}
The string field in the split-string formulation is identified with the
usual position representation $\Psi (x\left( \sigma \right) ),$ which is
written in terms of modes 
\begin{equation}
{\tilde{A}}(\bar{c},l_{o}^{b},l_{o}^{c},r_{o}^{b},r_{o}^{c})\sim \Psi
(c_{0},x_{n},y_{n}):=\langle c_{0},x_{n},y_{n}|\Psi \rangle \,.
\label{eq:APsi_corr}
\end{equation}
In order to map it to the Moyal formulation, we compare Eqs.(\ref{half_odd_b}%
)(\ref{half_odd_c}) with Eq.(\ref{anti-from half}), and note the
similarities (we add prime ${}^{\prime }$ to distinguish the variables with
anti-overlapping condition from the variables with overlapping conditions) 
\begin{eqnarray}
\bar{S}x_{e}+x_{o}\sim x+{\frac{y}{2}}\,,\quad -\bar{S}x_{e}+x_{o}\sim x-{%
\frac{y}{2}}\,,\quad Ry_{e}+y_{o}\sim -x^{\prime }+{\frac{y^{\prime }}{2}}%
\,,\quad Ry_{e}-y_{o}\sim x^{\prime }+{\frac{y^{\prime }}{2}}\,
\end{eqnarray}
or equivalently\footnote{%
We note that the \textit{odd} modes correspond to the ``$x$-variable''\ of
phase space in the Moyal formulation of ghosts. By contrast, in the matter
sector, the even modes played the corresponding r\^{o}le (see Eq.(30) in 
\cite{B1}).} 
\begin{eqnarray}
x_{o}\sim x\,,\quad \bar{S}x_{e}\sim {\frac{y}{2}}\,,\quad y_{o}\sim
-x^{\prime }\,,\quad Ry_{e}\sim {\frac{y^{\prime }}{2}}\,.
\label{eq:var_corr}
\end{eqnarray}
Thus, using Eqs.(\ref{anti-from half})(\ref{eq:APsi_corr}), we obtain the
map from the field in the position representation to the Moyal representation 
\begin{eqnarray}
A(\bar{c},x_{o},p_{o},y_{o},q_{o}):=2^{-2N}(1+\bar{w}w)^{-{\frac{1}{4}}}\int
\prod_{e>0}\left( i^{-1}dx_{e}dy_{e}\right) e^{-2p_{o}\bar{S}
x_{e}-2q_{o}Ry_{e}}\Psi (\bar{c}+\bar{w}y_{e},x_{n},y_{n})\,.~~
\end{eqnarray}
At this point, we used the MSFT regularization scheme, by truncating the
ghost modes $x_{n},y_{n}$ to $n\leq 2N$, and using the parameters $(N,\kappa
_{e},\kappa _{o})$ given in the previous section. Thus, $w,R,S$ are
redefined in Eqs.(\ref{wv_exp})(\ref{TR_exp})(\ref{eqs:defns}) and $2^{2N},%
\bar{w}w$ are finite. We fixed the normalization factor $\left( \det (16\bar{%
S}\bar{R})\right) ^{-{\frac{1}{2}}}=2^{-2N}(1+\bar{w}w)^{-{\frac{1}{4}}}$
consistently with the trace that will be given later. {}Hence, from Eqs.(\ref
{eq:anti-Moyal})(\ref{eq:var_corr}), the Moyal $\star $ product that
corresponds to Witten's $\ast $ product is 
\begin{eqnarray}
&&A\star B(\bar{c},x_{o},p_{o},y_{o},q_{o})  \notag \\
&=&A(\bar{c},x_{o},p_{o},y_{o},q_{o})\,e^{-{\frac{1}{2}}\sum_{o>0}\left( {%
\frac{\overleftarrow{\partial }}{\partial x_{o}}}{\frac{\overrightarrow{%
\partial }}{\partial p_{o}}}+{\frac{\overleftarrow{\partial }}{\partial y_{o}%
}}{\frac{\overrightarrow{\partial }}{\partial q_{o}}}+{\frac{\overleftarrow{%
\partial }}{\partial p_{o}}}{\frac{\overrightarrow{\partial }}{\partial x_{o}%
}}+{\frac{\overleftarrow{\partial }}{\partial q_{o}}}{\frac{\overrightarrow{%
\partial }}{\partial y_{o}}}\right) }B(\bar{c},x_{o},p_{o},y_{o},q_{o})\,.
\label{eq:Moyal_org}
\end{eqnarray}
We introduced an arbitrary parameter $\theta ^{\prime }$ which is the analog
of $\theta $ in the matter sector\cite{BM2} to absorb units. We note that
the product is local as a function of the midpoint $\bar{c}$, while $\bar{c}$
is related to the center or mass variable $c_{0}$ by 
\begin{equation}
c_{0}=\bar{c}+\bar{w}y_{e}\,.
\end{equation}
By rescaling variables and performing Fourier transformation with respect to 
$\bar{c}$, we arrive at the definition of the string field in MSFT 
\begin{eqnarray}
&&\hat{A}(\xi _{0},x_{o},p_{o},y_{o},q_{o})  \notag \\
&=&\int d\bar{c}e^{-\xi _{0}\bar{c}}A(\bar{c},x_{o},-p_{o}/{\theta ^{\prime }%
},y_{o},-q_{o}/{\theta ^{\prime }}) \nonumber\\
&=&2^{-2N}(1+\bar{w}w)^{-{\frac{1}{4}}}\int dc_{0}\prod_{e>0}\left(
i^{-1}dx_{e}dy_{e}\right) e^{-\xi _{0}c_{0}+\xi _{0}\bar{w}y_{e}+{\frac{2}{%
\theta ^{\prime }}}p_{o}\bar{S}x_{e}+{\frac{2}{\theta ^{\prime }}}%
q_{o}Ry_{e}}\Psi (c_{0},x_{n},y_{n})\,.~~  \label{eq:MSFTfield}
\end{eqnarray}
By this Fourier transformation with respect to zero mode, the Grassmann
parity of $\hat{A}(\xi _{0},x_{o},p_{o},y_{o},q_{o})$ and the corresponding $%
|\Psi \rangle $ coincide. The Moyal $\star $ product which is modified after
Eqs.(\ref{eq:Moyal_org})(\ref{eq:MSFTfield}) is 
\begin{eqnarray}
&&\star :=\exp\left({1\over 2}{\overleftarrow{\partial}\over \partial\xi}
\Sigma{\overrightarrow{\partial}\over \partial\xi}\right)
=\exp \left( {\frac{1}{2}}\left( {\frac{\overleftarrow{\partial }}{%
\partial \xi _{1}}}\sigma ^{\prime }{\frac{\overrightarrow{\partial }}{%
\partial \xi _{1}}}+{\frac{\overleftarrow{\partial }}{\partial \xi _{2}}}%
\sigma ^{\prime }{\frac{\overrightarrow{\partial }}{\partial \xi _{2}}}%
\right) \right) \,,  \label{eq:Moyal_dfn} \\
&&\xi =\left( 
\begin{array}{c}
\xi _{1} \\ 
\xi _{2}
\end{array}
\right) \,,~~\xi _{1}=\left( 
\begin{array}{c}
x_{o} \\ 
p_{o}
\end{array}
\right) \,,~~\xi _{2}=\left( 
\begin{array}{c}
y_{o} \\ 
q_{o}
\end{array}
\right) \,,\quad ~
\Sigma=\left(\begin{array}[tb]{cc}
	\sigma'& 0\\
 0&\sigma'
	     \end{array}
\right)\,,~~
\sigma ^{\prime }=\theta ^{\prime }\left( 
\begin{array}{cc}
0 & 1 \\ 
1 & 0
\end{array}
\right) \,.~~~~~~~
\end{eqnarray}
We define the trace in MSFT as integration over the ``phase space'': 
\begin{equation}
\mathrm{Tr}\,\hat{A}(\xi _{0},\xi )=\det \sigma ^{\prime }
\int d\xi \,\hat{A}(\xi _{0},\xi )=(-1)^{N}{\theta ^{\prime }}%
^{2N}\int \prod_{o>0}\left( dx_{o}dp_{o}dy_{o}dq_{o}\right) \,\hat{A}(\xi
_{0},\xi )\,.  \label{eq:tracedef}
\end{equation}
The Moyal field $\hat{A}(\xi _{0},\xi )$ 
which is mapped from $|\Psi \rangle $ by (%
\ref{eq:MSFTfield}) is normalized as 
\begin{equation}
\int d\xi _{0}\mathrm{Tr}\left( \left( \hat{A}(\xi _{0},\xi )\right)
^{\dagger }\star \left( {\frac{\partial }{\partial \xi _{0}}}-{\frac{\theta
^{\prime }}{2}}\bar{v}{\frac{\partial }{\partial q_{o}}}\right) \hat{A}(\xi
_{0},\xi )\right) =\langle \Psi |\hat{c}_{0}|\Psi \rangle \,,  \label{norm}
\end{equation}
where ${\frac{\partial }{\partial \xi _{0}}}-{\frac{\theta ^{\prime }}{2}}%
\bar{v}{\frac{\partial }{\partial q_{o}}}$ corresponds to $-\hat{c}_{0}$ (%
\ref{eq:oscill_xypq}).

It is convenient to introduce the bra $\langle \xi _{0},\xi |$ as a state in
Fock space such that the Moyal field is related directly to the Fock space
field via $\hat{A}(\xi _{0},\xi )=\langle \xi _{0},\xi |\Psi \rangle $. This
can be obtained from the bra state $\langle c_{0},x_{n},y_{n}|$ (\ref
{e_bra_xy}) by Fourier transformation (\ref{eq:MSFTfield}), 
\begin{eqnarray}
\langle \xi _{0},\xi | &=&\langle \xi _{0},\xi _{1},\xi _{2}|=\langle \xi
_{0},x_{o},p_{o},y_{o},q_{o}|  \notag \\
&=&2^{-2N}(1+\bar{w}w)^{-{\frac{1}{4}}}\int dc_{0}\prod_{e>0}\left(
i^{-1}dx_{e}dy_{e}\right) e^{-\xi _{0}c_{0}+\xi _{0}\bar{w}y_{e}+{\frac{2}{%
\theta ^{\prime }}}p_{o}\bar{S}x_{e}+{\frac{2}{\theta ^{\prime }}}%
q_{o}Ry_{e}}\langle c_{0},x_{n},y_{n}|  \notag \\
&=&-2^{-2N}(1+\bar{w}w)^{-{\frac{1}{4}}}\langle \Omega |\hat{c}_{-1}e^{-\xi
_{0}(\hat{c}_{0}-\sqrt{2}\bar{w}\hat{c}_{e})}e^{\hat{c}_{e}\hat{b}_{e}-\hat{c%
}_{o}\hat{b}_{o}-2i\bar{\xi _{1}}M_{0}^{\left( o\right) }\xi _{2}-\bar{\xi
_{1}}\lambda _{1}-\bar{\xi _{2}}\lambda _{2}}\quad
\label{eq:Moyal-basis-ghost}
\end{eqnarray}
where we used notation: 
\begin{equation}
M_{0}^{\left( o\right) }=\left( 
\begin{array}{cc}
{\frac{1}{2}} & 0 \\ 
0 & {\frac{2}{{\theta ^{\prime }}^{2}}}\bar{S}\bar{R}
\end{array}
\right) \,,~~\lambda _{1}=\left( 
\begin{array}{c}
-i\sqrt{2}\hat{c}_{o} \\ 
-{\frac{2\sqrt{2}i}{\theta ^{\prime }}}\bar{S}\hat{b}_{e}+{\frac{2i}{\theta
^{\prime }}}\bar{S}w\xi _{0}
\end{array}
\right) \,,~~\lambda _{2}=\left( 
\begin{array}{c}
-\sqrt{2}\hat{b}_{o} \\ 
-{\frac{2\sqrt{2}}{\theta ^{\prime }}}R\hat{c}_{e}
\end{array}
\right) \,.
\end{equation}
This is the result given in the summary at the beginning of this section.

\paragraph{Examples}

Here we give some examples of string fields in MSFT.

For the conventional ghost number 1 vacuum $\hat{c}_{1}|\Omega \rangle $ and 
$SL(2,R)$ invariant vacuum $|\Omega \rangle $, the corresponding fields are
given by 
\begin{eqnarray}
\hat{A}_{0}(\xi _{0},\xi ) &=&\langle \xi _{0},\xi |\hat{c}_{1}|\Omega
\rangle =2^{-2N}(1+\bar{w}w)^{-{\frac{1}{4}}}\xi _{0}\,e^{-ix_{o}y_{o}-i{%
\frac{4}{{\theta ^{\prime }}^{2}}}p_{o}\left( \bar{S}\bar{R}\right)
_{oo^{\prime }}q_{o^{\prime }}}\,,  \label{perturbative_vacuum_odd} \\
\hat{A}_{\Omega }(\xi _{0},\xi ) &=&\langle \xi _{0},\xi |\Omega \rangle
=-i\,2^{-2N+{\frac{1}{2}}}(1+\bar{w}w)^{-{\frac{1}{4}}}\,\xi
_{0}\,x_{1}\,e^{-ix_{o}y_{o}-i4p_{o}\left( \bar{S}\bar{R}\right)
_{oo^{\prime }}q_{o^{\prime }}}\,.
\end{eqnarray}
We note that in the open string limit $\kappa _{e}=e,$ $\kappa _{o}=o,$ $%
N=\infty $, these expressions become singular. For example, the coefficient
of $p_{o}q_{o^{\prime }}$ is divergent, 
\begin{equation}
(\bar{S}\bar{R})_{oo^{\prime }}={\frac{16\,i^{o+o^{\prime }+2}}{\pi
^{2}o^{\prime }}}\sum_{e=2}^{\infty }{\frac{e^{3}}{(e^2-(o^{\prime
})^{2})(e^2-o^{2})}}=\pm \infty \,.  \label{RR_divergence}
\end{equation}
Therefore, it is advisable not to take the open string limit at the level of
the state, but wait for the end of a computation. We note that the physical
quantities such as the scattering amplitude become regular in this limit 
\cite{BKM1}.

For the identity-like state in the Siegel gauge: $|\tilde{I}\rangle =%
\mathcal{N}_{\tilde{I}}\,e^{\sum_{n=1}^{\infty }(-1)^{n}\hat{c}_{-n}\hat{b}%
_{-n}}\hat{c}_{1}|\Omega \rangle $, which is a delta function with respect
to the even modes in position basis, $\Psi _{\tilde{I}}(c_{0},x_{n},y_{n})=%
\mathcal{N}_{\tilde{I}}\prod_{e>0}(-4i\delta (x_{e})\delta (y_{e}))$, the
corresponding field is 
\begin{equation*}
\hat{A}_{\tilde{I}}(\xi _{0},x_{o},p_{o},y_{o},q_{o})=(-1)^{N}(1+\bar{w}w)^{-%
{\frac{1}{4}}}\mathcal{N}_{\tilde{I}}\,\xi _{0}\,.
\end{equation*}
Except for the zero mode and the normalization factor, this $\hat{A}_{\tilde{%
I}}$ is the identity element with respect to the Moyal $\star $ product. The
conventional identity state $|I\rangle $ (\ref{eq:identityket}) \cite{GJ}
(which is BRST invariant, and not in the Siegel gauge) becomes more
complicated in MSFT.

\subsection{Oscillators
{\label{oscillators}}}

In this subsection, we obtain the Moyal images of the conventional
oscillators which are used in applications in MSFT. In this way we can write
various operators in oscillator language and in particular discuss the form
of $L_{0}$ and the butterfly state which came up in our work in \cite{BKM2}.

\subsubsection{Oscillators on the fields in MSFT
\label{sec:derivative}}

For an operator $\hat{\mathcal{O}}$ which consists of $\hat{b}_{n},\hat{c}%
_{n},$ acting on a state $|\Psi \rangle $ in Fock space, we define its Moyal
image $\hat{\beta}_{\hat{\mathcal{O}}},$ which is a differential operator
acting on the Moyal field $\hat{A}_{\Psi }(\xi _{0},\xi )=\langle \xi
_{0},\xi |\Psi \rangle $, as follows 
\begin{equation}
\hat{\beta}_{\hat{\mathcal{O}}}\hat{A}_{\Psi }(\xi _{0},\xi )=\langle \xi
_{0},\xi |{\hat{\mathcal{O}}}|\Psi \rangle \,.
\end{equation}
For the basic operators, $\hat{c}_{0},\hat{b}_{0},\hat{x}_{o},\hat{y}_{o},%
\hat{x}_{e},\hat{y}_{e},\hat{p}_{o},\hat{q}_{o},\hat{p}_{e},\hat{q}_{e}$,
this rule gives the corresponding operators in MSFT 
\begin{eqnarray}
&&\hat{\beta}_{\hat{c}_{0}}=-{\frac{\partial }{\partial \xi _{0}}}+{\frac{%
\theta ^{\prime }}{2}}\bar{v}{\frac{\partial }{\partial q_{o}}}\,,\quad \hat{%
\beta}_{\hat{b}_{0}}=-\xi _{0}\,,   \\
&&\hat{\beta}_{\hat{x}_{o}}=x_{o}\,,\qquad \hat{\beta}_{\hat{y}%
_{o}}=y_{o}\,,\qquad \hat{\beta}_{\hat{x}_{e}}={\frac{\theta ^{\prime }}{2}}S%
{\frac{\partial }{\partial p_{o}}}\,,\quad {\beta }_{\hat{y}_{e}}={\frac{%
\theta ^{\prime }}{2}}T{\frac{\partial }{\partial q_{o}}}\,,   \\
&&\hat{\beta}_{\hat{p}_{o}}={\frac{\partial }{\partial x_{o}}}\,,\qquad \hat{%
\beta}_{\hat{q}_{o}}={\frac{\partial }{\partial y_{o}}}\,,\qquad \hat{\beta}%
_{\hat{p}_{e}}={\frac{2}{\theta ^{\prime }}}Sp_{o}\,,\qquad \hat{\beta}_{%
\hat{q}_{e}}={\frac{2}{\theta ^{\prime }}}\bar{R}\,q_{o}+w_{e}\xi _{0}\,.
\label{eq:oscill_xypq}
\end{eqnarray}
The nonzero modes of the oscillators $\hat{b}_{n},\hat{c}_{n}$ become 
\begin{eqnarray}
&&\hat{\beta}_{e}^{b}={\frac{1}{\sqrt{2}}}\sum_{o>0}\left( {\frac{2}{\theta
^{\prime }}}q_{o}R_{o|e|}-i\epsilon (e){\frac{\theta ^{\prime }}{2}}
S_{|e|o}\,{\frac{\partial }{\partial p_{o}}}\right) +{\frac{1}{\sqrt{2}}}
w_{|e|}\xi _{0}=\sum_{o}\bar{\beta}_{o}^{b}U_{-o,e}^{-1}+w_{e}^{\prime }\xi
_{0}\,,   \label{eq:betahats_a}\\
&&\hat{\beta}_{o}^{b}={\frac{1}{\sqrt{2}}}\left( {\frac{\partial }{\partial
y_{|o|}}}-i\epsilon (o)x_{|o|}\right) \,,   \\
&&\hat{\beta}_{e}^{c}={\frac{1}{\sqrt{2}}}\sum_{o>0}\left( {\frac{\theta
^{\prime }}{2}}T_{|e|o}{\frac{\partial }{\partial q_{o}}}-i\epsilon (e){
\frac{2}{\theta ^{\prime }}}S_{|e|o}\,p_{o}\right) =\sum_{o}U_{e,-o}\bar{
\beta}_{o}^{c}\,,  \label{eq:betahats} \\
&&\hat{\beta}_{o}^{c}={\frac{1}{\sqrt{2}}}\left( y_{|o|}-i\epsilon (o){\frac{
\partial }{\partial x_{|o|}}}\right) \,.  
\end{eqnarray}
In the first and third equations we introduced the symbols $\bar{\beta}
_{o}^{b}$,$\bar{\beta}_{o}^{c}$ to denote the differential operators 
\begin{equation}
\bar{\beta}_{o}^{b}={\frac{1}{\sqrt{2}}}\left( {\frac{2}{\theta ^{\prime }}}
q_{|o|}-i{\frac{\theta ^{\prime }}{2}}\epsilon (o){\frac{\partial }{\partial
p_{|o|}}}\right) \,,\quad \bar{\beta}_{o}^{c}={\frac{1}{\sqrt{2}}}\left( {
\frac{\theta ^{\prime }}{2}}{\frac{\partial }{\partial q_{|o|}}}-i\epsilon
(o){\frac{2}{\theta ^{\prime }}}p_{|o|}\right) \,.
\end{equation}
The even and the odd sets satisfy canonical anti-commutation relations 
\begin{equation}
\{\hat{\beta}_{e}^{b},\hat{\beta}_{e^{\prime }}^{c}\}=\delta _{e+e^{\prime
}}\,,\quad \{\hat{\beta}_{o}^{b},\hat{\beta}_{o^{\prime }}^{c}\}=\delta
_{o+o^{\prime }}\,,\quad \{\bar{\beta}_{o}^{b},\bar{\beta}_{o^{\prime
}}^{c}\}=\delta _{o+o^{\prime }}\,.
\end{equation}
Using the above maps of operators, we can translate operators in the usual
oscillator representation into MSFT language. For example, the ghost number
operator (we assigned ghost number 1 to $\hat{c}_{1}|\Omega \rangle $) 
\begin{equation}
N_{\mathrm{gh}}=\sum_{n\geq 1}(\hat{c}_{-n}\hat{b}_{n}-\hat{b}_{-n}\hat{c}%
_{n})+\hat{c}_{0}\hat{b}_{0}+1\,,
\end{equation}
is mapped to its MSFT image 
\begin{equation}
N_{\mathrm{gh}}=\sum_{o>0}\left( y_{o}{\frac{\partial }{\partial y_{o}}}%
-x_{o}{\frac{\partial }{\partial x_{o}}}+p_{o}{\frac{\partial }{\partial
p_{o}}}-q_{o}{\frac{\partial }{\partial q_{o}}}\right) -\xi _{0}{\frac{%
\partial }{\partial \xi _{0}}}+2\,.
\end{equation}

\subsubsection{Oscillator as a field}

It is often useful to rewrite differential operators in Moyal space in terms
of the star product. The idea is to replace the derivative by the $\star $
-(super)commutator:\footnote{
We define the supercommutator as 
$[\hat{A}_{1},\hat{A}_{2}\}_{\star}:=\hat{A}_{1}\star
\hat{A}_{2}-(-1)^{|A_{1}||A_{2}|}\hat{A}_{2}\star
\hat{A_{1}}$. 
Note that $\hat{A}{\frac{\overleftarrow{\partial
}}{\partial \xi }}=-(-1)^{|A|}{\frac{\overrightarrow{\partial
}}{\partial \xi }}\hat{A}$ 
.}
\begin{eqnarray}
{\frac{\partial }{\partial \xi }}\hat{A}=\Sigma^{-1}
[\xi ,\hat{A}\}_{\star }\,.
\end{eqnarray}
More concretely 
\begin{equation}
{\frac{\partial }{\partial x_{o}}}\hat{A}={\frac{1}{{\theta ^{\prime }}}}
[p_{o},\hat{A}\}_{\star }\,,\quad {\frac{\partial }{\partial y_{o}}}\hat{A}={
\frac{1}{{\theta ^{\prime }}}}[q_{o},\hat{A}\}_{\star }\,,\quad {\frac{
\partial }{\partial p_{o}}}\hat{A}={\frac{1}{{\theta ^{\prime }}}}[x_{o},
\hat{A}\}_{\star }\,,\quad {\frac{\partial }{\partial q_{o}}}\hat{A}={\frac{1
}{{\theta ^{\prime }}}}[y_{o},\hat{A}\}_{\star }\,.
\end{equation}
This observation leads to the star product representation of the
differential operators as follows 
\begin{eqnarray}
&&\hat{\beta}_{o}^{b}\hat{A}={\frac{1}{\sqrt{2}}}\left( \beta _{o}^{b}\star 
\hat{A}-(-1)^{|A|}\hat{A}\star \beta _{-o}^{b}\right) \,,\quad \bar{\beta}
_{o}^{b}\hat{A}={\frac{1}{\sqrt{2}}}\left( \beta _{o}^{b}\star \hat{A}
+(-1)^{|A|}\hat{A}\star \beta _{-o}^{b}\right) \,,   \\
&&\hat{\beta}_{o}^{c}\hat{A}={\frac{1}{\sqrt{2}}}\left( \beta _{o}^{c}\star 
\hat{A}+(-1)^{|A|}\hat{A}\star \beta _{-o}^{c}\right) \,,\quad \bar{\beta}
_{o}^{c}\hat{A}={\frac{1}{\sqrt{2}}}\left( \beta _{o}^{c}\star \hat{A}
-(-1)^{|A|}\hat{A}\star \beta _{-o}^{c}\right) \,,~~~\label{eq:odd_osci}
\end{eqnarray}
where we defined the fields$\beta _{o}^{b,c}$ in Moyal space that play the
fundamental role of oscillators 
\begin{equation}
\beta _{o}^{b}:={\frac{1}{{\theta ^{\prime }}}}q_{|o|}-{\frac{i}{2}}\epsilon
(o)x_{|o|},\qquad \beta _{o}^{c}:={\frac{1}{2}}y_{|o|}-{\frac{i}{{\theta
^{\prime }}}}\epsilon (o)p_{|o|}\,.
\end{equation}
The odd $\bar{\beta}_{o}^{b,c}$ and even $\hat{\beta}_{e}^{b,c}$
differential operators are related to each other as in
Eqs.(\ref{eq:betahats_a},\ref{eq:betahats}). 
Therefore we also define fields with even labels via the Bogoliubov
transformation 
\begin{equation}
\beta _{e}^{b}:=\sum_{o}\beta _{o}^{b}\,U_{-o,e}^{-1}\,,\qquad \beta
_{e}^{c}:=\sum_{o}U_{e,-o}\beta _{o}^{c}\,  \label{evenmode_fields}
\end{equation}
where the sum runs over odd integers from $-2N+1$ to $2N-1$. These give the
star product representation of the even differential operators $\hat{\beta}%
_{e}^{b,c}$ 
\begin{eqnarray}
&&\hat{\beta}_{e}^{b}\hat{A}={\frac{1}{\sqrt{2}}}\left( \beta _{e}^{b}\star 
\hat{A}+(-1)^{|A|}\hat{A}\star \beta _{-e}^{b}\right) +w_{e}^{\prime }\xi
_{0}\hat{A}\,, \label{eq:even_osci_a} \\
&&\hat{\beta}_{e}^{c}\hat{A}={\frac{1}{\sqrt{2}}}\left( \beta _{e}^{c}\star 
\hat{A}-(-1)^{|A|}\hat{A}\star \beta _{-e}^{c}\right) \,.
\label{eq:even_osci}
\end{eqnarray}
The fields $\beta _{e}^{b,c}$ or $\beta _{o}^{b,c}$ satisfy the oscillator
anticommutation relations with respect to the star product 
\begin{equation}
\{\beta _{o}^{b},\beta _{o^{\prime }}^{c}\}_{\star }=\delta _{o+o^{\prime
}}\,,~~\{\beta _{e}^{b},\beta _{e^{\prime }}^{c}\}_{\star }=\delta
_{e+e^{\prime }}\,.
\end{equation}
But they do not anticommute with each other $~$%
\begin{equation}
\{\beta _{-e}^{b},\beta _{o}^{c}\}_{\star }=U_{o,-e}^{-1}\,,~~\{\beta
_{o}^{b},\beta _{-e}^{c}\}_{\star }=U_{-e,o}
\end{equation}
since they are related to each other by the Bogoliubov transformation given
above.

One may regard the fields $\beta ^{b,c}$ as the harmonic oscillators in
Moyal space which act on the string field $A$ from either side by the star
product. It is then natural to introduce the vacuum field associated with
the odd (and similarly the even) oscillators, as the field that satisfies
the following conditions under the star product 
\begin{equation}
\beta _{o}^{b}\star \hat{A}_{B}=\beta _{o}^{c}\star \hat{A}_{B}=\hat{A}%
_{B}\star \beta _{-o}^{b}=\hat{A}_{B}\star \beta _{-o}^{c}=0\,,\quad \forall
o>0\,.  \label{oddbuttterfly}
\end{equation}
By definition, this field is the Moyal image of the Fock space operator 
$\hat{A}_{B}\sim |0\rangle \langle 0|,$
 where $|0\rangle $ is the vacuum state with
respect to the oscillators $\beta _{o}^{b,c}$. These are first order
differential equations whose solution is the gaussian 
\begin{equation}
\hat{A}_{B}=\xi _{0}\,2^{-2N}\exp \left( -\sum_{o>0}\left( ix_{o}y_{o}+{%
\frac{4i}{{\theta ^{\prime }}^{2}}}p_{o}q_{o}\right) \right) \,.
\label{butterodd}
\end{equation}
If we write $\hat{A}_{B}=\xi _{0}A_{B}$, we see that $A_{B}$ is a projector (%
\ref{eq:projector}) with respect to the Moyal $\star $ product: $A_{B}\star
A_{B}=A_{B}$. It turns out this is the butterfly projector that came up in
other formulations of string field theory
\cite{Gaiotto:2001ji,Schnabl:2002ff,Gaiotto:2002kf} as shown in appendix
\ref{sec:butterfly}.

\subsubsection{$L_0$ and $\mathcal{L}_0$ \label{sec:L0L0}}

In string field theory computations the zeroth Virasoro operator $L_{0}$
plays a critical role since it defines the propagator. In this section, we
derive various forms of $L_{0}$ in MSFT. In the usual oscillator
representation acting on Fock space, $L_{0}$ is given by\footnote{%
We take the convention such that $L_{0}(\hat{c}_{1}|\Omega \rangle )=0$
fixes the constant that comes from normal ordering.} 
\begin{equation}
L_{0}=\sum_{k=1}^{\infty }k(\hat{b}_{-k}\hat{c}_{k}+\hat{c}_{-k}\hat{b}
_{k})\,.
\end{equation}
In MSFT which is regularized by $(N,\kappa _{e},\kappa _{o})$, we truncate
the number of oscillators to $2N$ and replace the frequencies by $\kappa
_{e,o}$. Then the Moyal image of $L_{0}$ becomes the following differential
operator 
\begin{eqnarray}
L_{0} &=&\sum_{k=1}^{2N}\kappa _{k}(\hat{\beta}_{-k}^{b}\hat{\beta}_{k}^{c}+%
\hat{\beta}_{-k}^{c}\hat{\beta}_{k}^{b})   \\
&=&\sum_{k=1}^{2N}\kappa _{k}+i\sum_{o>0}\kappa _{o}\left( x_{o}y_{o}+{\frac{%
\partial }{\partial x_{o}}}{\frac{\partial }{\partial y_{o}}}+{\frac{4}{{%
\theta ^{\prime }}^{2}}}p_{o}q_{o}+{\frac{{\theta ^{\prime }}^{2}}{4}}{\frac{%
\partial }{\partial p_{o}}}{\frac{\partial }{\partial q_{o}}}\right)   \notag
\\
&&+{\frac{4i}{{\theta ^{\prime }}^{2}}}(1+\bar{w}w)\left( \sum_{o>0}\kappa
_{o}v_{o}p_{o}\right) \left( \sum_{o^{\prime }>0}v_{o^{\prime }}q_{o^{\prime
}}\right) +{\frac{2i}{{\theta ^{\prime }}}}(1+\bar{w}w)\left(
\sum_{o>0}v_{o}\kappa _{o}p_{o}\right) \xi _{0}\,.
\end{eqnarray}
The last two terms come from the identities 
\begin{equation}
\bar{S}\kappa _{e}\bar{R}=\kappa _{o}R\bar{R}=\kappa _{o}+(1+\bar{w}w)\kappa
_{o}v\bar{v}\,,\quad \sum_{e>0}S_{eo}\kappa _{e}w_{e}=\kappa
_{o}Rw_{e}=\kappa _{o}v(1+\bar{w}w)\,.
\end{equation}
The similarity to the matter sector is enhanced by introducing an even basis 
$\left( x_{e}^{b},p_{e}^{b},x_{e}^{c},p_{e}^{c}\right) $ which is related to
the odd basis through the following linear canonical transformation\footnote{%
We used these even modes in \cite{BKM2}. They are just a linear
transformation of the odd modes in \S \ref{sec:Moyalbc} and different from
the even modes which are considered in appendix \ref{DDNN}.} 
\begin{equation}
x_{e}^{b}:=\kappa _{e}^{-1}Sx_{o}\,,\quad p_{e}^{b}:=\kappa
_{e}Sp_{o}\,,\quad x_{e}^{c}:=Ty_{o}\,,\quad p_{e}^{c}:=\bar{R}q_{o}\,.
\label{bc_even_variable}
\end{equation}
The Moyal $\star $ product (\ref{eq:Moyal_dfn}) and trace (\ref{eq:tracedef}%
) are invariant under this canonical transformation. With the new variables, 
$L_{0}$ is rewritten as 
\begin{eqnarray}
L_{0} &=&\sum_{k=1}^{2N}\kappa _{k}+i\sum_{e>0}\left( \kappa
_{e}^{2}\,x_{e}^{b}x_{e}^{c}+{\frac{\partial }{\partial x_{e}^{b}}}{\frac{%
\partial }{\partial x_{e}^{c}}}+{\frac{4}{{\theta ^{\prime }}^{2}}}%
p_{e}^{b}p_{e}^{c}+{\frac{{\theta ^{\prime }}^{2}}{4}}\kappa _{e}^{2}{\frac{%
\partial }{\partial p_{e}^{b}}}{\frac{\partial }{\partial p_{e}^{c}}}\right) 
\notag \\
&&-{\frac{i}{1+\bar{w}w}}\left( \sum_{e>0}w_{e}{\frac{\partial }{\partial
x_{e}^{b}}}\right) \left( \sum_{e^{\prime }>0}w_{e^{\prime }}{\frac{\partial 
}{\partial x_{e^{\prime }}^{c}}}\right) +{\frac{2i}{{\theta ^{\prime }}}}%
\left( \sum_{e>0}w_{e}p_{e}^{b}\right) \xi _{0}\,.  \label{eq:L0_bc_even}
\end{eqnarray}
Under this change of variables, the usual perturbative vacuum that was given
in the odd basis (\ref{perturbative_vacuum_odd}) becomes: 
\begin{equation}
\hat{A}_0=2^{-2N}(1+\bar{w}w)^{-{1\over 4}}\xi_0\,
e^{-ix_{e}^{b}\bar{R}\kappa _{o}Rx_{e}^{c}
-i{\frac{4}{{\theta ^{\prime }}^{2}}}p_{e}^{b}\kappa _{e}^{-1}p_{e}^{c}}\,.
\end{equation}
Then the apparent divergence (\ref{RR_divergence}) of the coefficient at the
limit $\kappa _{e}=e,\kappa _{o}=o,N=\infty $ does not occur, since 
\begin{equation}
|(\bar{R}\kappa _{o}R)_{ee^{\prime }}|=\left|{\frac{16\,i^{e+e^{\prime }}
\left(ee^{\prime }\right)^2 }{\pi ^{2}}}\sum_{o=1}^{\infty }{\frac{1}{
o(e^2-o^{2})(e^{\prime 2}-o^2)}}\right|\quad <\infty \,.
\end{equation}

Following the ideas of the previous subsection, the differential operator $%
L_{0}$ can be represented in terms of star products by introducing the field 
$\mathcal{L}_{0}$ and the ``midpoint correction'' $\gamma $ as follows 
\begin{eqnarray}
L_{0}\hat{A} &=&\mathcal{L}_{0}\star \hat{A}+\hat{A}\star \mathcal{L}%
_{0}+\gamma \hat{A}\,,   \\
\mathcal{L}_{0} &=&i\sum_{e>0}\left( {\frac{\kappa _{e}^{2}}{2}}%
x_{e}^{b}x_{e}^{c}+{\frac{2}{{\theta ^{\prime }}^{2}}}p_{e}^{b}p_{e}^{c}%
\right) +{\frac{1}{2}}\left( \sum_{e>0}\kappa _{e}+\sum_{o>0}\kappa
_{o}\right) +{\frac{i}{{\theta ^{\prime }}}}\left(
\sum_{e>0}w_{e}p_{e}^{b}\right) \xi _{0}   \\
&=&\sum_{e>0}\kappa _{e}\left( \beta _{-e}^{b}\star \beta _{e}^{c}+\beta
_{-e}^{c}\star \beta _{e}^{b}\right) -{\frac{1}{2}}\left( \sum_{e>0}\kappa
_{e}-\sum_{o>0}\kappa _{o}\right) +{\frac{i}{{\theta ^{\prime }}}}\left(
\sum_{e>0}w_{e}p_{e}^{b}\right) \xi _{0}\,,   \\
\gamma &=&-\frac{i}{1+\bar{w}w}\left( \sum_{e>0}w_{e}{\frac{\partial }{%
\partial x_{e}^{b}}}\right) \left( \sum_{e^{\prime }>0}w_{e^{\prime }}{\frac{%
\partial }{\partial x_{e^{\prime }}^{c}}}\right)
\end{eqnarray}
where $\beta _{e}^{b,c}$ can be rewritten in terms of the
even mode variables in Eq.(\ref{bc_even_variable}) 
\begin{equation}
\beta _{e}^{b}={\frac{1}{\theta ^{\prime }}}p_{|e|}^{c}-{\frac{i}{2}}%
\epsilon (e)\kappa _{e}x_{|e|}^{b}\,,\qquad \beta _{e}^{c}={\frac{1}{2}}%
x_{|e|}^{c}-{\frac{i}{{\theta ^{\prime }}}}\epsilon (e)\kappa
_{e}^{-1}p_{|e|}^{b}\,.\label{evenmode_fields2}
\end{equation}
The $\xi _{0}$-dependent term vanishes when $L_{0}$ acts on the fields in
the Siegel gauge. The field $\mathcal{L}_{0}$ is multiplied with the star
product, but $\gamma $ is still a differential operator. It is possible to
rewrite $\gamma $ in terms of star products (a double supercommutator)
\begin{eqnarray}
 \gamma\,\hat{A}=-{i\over {\theta'}^2(1+\bar{w}w)}
\sum_{e,e'>0}w_ew_{e'}[p^b_e,[p^c_{e'},\hat{A}\}_{\star}\}_{\star}\,,
\end{eqnarray}
but this does not have the single star product structure as the star products
with $\mathcal{L}_{0}$ and therefore $\gamma $ cannot be absorbed into a
redefinition of $\mathcal{L}_{0}$. As discussed in \cite{BKM2}, $\gamma $
depends only on a single combination of modes in the direction of the vector 
$w_{e},$ which is closely related to the midpoint. If it were not for the
``midpoint correction'' term $\gamma $, string field theory would reduce to
a matrix-like theory that would be exactly solvable, as shown in \cite{BKM2}%
. The above form of writing $L_{0}$ focuses on the $\gamma $ term as the
remaining difficult aspect of string field theory.

A similar structure exists in the canonically equivalent odd basis by
using $\mathcal{L}_{0}^{\prime },\gamma ^{\prime }$ 
\begin{eqnarray}
L_{0}\hat{A} &=&\mathcal{L}_{0}^{\prime }\star \hat{A}+\hat{A}\star \mathcal{
L}_{0}^{\prime }+\gamma ^{\prime }\hat{A}\,,   \notag\\
\mathcal{L}_{0}^{\prime } &=&i\sum_{o>0}\kappa _{o}\left( {\frac{1}{2}}
x_{o}y_{o}+{\frac{2}{{\theta ^{\prime }}^{2}}}p_{o}q_{o}\right) +{\frac{1}{2}
}\left( \sum_{e>0}\kappa _{e}+\sum_{o>0}\kappa _{o}\right) +(1+\bar{w}w){
\frac{i}{{\theta ^{\prime }}}}\left( \sum_{o>0}v_{o}\kappa _{o}p_{o}\right)
\xi _{0}  \notag \\
&=&\sum_{o>0}\kappa _{o}\left( \beta _{-o}^{b}\star \beta _{o}^{c}+\beta
_{-o}^{c}\star \beta _{o}^{b}\right) +{\frac{1}{2}}\left( \sum_{e>0}\kappa
_{e}-\sum_{o>0}\kappa _{o}\right) +(1+\bar{w}w){\frac{i}{{\theta ^{\prime }}}
}\left( \sum_{o>0}v_{o}\kappa _{o}p_{o}\right) \xi _{0}\,,  \notag \\
\gamma ^{\prime } &=&{\frac{4i}{{\theta ^{\prime }}^{2}}}(1+\bar{w}w)\left(
\sum_{o>0}\kappa _{o}v_{o}p_{o}\right) \left( \sum_{o^{\prime
}>0}v_{o^{\prime }}q_{o^{\prime }}\right) \,\,.
\end{eqnarray}

\section{Monoid and Neumann coefficients \label{sec:partII}}

The perturbative states and nonperturbative squeezed states in Fock space
(perturbative vacuum, sliver state, butterfly state and so on) are mapped to
gaussian functions in Moyal space \cite{BM2}. In fermionic Moyal space, such
as the basis $\xi =\left( x_{e}^{b},p_{e}^{b},x_{e}^{c},p_{e}^{c}\right) ,$
the generic form of such a gaussian is written as, 
\begin{equation}
A_{{\mathcal{N}},M,\lambda }(\xi )=\mathcal{N}e^{-{\bar{\xi}M\xi -\bar{\xi}%
\lambda }}\,,\qquad \bar{M}=-M\,.  \label{e_gh_Monoid}
\end{equation}
As in the bosonic case, such shifted gaussians form a monoid algebra 
\cite{BM2}.\footnote{
Some issues on Moyal product are discussed in \cite{OMMY}.
} A monoid is almost a group except for the property of inverse. This
implies that under star products the shifted gaussians satisfy the
properties of closure, identity and associativity. Although the generic
gaussian has also an inverse under star products, not all of them do (for
example, projectors such as Eq.(\ref{butterodd}) do not have an
inverse). 
The identity element under star products is the natural number $1,$ which
corresponds to the trivial gaussian.

The monoid structure is an effective tool for computations in MSFT. In
particular it was used to calculate the product of $n$ gaussian functions,
whose trace gives the $n$-point vertex. A corollary of this result is the
determination of the Neumann coefficients for the $n$-string vertex. These
coefficients were computed in \cite{GJ} from conformal field theory, but
now they can be determined by using only the Moyal product. The MSFT
approach gives simple expressions for all Neumann coefficients in terms of a
single matrix $t_{eo}=\kappa _{e}^{1/2}T_{eo}\kappa_o^{-1/2}$. Neumann
coefficients are not needed for computations in MSFT, but the computation
can be used to test the MSFT formalism. This was used as successful test of
MSFT in the matter sector \cite{BM2}.

In this section we will carry out a similar program in the ghost sector.
While the treatment of the unity elements becomes more subtle because of the
zero mode, the closure of gaussian functions (\ref{e_gh_Monoid}) under the
star product will be proved exactly the same way as in the bosonic sector.
In particular, the algebraic structure is formally the same. With this
information, one can compute the product of $n$ fermionic gaussians and
derive the Neumann coefficients by using this algebraic machinery.

In particular, we verify consistency of the Moyal $\star $ product (\ref
{eq:Moyal_dfn}) with the conventional reflector and the 3-string vertex in
oscillator language. We will show the correspondence by including the zero
mode part, 
\begin{eqnarray}
\langle \Psi _{1}|\Psi _{2}\rangle &\leftrightarrow &
\int d\xi_0\, {\rm Tr}\left(\hat{A}_{1}(\xi_0,\xi)
\star \hat{A}_{2}(\xi_0,\xi)\right)\,,\\
\langle \Psi _{1}|\Psi _{2}\star ^{W}\Psi _{3}\rangle &\leftrightarrow &\int
d\xi _{0}^{(3)}d\xi _{0}^{(2)}d\xi _{0}^{(1)}\,
\mathrm{Tr}\left(\hat{A}_{1}(\xi_0^{(1)},\xi)\star 
\hat{A}_{2}(\xi_0^{(2)},\xi)\star \hat{A}_{3}(\xi_0^{(3)},\xi)\right)\,.
\end{eqnarray}

\subsection{Monoid structure for gaussian elements}

We first derive the structure of the monoid (\ref{e_gh_Monoid}) under the
fermionic star product defined by $\left\{ \xi _{a},\xi _{b}\right\} =\Sigma
_{ab}$ for a general symmetric $\Sigma _{ab}.$ This matrix takes a block
diagonal form $\Sigma =diag\left( \sigma ^{\prime },\sigma ^{\prime }\right) 
$ in the bases we discussed so far, but it is useful to develop the
formalism for any $\Sigma $. For the moment, we suppress the $\xi _{0}$%
-dependence because it is not relevant to the Moyal $\star $-product (\ref
{eq:Moyal_dfn}). The structure of the monoid is summarized in the following
algebra, 
\begin{eqnarray}
&&A_{\mathcal{N}_{1},M_{1},\lambda _{1}}\star A_{\mathcal{N}
_{2},M_{2},\lambda _{2}}=A_{\mathcal{N}_{12},M_{12},\lambda _{12}}\,, 
\label{e_gh_structure_a}
\\
&&m_{i}:=M_{i}\Sigma \,,\qquad \bar{m}=-\Sigma m\Sigma ^{-1}\,,   \\
&&m_{12}=M_{12}\Sigma
=(1+m_{2})m_{1}(1+m_{2}m_{1})^{-1}+(1-m_{1})m_{2}(1+m_{1}m_{2})^{-1}\,,  \\
&&\lambda _{12}=(1-m_{1})(1+m_{2}m_{1})^{-1}\lambda
_{2}+(1+m_{2})(1+m_{1}m_{2})^{-1}\lambda _{1}\,,   \\
&&\mathcal{N}_{12}=\mathcal{N}_{1}\mathcal{N}_{2}\,{\det }^{\frac{1}{2}
}(1+m_{2}m_{1})\,e^{-{\frac{1}{4}}\sum_{a,b=1}^{2}\bar{\lambda}_{a}\Sigma
K_{ab}\lambda _{b}}\,,   \\
&&K_{ab}=\left( 
\begin{array}{cc}
(m_{2}^{-1}+m_{1})^{-1} & (1+m_{2}m_{1})^{-1} \\ 
-(1+m_{1}m_{2})^{-1} & (m_{2}+m_{1}^{-1})^{-1}
\end{array}
\right) \,.  \label{e_gh_structure}
\end{eqnarray}
To prove this formula, it is convenient to use Fourier transformation. We
define 
\begin{equation}
\tilde{A}(\eta ):=\int d\xi \,e^{\bar{\xi}\eta }A(\xi )\,,\qquad A(\xi
)=\int d\eta \,e^{-\bar{\xi}\eta }\tilde{A}(\eta )\,.
\end{equation}
The $\star $ product is rewritten in terms of Fourier coefficients, 
\begin{equation}
A_{1}\star A_{2}(\xi )=\int d\eta _{1}d\eta _{2}e^{-{\frac{1}{2}}\bar{\eta
_{1}}\Sigma \eta _{2}-\bar{\xi}\eta _{1}-\bar{\xi}\eta _{2}}\tilde{A}%
_{1}(\eta _{1})\tilde{A}_{2}(\eta _{2})\,.
\end{equation}
For the gaussian $A_{\mathcal{N},M,\lambda }$ (\ref{e_gh_Monoid}), the
Fourier transform is also gaussian: 
\begin{equation}
\widetilde{A_{{\mathcal{N}},M,\lambda }}(\eta )=\mathcal{N}(\det (2M))^{%
\frac{1}{2}}\,e^{-{\frac{1}{4}}\bar{\lambda}M^{-1}\lambda }e^{-{\frac{1}{4}}%
\bar{\eta}M^{-1}\eta +{\frac{1}{2}}\bar{\eta}M^{-1}\lambda }\,.
\end{equation}
The main result (\ref{e_gh_structure_a}--\ref{e_gh_structure}) follows by 
carrying out the gaussian integration over fermionic variables.

We note that Eq.(\ref{e_gh_Monoid}) have \emph{exactly} the same form as the
bosonic case (Eqs.(3.11--3.17) in \cite{BM2}) if we put $d=-2$. Similarly,
the formula for the trace also related to the bosonic case as if $d=-2$ 
\begin{equation}
\mathrm{Tr}(A(\xi ))=\mathcal{N}(\det (2m))^{\frac{1}{2}}e^{-{\frac{1}{4}}%
\bar{\lambda}M^{-1}\lambda }\,.  \label{eq:gaussian-trace}
\end{equation}
With this observation, we find that all the algebraic manipulations in \cite
{BM2} which use the structure of the star product for monoids also apply in
the ghost sector with no other modification.

In particular the product of $n$ monoids with the same $M,$ but different $%
\lambda _{i},\mathcal{N}_{i}$ is one of the useful results that is used to
compute the $n$-point vertex. The result in \cite{BM2} is now adopted to the
ghost sector as follows 
\begin{eqnarray}
&&A_{\mathcal{N}_{12\cdots n},M^{(n)},\lambda _{12\cdots n}}:=A_{\mathcal{N}
_{1},M,\lambda _{1}}\star A_{\mathcal{N}_{2},M,\lambda _{2}}\star \dots
\star A_{\mathcal{N}_{n},M,\lambda _{n}}\,,  \label{eq:star_a} \\
&&m^{(n)}=M^{(n)}\Sigma ={\frac{J_{n}^{-}}{J_{n}^{+}}}\,,\qquad J_{n}^{\pm
}:={\frac{(1+m)^{n}\pm (1-m)^{n}}{2}}\,,  \label{eq:star_a2} \\
&&\lambda _{12\cdots n}={\frac{1}{J_{n}^{+}}}
\sum_{r=1}^{n}(1-m)^{r-1}(1+m)^{n-r}\lambda _{r}\,,   \\
&&\mathcal{N}_{12\cdots n}=\mathcal{N}_{1}\mathcal{N}_{2}\cdots \mathcal{N}
_{n}(\det {J_{n}^{+}})^{\frac{1}{2}}\exp \left( -{\frac{1}{4}}K_{n}(\lambda
)\right) \,,   \\
&&K_{n}(\lambda )=\sum_{r=1}^{n}\bar{\lambda}_{r}\Sigma {\frac{J_{n-1}^{-}}{
J_{n}^{+}}}\lambda _{r}+2\sum_{r<s}^{n}\bar{\lambda}_{r}\Sigma {\frac{
(1-m)^{s-r-1}(1+m)^{n+r-s-1}}{J_{n}^{+}}}\lambda _{s}\,,   \\
&&\mathrm{Tr}(A_{\mathcal{N}_{12\cdots n},M^{(n)},\lambda _{12\cdots n}})=
\mathcal{N}_{1}\mathcal{N}_{2}\cdots \mathcal{N}_{n}{\det }^{\frac{1}{2}
}(2J_{n}^{-})\exp \left( -{\frac{1}{4}}\sum_{r,s=1}^{n}\bar{\lambda}
_{r}\Sigma \mathcal{O}_{(s-r)}^{(n)}(m)\lambda _{s}\right) \,,   \\
&&\mathcal{O}_{(r)}^{(n)}(m):=\mathcal{O}_{r\ {\text{mod}\ n}}^{(n)}\,, 
 \\
&&\mathcal{O}_{0}^{(n)}(m)={\frac{J_{n-1}^{+}}{J_{n}^{-}}}={\frac{
(1+m)^{n-1}+(1-m)^{n-1}}{(1+m)^{n}-(1-m)^{n}}}\,,   \\
&&\mathcal{O}_{i}^{(n)}(m)={\frac{(1+m)^{n-i-1}(1-m)^{i-1}}{J_{n}^{-}}}
={\frac{2(1+m)^{n-i-1}(1-m)^{i-1}}{(1+m)^{n}-(1-m)^{n}}}\,,
\quad (1\leq i\leq n-1)\,.~~~~
\label{eq:star}
\end{eqnarray}
This will be used in the next section to compute the Neumann coefficients as
a by-product.

The algebraic structure is also used to construct projectors (such as sliver
state or butterfly). As in \cite{BM2}, the generic form of the projector in
the ghost sector can be written as a particular class of gaussian functions 
\begin{eqnarray}
&&A_{D,\lambda }(\xi )=2^{-2N}\,e^{{\frac{1}{4}}\bar{\lambda}\Sigma D\Sigma
\lambda }e^{-\bar{\xi}D\xi -\bar{\xi}\lambda }\,,\qquad (D\Sigma )^{2}=1\,, 
 \\
&&A_{D,\lambda }\star A_{D,\lambda }(\xi )=A_{D,\lambda }(\xi )\,,\qquad 
\text{Tr}(A_{D,\lambda }(\xi ))=1\,.  \label{eq:projector}
\end{eqnarray}

\subsection{Neumann coefficients\label{sec:Neumann coef}}

In this section, we construct the Neumann coefficients by using the MSFT
formalism for the $n$-point vertices given above. The basic idea is to use
the correspondence of vertices in the operator formalism and in MSFT given
by \cite{BM2}, 
\begin{equation}
{}_{1}\langle \Psi_1|\otimes \cdots \otimes {}_{n}\langle \Psi_n
|V_{n}\rangle \sim \mbox{Tr}(\hat{A}_1\star \cdots \star \hat{A}_n)
\label{e_basic_corresp}
\end{equation}
where $\hat{A}_r(\xi _{0},\xi )=
\langle \xi _{0},\xi |\tilde{\Psi}_r\rangle$
 and $|\tilde{\Psi}_r \rangle=\langle\Psi_r|V_2\rangle $. For the ghost
sector, we have to be careful in the treatment of the zero mode in
(\ref{e_basic_corresp}).
We use this identification to express Neumann coefficients by taking the
following steps.
\begin{enumerate}
\item  In Fock space we choose $n$ coherent states for $\langle \Psi_r|,$ 
$r=1,2,\cdots n,$ labelled by parameters $\mu ^{*\left( r\right) }$. The
left hand side of (\ref{e_basic_corresp}) can be computed in the operator
formalism. The result takes the form of an exponential that contains a
quadratic form in the parameters of the coherent states $\mu ^{\left(
r\right) }$. The Neumann coefficients that define $|V_{n}\rangle $ appear as
the coefficients in the quadratic form. We treat the Neumann coefficients as
unknown matrices.

\item  We calculate 
$\hat{A}_r(\xi _{0},\xi )=
\langle \xi _{0},\xi |\tilde{\Psi}_r\rangle $ 
which gives the Moyal image of the coherent states in the form of  monoids, 
with the $\lambda^{(r)}$ related to the parameters $\mu ^{*\left(r\right) }$ 
of the coherent state.

\item  We compute the right hand side of (\ref{e_basic_corresp}) by using
the result in Eqs.(\ref{eq:star_a}--\ref{eq:star}). 
As in item (1) this is also an exponential
containing a quadratic form in the coherent state parameters $\mu ^{*\left(
r\right) }$, but with the coefficients determined by the monoid algebra
given above.

\item  We compare the coefficients of the parameters in both sides and thus
determine the Neumann coefficients completely. 
They turn out to be simple functions of a single matrix 
$t_{eo}=\kappa _{e}^{1/2}T_{eo}\kappa_{o}^{-1/2}.$
\end{enumerate}

Throughout this subsection, we use the regularized framework ($N,\kappa
_{e},\kappa _{o}$) to make the algebraic manipulation consistent. This gives
a new generalization of Neumann coefficients since the new expression
includes arbitrary spectral parameters and arbitrary $N$. To compare to the
Neumann coefficients computed through conformal field theory, the open
string limit ($N=\infty ,\kappa _{e}=e,\kappa _{o}=o$) is taken at the end.
Through analytic and numerical methods it is shown that these very different
looking forms of Neumann coefficients are indeed the same. This successful
test of MSFT provides confidence about its correctness and shows that MSFT
is an alternative tool for computation in string theory.

\paragraph{Coherent states}

Coherent states $\langle \Psi |$ are defined by 
\begin{equation}
\langle \Psi |\hat{b}^{\dagger }=\langle \Psi |\mu _{b}^{\ast }\,,\quad
~\langle \Psi |\hat{c}^{\dagger }=\langle \Psi |\mu _{c}^{\ast }\,,\quad
\langle \Psi |\hat{b}_{0}=\langle \Psi |\mu _{0}^{\ast }\,,
\end{equation}
They have the following explicit form 
\begin{equation}
\langle \Psi |=\langle \Omega |\hat{c}_{-1}e^{\mu _{b}^{\ast }\hat{c}+\mu
_{c}^{\ast }\hat{b}+\mu _{0}^{\ast }\hat{c}_{0}}\,.  \label{eq:coherent}
\end{equation}
The inner product between the standard $n$-string vertices $|V_{n}\rangle $
(appendix \ref{sec:functional_rep}, \cite{GJ}) and the coherent states $\langle
\Psi _{r}|$ is given as follows for $n=1,2,3$ 
\begin{eqnarray}
n=1 &:&\langle \Psi |I\rangle ={\frac{\pi }{2\sqrt{2}}}\bar{v}_{o}\kappa
_{o}\mu _{bo}^{\ast }\left( \mu _{0}^{\ast }-\sqrt{2}\bar{w}\mu _{be}^{\ast
}\right) e^{\bar{\mu}_{c}^{\ast }C\mu _{b}^{\ast }}\,,  \label{eq:1coherent} \\
n=2 &:&{}_1\langle \Psi_1|{}_2\langle \Psi_2|V_{2}\rangle_{12}
=(\mu _{0}^{\ast(1) }-\mu _{0}^{\ast (2)})e^{\bar{\mu}_{c}^{\ast
(1)}C\mu _{b}^{\ast (2)}+\bar{\mu}_{c}^{\ast (2)}C\mu _{b}^{\ast (1)}} 
\notag \\
&&\qquad =(\mu _{0}^{\ast (1)}-\mu _{0}^{\ast (2)})e^{{\frac{1}{2}}\bar{\mu}%
^{\ast (1)}\varepsilon C\mu ^{\ast (2)}+{\frac{1}{2}}\bar{\mu}^{\ast
(2)}\varepsilon C\mu ^{\ast (1)}}\,,  \label{eq:2coherent} \\
n=3 &:&{}_1\langle \Psi_1|{}_2\langle \Psi_2|{}_3\langle
\Psi_3|V_{3}\rangle_{123}=\exp \left( -\bar{\mu}_{c}^{\ast(r)
}X^{rs}\mu _{b}^{\ast(s) }-\bar{\mu}_{c}^{\ast(r) }X^{r}{}_{0}^{s}\mu
_{0}^{\ast(s) }\right)  \notag \\
&&\qquad =\exp \left( -{\frac{1}{2}}\bar{\mu}^{\ast(r)}\varepsilon \mathcal{X}%
^{rs}\mu ^{\ast(s)}-\bar{\mu}_{c}^{\ast(r)}X^{r}{}_{0}^{s}\mu _{0}^{\ast(s)
}\right) \,,  \label{eq:3coherent}
\end{eqnarray}
where 
\begin{equation}
\mu ^{\ast }=\left( 
\begin{array}{c}
\mu _{c}^{\ast } \\ 
\mu _{b}^{\ast }
\end{array}
\right) \,,~\varepsilon =\left( 
\begin{array}{cc}
0 & 1 \\ 
-1 & 0
\end{array}
\right) \,,~\mathcal{X}^{rs}=\left( 
\begin{array}{cc}
\bar{X}^{sr} & 0 \\ 
0 & X^{rs}
\end{array}
\right) \,,~C=\left( 
\begin{array}{cc}
C & 0 \\ 
0 & C
\end{array}
\right) \,,~C_{nm}=(-1)^{n}\delta _{n,m}\,.
\end{equation}
As we noted, the Neumann coefficients appear as the coefficients of
quadratic functions of $\mu ^{*\left( r\right) }$ in the exponential.

\paragraph{Moyal image of coherent state}

We define a Moyal field which corresponds to the coherent state $\langle
\Psi |$. First, we have the corresponding ket $|\tilde{\Psi}\rangle $ by
using the reflector (\ref{eq:V_2ket}): 
\begin{equation}
|\tilde{\Psi}\rangle_1 :={}_2\langle \Psi |V_{2}\rangle_{12} 
=e^{\hat{c}_{0}^{(1)}\mu_{0}^{\ast }+\hat{c}^{\dagger (1)}C\mu _{b}^{\ast }
+\mu _{c}^{\ast }C\hat{b}^{\dagger(1)}}\hat{c}_{1}^{(1)}|\Omega \rangle_1 \,.
\end{equation}
Then we get the Moyal field by using the ket $\langle \xi _{0},\xi |$ (\ref
{eq:Moyal-basis-ghost}) which defines the Moyal basis: 
\begin{equation}
\hat{A}(\xi _{0},\xi ):=\langle \xi _{0},\xi |\tilde{\Psi}\rangle
=-2^{-2N}(1+\bar{w}w)^{-{\frac{1}{4}}}(\xi _{0}e^{-\sqrt{2}\bar{\mu}
_{c}^{\ast }w\mu _{0}^{\ast }}+\mu _{0}^{\ast })e^{-{\frac{1}{2}}\bar{\mu}
^{\ast }\varepsilon C\mu ^{\ast }-\bar{\xi}M_0\xi 
-\bar{\xi}\lambda }
\end{equation}
where we denoted 
\begin{eqnarray}
&&M_0=\left(\begin{array}[tb]{cc}
       0& iM_0^{(o)}\\
-i\bar{M}_0^{(o)}&0
	    \end{array}
\right)\,,\qquad \lambda =\left( 
\begin{array}{c}
-i\sqrt{2}\mu _{co}^{\ast } \\ 
{\frac{-2\sqrt{2}i}{{\theta ^{\prime }}}}\bar{S}\mu _{be}^{\ast }+{\frac{2i}{%
\theta ^{\prime }}}\bar{S}w\xi _{0}  \\ 
\sqrt{2}\mu _{bo}^{\ast } \\ 
{\frac{2\sqrt{2}}{\theta ^{\prime }}}R\mu _{ce}^{\ast }
\end{array}
\right) =2K^{\ast }(\mu ^{\ast }+W\xi _{0})\,, ~~~~\\
&&K^{\ast }=\left( 
\begin{array}{cccc}
0 & -{\frac{i}{\sqrt{2}}} & 0 & 0 \\ 
0 & 0 & -{\frac{\sqrt{2}i}{{\theta ^{\prime }}}}\bar{S} & 0 \\ 
0 & 0 & 0 & {\frac{1}{\sqrt{2}}} \\ 
{\frac{\sqrt{2}}{{\theta ^{\prime }}}}R & 0 & 0 & 0
\end{array}
\right) \,,\qquad W=\left( 
\begin{array}{c}
0 \\ 
0 \\ 
-{\frac{1}{\sqrt{2}}}w \\ 
0
\end{array}
\right) \,.
\end{eqnarray}
It takes the form of the standard element of monoid although with the
pre-factor $\mathcal{N}$ and the $\lambda $ in the exponent depend on the
zero mode $\xi _{0}$. We can apply the Moyal $\star $ product formula for
monoids which was developed in the previous subsection.

\paragraph{Explicit form of $n$-th product}

After using the results of (\ref{eq:star_a}--\ref{eq:star}), 
we have obtained the trace
formula for $n$-th product of Moyal fields which correspond to coherent
states 
\begin{eqnarray}
&&\mathrm{Tr}(\hat{A}_1(\xi _{0}^{(1)},\xi )\star \hat{A}_2(\xi
_{0}^{(2)},\xi )\star \cdots \star \hat{A}_n(\xi _{0}^{(n)},\xi )) 
\notag \\
&=&(-1)^{n}2^{-2nN}(1+\bar{w}w)^{-{\frac{n}{4}}}{\det }^{\frac{1}{2}%
}(2J_{n}^{-})\prod_{r=1}^{n}(\xi _{0}^{(r)}e^{-\sqrt{2}\bar{\mu}_{c}^{\ast
(r)}w\mu _{0}^{\ast (r)}}+\mu _{0}^{\ast (r)})  \notag \\
&&\qquad \times \,e^{-{\frac{1}{2}}\sum_{r=1}^{n}\bar{\mu}^{\ast
(r)}\varepsilon C\mu ^{\ast (r)}-{\frac{1}{4}}\sum_{r,s=1}^{n}
\bar{\lambda}^{(r)}\Sigma \mathcal{O}_{(s-r)}(m_{0})\lambda^{(s)}}
\qquad   \notag\\
&=&(-1)^{n}2^{-2nN}(1+\bar{w}w)^{-{\frac{n}{4}}}{\det }^{\frac{1}{2}%
}(2J_{n}^{-})\prod_{r}(\xi _{0}^{(r)}e^{-\sqrt{2}\bar{\mu}_{c}^{\ast
(r)}w\mu _{0}^{\ast (r)}}+\mu _{0}^{\ast (r)})  \notag \\
&&\quad \times \,\exp \left( {\frac{1}{2}}\sum_{r,s=1}^{n}\bar{\mu}^{\ast
(r)}\varepsilon C(K^{\ast -1}2m_{0}\mathcal{O}_{(s-r)}(m_{0})K^{\ast
}-\delta ^{r,s})\mu ^{\ast (s)}\right)   \notag \\
&&\quad \times \,\exp \left( \sum_{r,s=1}^{n}\bar{\mu}^{\ast (r)}\varepsilon
CK^{\ast -1}2m_{0}\mathcal{O}_{(s-r)}(m_{0})K^{\ast }W\xi _{0}^{(s)}\right)
\,,  \label{eq:n-coherent_nonzero}
\end{eqnarray}
where we used the relations\footnote{
Similar relations were used to obtain the Neumann coefficients in matter
sector. (\S \ref{sec:Neumann-matter} \cite{BM2})} 
\begin{eqnarray}
&&\bar{K}^{\ast }\Sigma =-\varepsilon CK^{\ast -1}m_{0}\,,\quad
m_{0}:=M_{0}\Sigma \,,   \\
&&K^{\ast -1}m_{0}K^{\ast }=\left( 
\begin{array}{cc}
\bar{\hat{m}}_{0}^{\ast -1} & 0 \\ 
0 & \hat{m}_{0}^{\ast -1}
\end{array}
\right) \,,~~~\hat{m}_{0}^{\ast }:=\sqrt{\kappa }\tilde{m}_{0}^{\ast }{\frac{%
1}{\sqrt{\kappa }}}=\left( 
\begin{array}{cc}
0 & -S \\ 
-\bar{T} & 0
\end{array}
\right) \,,   \\
&&\bar{W}\varepsilon CK^{\ast -1}2m_{0}\mathcal{O}_{(s-r)}(m_{0})K^{\ast
}W=0\,.
\end{eqnarray}
Here $\tilde{m}_{0}^{\ast }$ which defines $\hat{m}_{0}^{\ast }$ has
appeared in Eq.(\ref{eq:tilm0}) in the context of the matter sector in MSFT.
We should emphasize that although we are using a notation similar to the one
in the matter sector, the meaning of $m_{0}$ here is \textit{not} the same
as the one in the matter sector which is defined by Eq.(\ref{eq:m0matterdef}%
). However, because there are some relationships between the $m_{0}$ in the
matter and ghost sectors there are some relations among Neumann coefficients
in these sectors. Note that in Eq.(\ref{eq:n-coherent_nonzero}) we assigned
different $\xi _{0}$'s for each Moyal field in the trace. This prescription
is necessary to find agreement with Witten's star product as we will see
soon. Noting 
\begin{equation}
\xi _{0}^{(r)}e^{-\sqrt{2}\bar{\mu}_{c}^{\ast (r)}w\mu _{0}^{\ast (r)}}+\mu
_{0}^{\ast (r)}=\delta (\xi _{0}^{(r)}+\mu _{0}^{\ast (r)})\,e^{-\sqrt{2}%
\bar{\mu}_{c}^{\ast (r)}w\mu _{0}^{\ast (r)}}=\delta (\xi _{0}^{(r)}+\mu
_{0}^{\ast (r)})\,e^{2\bar{\mu}^{\ast (r)}\varepsilon CW\mu _{0}^{\ast
(r)}}\,,
\end{equation}
we perform the $\xi _{0}^{\left( r\right) }$-integrations for all zero
modes, and find 
\begin{eqnarray}
&&\int d\xi _{0}^{(n)}d\xi _{0}^{(n-1)}\cdots d\xi _{0}^{(1)}\,\mathrm{Tr}(
\hat{A}_1(\xi _{0}^{(1)},\xi )\star \hat{A}_2(\xi_{0}^{(2)},\xi )\star 
\cdots \star \hat{A}_n(\xi _{0}^{(n)},\xi )) \notag \\
&=&(-1)^{n}2^{-2nN}(1+\bar{w}w)^{-{\frac{n}{4}}}{\det }^{\frac{1}{2}%
}(2J_{n}^{-})  \notag \\
&&\quad \times \exp \left( {\frac{1}{2}}\sum_{r,s=1}^{n}\bar{\mu}^{\ast
(r)}\varepsilon C(K^{\ast -1}2m_{0}\mathcal{O}_{(s-r)}(m_{0})K^{\ast
}-\delta ^{r,s})\mu ^{\ast (s)}\right)   \notag \\
&&\quad \times \exp \left( \sum_{r,s=1}^{n}\bar{\mu}^{\ast (r)}\varepsilon
C(2\delta ^{r,s}-K^{\ast -1}2m_{0}\mathcal{O}_{(s-r)}(m_{0})K^{\ast })W\mu
_{0}^{\ast (s)}\right) \,.  \label{eq:n-coherent}
\end{eqnarray}

\paragraph{Comparison of coefficients}

Now, we consider the above formulas in the cases $n=1,2,3$.

\noindent \underline{$n=1$ case}: Eq.(\ref{eq:n-coherent}) becomes 
\begin{equation}
\int d\xi _{0}\,\mathrm{Tr}\,\hat{A}(\xi _{0},\xi )=-(1+\bar{w}w)^{\frac{1}{4%
}}e^{\mu _{c}^{\ast }C\mu _{b}^{\ast }}\,.
\end{equation}
In this case from Eq.(\ref{eq:1coherent}) there is a correspondence 
\begin{equation}
\int d\xi _{0}\,\mathrm{Tr}\,\hat{A}(\xi _{0},\xi )\sim \langle \Psi
|I\rangle \,.
\end{equation}
\textit{up to pre-factor} which comes from the $b$-ghost insertion in
conventional operator formalism.\cite{GJ} On the other hand, up to a
constant factor, we have 
\begin{equation}
\int d\xi _{0}\delta (\xi _{0})\,\mathrm{Tr}\,\hat{A}(\xi _{0},\xi )\sim
\langle \Psi |\tilde{I}\rangle =\mu _{0}^{\ast }\,e^{\mu _{c}^{\ast }C\mu
_{b}^{\ast }}\,,
\end{equation}
for the identify-like state $|\tilde{I}\rangle $ which corresponds to the 
\textit{identity} element for the reduced product \cite{IK,Oku1}. In
MSFT, the Moyal field $\langle \xi _{0},\xi |\tilde{I}\rangle $ is the
identity element in the Siegel gauge. In this sense, $|\tilde{I}\rangle $
appears naturally rather than the BRST-invariant $|I\rangle $ in the context
of MSFT.

\noindent \underline{$n=2$ case}: Using Eq.(\ref{eq:n-coherent_nonzero}),
we get
\begin{eqnarray}
&&\int d\xi _{0}^{(2)}d\xi _{0}^{(1)}\delta (\xi _{0}^{(1)}-\xi _{0}^{(2)})%
\mathrm{Tr}\,(\hat{A}_1(\xi _{0}^{(1)},\xi )\star \hat{A}_2(\xi
_{0}^{(2)},\xi ))  \notag \\
&=&(\mu _{0}^{\ast (1)}-\mu _{0}^{\ast (2)})\,e^{{\frac{1}{2}}\bar{\mu}%
^{\ast (1)}\varepsilon C\mu ^{\ast (2)}+{\frac{1}{2}}\bar{\mu}^{\ast
(2)}\varepsilon C\mu ^{\ast (1)}}\,.
\end{eqnarray}
From Eq.(\ref{eq:2coherent}) we have obtained the correspondence 
\begin{equation}
\int d\xi _{0}^{(2)}d\xi _{0}^{(1)}\delta (\xi _{0}^{(1)}-\xi _{0}^{(2)})%
\mathrm{Tr}(\hat{A}_1(\xi _{0}^{(1)},\xi )\star \hat{A}_2(\xi
_{0}^{(2)},\xi ))={}_1\langle \Psi_1|{}_2\langle \Psi_2|V_{2}\rangle_{12}\,.
\label{eq:V2n=2}
\end{equation}
In this case the normalization also coincides. 
 We can interpret $\int d\xi _{0}^{(2)}d\xi _{0}^{(1)}\delta
(\xi _{0}^{(1)}-\xi _{0}^{(2)})$ as the pre-factor $
(c_{0}^{(1)}+c_{0}^{(2)}) $ in the form of ${}_{1}\langle
c_{0}^{(1)},x_{n}^{(1)},y_{n}^{(1)}|{}_{2}\langle
c_{0}^{(2)},x_{n}^{(2)},y_{n}^{(2)}|V_{2}\rangle _{12}$ (\ref{eq:CV2pos}).
Eq.(\ref{eq:V2n=2}) can be rewritten as
\begin{eqnarray}
 \int d\xi_0 {\rm Tr}(\hat{A}_1(\xi_0,\xi )\star \hat{A}_2(\xi_0,\xi ))
=\langle \Psi_1| \tilde{\Psi}_{2}\rangle\,,
\end{eqnarray}
for $|\tilde{\Psi}_{2}\rangle_1:={}_2\langle \Psi_2|V_{2}\rangle_{12}$.
This is consistent with the normalization (\ref{norm}) 
which we adopted to fix
the map from the conventional field to the Moyal field (\ref{eq:MSFTfield}). 

\noindent \underline{$n=3$ case}: We can identify the Neumann coefficients
for the nonzero modes by comparing 
\begin{equation}
\mathrm{Tr}(\hat{A}_1(\xi _{0}^{(1)},\xi )\star \hat{A}_2(\xi
_{0}^{(2)},\xi )\star \hat{A}_3(\xi _{0}^{(3)},\xi ))\sim \langle
\Psi_1|\langle \Psi_2|\langle \Psi_3|V_{3}\rangle\,.
\end{equation}
From Eqs.(\ref{eq:n-coherent_nonzero},\ref{eq:3coherent}) we get the
Neumann coefficients in MSFT: 
\begin{equation*}
\mathcal{X}^{rs}=-C(K^{\ast -1}2m_{0}\mathcal{O}_{(s-r)}(m_{0})K^{\ast
}-\delta ^{r,s}).
\end{equation*}
More explicitly 
\begin{equation}
X^{(0)}=-C{\frac{\hat{m}_{0}^{\ast -2}-1}{\hat{m}_{0}^{\ast -2}+3}}\,,\quad
X^{(+)}=-C{\frac{2(1+\hat{m}_{0}^{\ast -1})}{\hat{m}_{0}^{\ast -2}+3}}%
\,,\quad X^{(-)}=-C{\frac{2(1-\hat{m}_{0}^{\ast -1})}{\hat{m}_{0}^{\ast -2}+3%
}}\,.\quad   \label{eq:Neumann_X}
\end{equation}
To identify the Neumann coefficients including the zero mode part, we should
perform the $\xi _{0}$ integration: $\int d\xi _{0}^{(3)}d\xi _{0}^{(2)}d\xi
_{0}^{(1)}$. We can interpret that this comes from the
pre-factor:
$$
(c_{0}^{(1)}-\bar{w}y_{e}^{(1)})(c_{0}^{(2)}-\bar{w}%
y_{e}^{(2)})(c_{0}^{(3)}-\bar{w}y_{e}^{(3)})=\bar{c}^{(1)}\bar{c}^{(2)}\bar{c%
}^{(3)}$$
 in the form
$$
{}_{1}\langle c_{0}^{(1)},x_{n}^{(1)},y_{n}^{(1)}|{}_{2}\langle
c_{0}^{(2)},x_{n}^{(2)},y_{n}^{(2)}|{}_{1}\langle
c_{0}^{(3)},x_{n}^{(3)},y_{n}^{(3)}|V_{2}\rangle
_{123}\,
$$
(\ref{eq:CV3pos}). 
In fact, by identifying\footnote{
The ghost zeromode $\xi _{0}$ dependence is similar to momentum $p_{0}$
dependence in matter sector. But in this case, we do not need ``momentum
conservation factor'' $\delta (\xi _{0}^{(1)}+\xi _{0}^{(2)}+\xi
_{0}^{(3)})$.  This fact correspond to the lack of $\delta
(b_{0}^{(1)}+b_{0}^{(2)}+b_{0}^{(3)})$ factor in Ref.\cite{IOS} Eq.(2.18)
which gives the correct 3-string vertex in oscillator representation.} 
Eq.(\ref{eq:n-coherent}) with Eq.(\ref{eq:3coherent}): 
\begin{equation}
\int d\xi _{0}^{(3)}d\xi _{0}^{(2)}d\xi _{0}^{(1)}\mathrm{Tr}(\hat{A}_1
(\xi _{0}^{(1)},\xi )\star \hat{A}_2(\xi _{0}^{(2)},\xi )\star 
\hat{A}_3(\xi _{0}^{(3)},\xi ))\sim \langle \Psi_1|
\langle \Psi_2|\langle \Psi_3|V_{3}\rangle
\end{equation}
up to constant factor, we obtain\footnote{
We used the notation: $w=\left( 
\begin{array}{c}
w_{e} \\ 
0
\end{array}
\right) $.} 
\begin{eqnarray}
&&X_{~\,0}^{(0)}=-2\varepsilon CW+\varepsilon CK^{\ast -1}{\frac{2+2m_{0}^{2}%
}{m_{0}^{2}+3}}K^{\ast }W={\frac{4}{\hat{m}_{0}^{\ast -2}+3}}{\frac{w}{\sqrt{%
2}}}\,,  \notag \\
&&X_{~~0}^{(+)}=\varepsilon CK^{\ast -1}{\frac{2+2m_{0}}{m_{0}^{2}+3}}%
K^{\ast }W=-{\frac{2-2\hat{m}_{0}^{\ast -1}}{\hat{m}_{0}^{\ast -2}+3}}{\frac{%
w}{\sqrt{2}}}\,,  \label{eq:Neumann_zero}\\
&&X_{~~0}^{(-)}=\varepsilon CK^{\ast -1}{\frac{2-2m_{0}}{m_{0}^{2}+3}}%
K^{\ast }W=-{\frac{2+2\hat{m}_{0}^{\ast -1}}{\hat{m}_{0}^{\ast -2}+3}}{\frac{%
w}{\sqrt{2}}}\,.  \notag
\end{eqnarray}
The Neumann coefficients $X^{(0,\pm )},X_{~~~0}^{(0,\pm )}$ agree with Eq.(%
\ref{eq:Neumann-3-ghost}) which was obtained by using the trace of 6
coherent states in the \textit{matter} sector. This implies that the
Gross-Jevicki nonlinear relations for Neumann coefficients in MSFT are all
satisfied for arbitrary $(N,\kappa _{e},\kappa _{o}),$ as was shown in \cite
{BM2}. Namely, our Moyal star product is consistent with the conventional
Witten star product in both the matter and ghost sectors.\footnote{%
This also implies that Moyal $\star $ product (\ref{eq:Moyal_dfn}) is
essentially the same as the \textit{reduced} product in
\cite{Erler,IK,Oku1}
 which was defined by omitting ghost zero mode-dependence in
original Witten's star product.} We have confirmed the correspondence
between Moyal $\star $ product in MSFT and Witten's one ($\star ^{W}$) : 
\begin{equation}
\int d\xi _{0}^{(3)}d\xi _{0}^{(2)}d\xi _{0}^{(1)}\mathrm{Tr}(\hat{A}_1
(\xi _{0}^{(1)},\xi )\star \hat{A}_2(\xi _{0}^{(2)},\xi )\star 
\hat{A}_3(\xi _{0}^{(3)},\xi ))~~~\leftrightarrow ~~~\langle \Psi_1|
\Psi_2\star ^{W}\Psi_3\rangle 
\end{equation}
up to constant factor for$~~\kappa _{e}=e,\kappa _{o}=o,N=\infty $. As we
will show in the next subsection we have numerical confirmation that the
generalized Neumann coefficients in MSFT for arbitrary $\kappa _{e},\kappa
_{o},N$ converge to the conventional one in the operator formalism when we
take the limit.

\subsection{Numerical comparison of Neumann coefficients \label%
{sec:Neumann_numerical}}

In this subsection we compare the generalized Neumann coefficients derived
algebraically in the Moyal star formalism for any $\kappa _{e},\kappa _{o},N,
$ \ with the independent computation from the point of view of conformal
field theory \cite{GJ} valid at $\kappa _{e}=e,\kappa _{o}=o,N=\infty $. We
summarize the Neumann coefficients computed from CFT in appendix \ref
{sec:Neumann_GJ}, together with some differences in the convention.

In \cite{BM2} we have already given an analytic proof that our algebraic
expression of Neumann coefficients coincides with the exact value in \cite
{GJ} in the limit, by comparing the spectroscopy of Neumann coefficients.
Namely, in our case by diagonalizing the matrix $t_{eo}=\kappa
_{e}^{1/2}T_{eo}\kappa _{o}^{-1/2}$ we diagonalize the Neumann coefficients
for $n$-point vertices, since they all depend on the same matrix $t.$ The
eigenvalues obtained in this way for the case of the 3-point vertex in the
limit $\kappa _{e}=e,\kappa _{o}=o,N=\infty $ coincide with the
corresponding eigenvalues obtained from Neumann spectroscopy in \cite
{spectro}.

A numerical study provides another approach to confirm that the Moyal star
and CFT calculations agree in the limit. In the following numerical analysis
we show that there is agreement in the limit, and furthermore that there is
a clear universal behavior of the approach to the limit as a function of $N$%
. In the tables given in appendix \ref{sec:Neumann_ratio} we give the MSFT
results for the numerical values of the Neumann coefficient $\mathcal{M}%
_{ee^{\prime }}^{(0)}\left( N\right) $ in the matter sector, and the Neumann
coefficient $X_{ee^{\prime }}^{(0)}\left( N\right) $ in the ghost sector for 
$e,e^{\prime }=2,4,6,8$, at different values of the cut-off parameter $N$.
We set the spectral parameters as $\kappa _{e}=e,\kappa _{o}=o$. The
expression of the Neumann coefficients in the Moyal star computation is
given in (5.32--5.34) in \cite{BM2} for the matter sector, and Eqs.(\ref
{eq:Neumann_X},\ref{eq:Neumann_zero}) in this paper for the ghost sector. In
the tables we write the ratio with their limiting value, $\mathcal{M}%
_{ee^{\prime }}^{(0)}(N)/\mathcal{M}_{ee^{\prime }}^{(0)}\left( cft\right) $
and $X_{ee^{\prime }}^{(0)}(N)/X_{ee^{\prime }}^{(0)}\left( cft\right) $,
where the limiting value is taken as the CFT value given in \cite{GJ}. The
tables at different values of $N$ clearly show the convergence 
\begin{equation}
\lim_{N\rightarrow \infty }\frac{\mathcal{M}_{ee^{\prime }}^{(0)}(N)}{%
\mathcal{M}_{ee^{\prime }}^{(0)}\left( cft\right) }=1\,,\quad
\lim_{N\rightarrow \infty }\frac{X_{ee^{\prime }}^{(0)}(N)}{X_{ee^{\prime
}}^{(0)}\left( cft\right) }=1\,\,.
\end{equation}
Namely in the open string limit, the Neumann coefficients derived
algebraically in MSFT becomes identical with their analytic value computed
in CFT.

We note that the convergence of the Neumann coefficients of the ghost sector
is much slower than those of matter sector. However log-log plot of $|%
\mathcal{M}(N)/\mathcal{M}(cft)-1|$ against $N$ clearly shows that the
deviation scales as power of $N$ with a very good accuracy.

As examples, we write the fitting of $(2,2)$ and $(2,4)$ components of above
ratios as\footnote{
These are based on the numerical data for $N=20,50,100,200,400$.
} 
\begin{eqnarray}
&&\frac{\mathcal{M}_{22}^{(0)}(N)}{\mathcal{M}_{22}^{(0)}(cft)}\sim
1+1.33\cdot N^{-1.34}\,,\qquad \frac{\mathcal{M}_{24}^{(0)}(N)}{\mathcal{M}%
_{24}^{(0)}(cft)}\sim 1+2.38\cdot N^{-1.36}\,,  \\
&&\frac{X_{22}^{(0)}(N)}{X_{22}^{(0)}(cft)}\sim 1+0.834\cdot
N^{-0.669}\,\,,\qquad \frac{X_{24}^{(0)}(N)}{X_{24}^{(0)}(cft)}\sim
1+1.22\cdot N^{-0.684}\,\,.  
\end{eqnarray}
We have numerically checked that all the Neumann coefficients including the
zero mode behave exactly the same way as above, 
\begin{equation}
\frac{\mathcal{M}^{(0,\pm )}(N)}{\mathcal{M}^{(0,\pm )}(cft)}-1\sim \alpha
_{nm}^{(0,\pm ){m}}N^{-\beta _{m}}\,,\qquad \frac{{X}^{(0,\pm )}(N)}{{X}%
^{(0,\pm )}(cft)}-1\sim \alpha _{nm}^{(0,\pm ){gh}}N^{-\beta _{gh}},
\label{eq_deviation}
\end{equation}
where the coefficients $\alpha $'s are order one quantity which depends on
the type of the Neumann coefficients. 
On the other hand, the power $\beta _{m}$ and $\beta _{gh}$ are universal for
matter and ghost sector. In the numerical study so far, $\beta _{m}\sim 1.33$
and $\beta _{gh}\sim 0.67$ for all types of Neumann coefficients. We suspect
that there may be an analytic evaluation of the deviations which will prove
such a systematic behavior. In any case, Eq.(\ref{eq_deviation}) gives a
useful numerical estimate of the deviation at finite $N$ from the $N=\infty $
values.

\section{Applications \label{sec:partIII}}

In this section, we consider the applications of the Moyal star formulation
in the ghost sector. We discuss two topics which are essential in the
development of MSFT.

The first issue is the derivation of the \textit{regularized} string field
theory action in the Siegel gauge including ghosts 
\begin{equation}
S=-\int d^{d}\bar{x}\,\mathrm{Tr}\left( {\frac{1}{2\alpha ^{\prime }}}%
\,A\star (L_{0}-1)A+{\frac{g}{3}}\,A\star A\star A\right)\,.
\label{eq:MSFT_action}
\end{equation}
The regularized version was the starting point of our recent discussions in 
\cite{BKM1,BKM2}.

The second issue is the derivation of the Feynman rules in the ghost sector.
Together with our previous work on Feynman diagrams in the matter sector 
\cite{BKM1} this provides the complete set of Feynman rules. We show some
explicit examples of computations of amplitudes.

\subsection{Regularized MSFT action and equation of motion}

We start from Witten's string field theory action in the operator
formulation 
\begin{equation}
S={\frac{1}{2}}\langle \Psi |Q_{B}|\Psi \rangle +{\frac{1}{3}}\langle \Psi
|\Psi \star ^{W}\Psi \rangle\, .  \label{eq:Witten's action}
\end{equation}
$Q_{B}$ is BRST operator, which may be written by separating out the $%
b_{0},c_{0}$ zero modes 
\begin{equation}
Q_{B}=\hat{c}_{0}(L_{0}-1)+2\hat{X}\hat{b}_{0}+\tilde{Q}\,,
\end{equation}
with 
\begin{eqnarray}
L_{0} &=&L_{0}^{matter}+L_{0}^{ghost}:={\frac{1}{2}}\alpha
_{0}^{2}+\sum_{n=1}^{\infty }\alpha _{-n}\alpha _{n}+\sum_{n=1}^{\infty }n(%
\hat{b}_{-n}\hat{c}_{n}+\hat{c}_{-n}\hat{b}_{n})\,,   \\
\hat{X} &=&-\sum_{n=1}^{\infty }n\hat{c}_{-n}\hat{c}_{n}\,,   \\
\tilde{Q} &=&\sum_{n\neq 0}\hat{c}_{-n}L_{n}^{matter}
+\sum_{m,n,m+n\neq 0}{\frac{
m-n}{2}}\hat{c}_{m}\hat{c}_{n}\hat{b}_{-m-n}\,,   \\
L_{n}^{matter} &=&{\frac{1}{2}}\sum_{m=-\infty }^{\infty }\alpha _{-m}\alpha
_{n+m}\,.
\end{eqnarray}
By imposing the Siegel gauge condition $\hat{b}_{0}|\Psi \rangle =0$ we
obtain the gauge fixed action 
\begin{equation}
S={\frac{1}{2}}\langle \Psi |\hat{c}_{0}(L_{0}-1)|\Psi \rangle +{\frac{1}{3}}%
\langle \Psi |\Psi \star ^{W}\Psi \rangle \,.  \label{eq:action_bra_ket}
\end{equation}
In the regularized version of MSFT with cut-off parameters $(N,\kappa
_{e},\kappa _{o})$, we cannot write a nilpotent $Q_{B}$ operator, at least
technically for the time being, because the conformal symmetry is explicitly
broken when the parameters $(N,\kappa _{e},\kappa _{o})$ are not at their
limiting values. Of course, any other approach that attempts to work with a
finite number of modes (such as level truncation) suffers from the same
problem. For complete control, what seems to be desirable is the
construction of a finite dimensional Lie algebra that would be a substitute
for the Virasoro algebra at finite $N,$ and which would tend to the Virasoro
algebra at infinite $N.$ If such an algebra could be constructed, then a
regulated version of $Q_{B}$ at finite $N$ would be straightforward, at
least in MSFT.

On the other hand we have seen in numerous cases by now that the regulator
is indispensable. With this restriction, we are forced to work with the
gauge fixed action Eq.(\ref{eq:action_bra_ket}) where the truncation of the
oscillators can be made self-consistently. In the open string limit, we
recover the original gauge fixed action which is equivalent to the original
gauge invariant action Eq.(\ref{eq:Witten's action}).

In the following we rewrite the action (\ref{eq:action_bra_ket}) in the
Moyal language. We use a field 
$\hat{A}(\bar{x},\xi _{0},\xi )=\xi _{0}\,A(\bar{x},\xi )$ 
in the Siegel gauge which is related to a conventional string field 
$\Psi $ by Fourier transformation (\ref{eq:MSFTfield_matter})(\ref
{eq:MSFTfield}). The kinetic term is rewritten as, 
\begin{eqnarray}
\langle \Psi |\hat{c}_{0}(L_{0}-1)|\Psi \rangle &=&\int (-d\xi _{0})\int
d^{d}\bar{x}\,\mathrm{Tr}\left( \hat{A}(\bar{x},\xi _{0},\xi )\star \hat{%
\beta}_{\hat{c}_{0}}(L_{0}-1)\hat{A}(\bar{x},\xi _{0},\xi )\right)  \notag \\
&=&\int d^{d}\bar{x}\,\mathrm{Tr}\left( A(\bar{x},\xi )\star (L_{0}-1)A(\bar{%
x},\xi )\right)
\end{eqnarray}
where\footnote{%
In this section, we use the variable $\bar{\xi}^{gh}=(\bar{\xi}_{e}^{b},\bar{%
\xi}_{e}^{c})=(x_{e}^{b},p_{e}^{b},x_{e}^{c},p_{e}^{c})$ which was
introduced in Eq.(\ref{bc_even_variable}) because this makes $L_{0}^{ghost}$
most similar to the $L_{0}^{matter}$.} $L_{0}=L_{0}^{matter}+L_{0}^{ghost}$
is given by 
\begin{eqnarray}
&&L_{0}^{matter}={\frac{1}{2}}\beta _{0}^{2}-{\frac{d}{2}}\mathrm{Tr}\,%
\tilde{\kappa}-{\frac{1}{4}}\bar{D}_{\xi }M_{0}^{-1}{\tilde{\kappa}}D_{\xi }+%
\bar{\xi}\tilde{\kappa}M_{0}\xi \,,  \notag \\
&&L_{0}^{ghost}=\mathrm{Tr}\,\tilde{\kappa}^{gh}-{\frac{1}{2}}\bar{\frac{%
\partial }{\partial \xi ^{b}}}\left( M_{0}^{gh}\right) ^{-1}\tilde{\kappa}%
^{gh}{\frac{\partial }{\partial \xi ^{c}}}+2\bar{\xi}^{b}\tilde{\kappa}%
^{gh}M_{0}^{gh}\xi ^{c}  \notag \\
&&~~~~~~~~=\mathrm{Tr}\,\tilde{\kappa}^{gh}-{\frac{1}{4}}\bar{\frac{\partial 
}{\partial \xi ^{gh}}}\varepsilon \left( M_{0}^{gh}\right) ^{-1}\tilde{\kappa%
}^{gh}{\frac{\partial }{\partial \xi ^{gh}}}+\bar{\xi}^{gh}\varepsilon 
\tilde{\kappa}^{gh}M_{0}^{gh}\xi ^{gh}\,,  \label{eq:total_L0} \\
&&\beta _{0}=-il_{s}{\frac{\partial }{\partial \bar{x}}}\,,~D_{\xi }=\left( 
\begin{array}{c}
{\frac{\partial }{\partial x_{e}}}-i{\frac{w_{e}}{l_{s}}}\beta _{0} \\ 
{\frac{\partial }{\partial p_{e}}}
\end{array}
\right) \,,~\tilde{\kappa}=\left( 
\begin{array}{cc}
\kappa _{e} & 0 \\ 
0 & T\kappa _{o}R
\end{array}
\right) \,,~M_{0}=\left( 
\begin{array}{cc}
{\frac{\kappa _{e}}{2l_{s}^{2}}} & 0 \\ 
0 & {\frac{2l_{s}^{2}}{\theta ^{2}}}T\kappa _{o}^{-1}\bar{T}
\end{array}
\right) \,,  \notag \\
&&\tilde{\kappa}^{gh}=\left( 
\begin{array}{cc}
\bar{R}\kappa _{o}\bar{T} & 0 \\ 
0 & \kappa _{e}
\end{array}
\right) \,,\quad \varepsilon =\left( 
\begin{array}{cc}
0 & 1 \\ 
-1 & 0
\end{array}
\right) \,,\quad M_{0}^{gh}=\left( 
\begin{array}{cc}
{\frac{i}{2}}\bar{R}\kappa _{o}R & 0 \\ 
0 & {\frac{2i}{{\theta ^{\prime }}^{2}}}\kappa _{e}^{-1}
\end{array}
\right)  \notag
\end{eqnarray}
and the Moyal $\star $ product and the trace are 
\begin{eqnarray}
&&\star =\exp \left( {\frac{1}{2}}{\frac{\overleftarrow{\partial }}{\partial
\xi }}\sigma {\frac{\overrightarrow{\partial }}{\partial \xi }}+{\frac{1}{2}}%
{\frac{\overleftarrow{\partial }}{\partial \xi ^{gh}}}\Sigma {\frac{%
\overrightarrow{\partial }}{\partial \xi ^{gh}}}\right) \,,~~~\mathrm{Tr}={%
\frac{\det \sigma'}{|\det (2\pi \sigma )|^{d/2}}}\int d^{2Nd}\xi
\,~d^{4N}\xi ^{gh}\,,  \label{eq:total_star} \\
&&~~~~~~\sigma =i\theta \left( 
\begin{array}{cc}
0 & 1 \\ 
-1 & 0
\end{array}
\right) \,,~~~~~~~\Sigma =\left( 
\begin{array}{cc}
\sigma' & 0 \\ 
0 & \sigma'
\end{array}
\right) \,,~~~\sigma'=\theta ^{\prime }\left( 
\begin{array}{cc}
0 & 1 \\ 
1 & 0
\end{array}
\right) \,.  \notag
\end{eqnarray}
On the other hand, the cubic term of the action becomes 
\begin{eqnarray}
\langle \Psi |\Psi \star ^{W}\Psi \rangle &=&\mu _{3}^{-1}\int d\xi
_{0}^{(3)}d\xi _{0}^{(2)}d\xi _{0}^{(1)}\int d^{d}\bar{x}\,\mathrm{Tr}\left( 
\hat{A}\left( \bar{x},\xi _{0}^{(1)},\xi \right) \star \hat{A}\left( \bar{x}%
,\xi _{0}^{(2)},\xi \right) \star \hat{A}\left( \bar{x},\xi _{0}^{(3)},\xi
\right) \right)  \notag \\
&=&\mu _{3}^{-1}\int d^{d}\bar{x}\,\mathrm{Tr}\left( A\left( \bar{x},\xi
\right) \star A\left( \bar{x},\xi \right) \star A\left( \bar{x},\xi \right)
\right)
\end{eqnarray}
where 
\begin{equation}
\mu _{3}=-2^{2N(d-2)}(1+\bar{w}w)^{-{\frac{d}{8}}+{\frac{3}{4}}}(\det (3+t%
\bar{t}))^{-d}(\det (1+3t\bar{t}))^{2}\,,~~~~~t:=\kappa
_{e}^{1/2}T\kappa _{o}^{-1/2}\,.  \label{normalization}
\end{equation}
After an appropriate rescaling of $A$, we obtain the gauge fixed action (\ref
{eq:MSFT_action}) in MSFT language: 
\begin{eqnarray}
S &=&-\int d^{d}\bar{x}\,\mathrm{Tr}\biggl({\frac{1}{2\alpha ^{\prime }}}%
\int (-d\xi _{0})\hat{A}(\bar{x},\xi _{0},\xi )\star \hat{\beta}_{\hat{c}%
_{0}}(L_{0}-1)\hat{A}(\bar{x},\xi _{0},\xi )  \notag \\
&&~~~~~~~~~~~~~~+{\frac{g}{3}}\int d\xi _{0}^{(3)}d\xi _{0}^{(2)}d\xi
_{0}^{(1)}\hat{A}\left( \bar{x},\xi _{0}^{(1)},\xi \right) \star \hat{A}%
\left( \bar{x},\xi _{0}^{(2)},\xi \right) \star \hat{A}\left( \bar{x},\xi
_{0}^{(3)},\xi \right) \biggr)  \notag \\
&=&-\int d^{d}\bar{x}\,\mathrm{Tr}\left( {\frac{1}{2\alpha ^{\prime }}}\,A(%
\bar{x},\xi )\star (L_{0}-1)A(\bar{x},\xi )+{\frac{g}{3}}\,A(\bar{x},\xi
)\star A(\bar{x},\xi )\star A(\bar{x},\xi )\right)
\label{eq:action_in_MSFT0}
\end{eqnarray}
where $\hat{A}\left( \bar{x},\xi _{0},\xi \right) =\xi _{0}\,A(\bar{x},\xi )$
is a Grassmann odd field in the Siegel gauge. The conventional reality
condition of the string field in Fock space $\langle V_{2}|\Psi \rangle
=(|\Psi \rangle )^{\dagger }$ is simply given by the usual reality of the
field in Moyal space $A\left( \bar{x},\xi \right) ^{\dagger }=A\left( \bar{x}%
,\xi \right) $.

The equation of motion in MSFT becomes 
\begin{equation}
(L_{0}-1)\,A(\bar{x},\xi )+\alpha ^{\prime }g\,A(\bar{x},\xi )\star A(\bar{x}%
,\xi )=0,  \label{eq:EOM_MSFT}
\end{equation}
which corresponds to the equation of motion in the Siegel gauge $%
(L_{0}-1)\Psi +b_{0}\Psi \star ^{W}\Psi =0$ in conventional language. The
counterpart of the usual classical equation of motion $Q_{B}\Psi +\Psi \star
^{W}\Psi =0$ is difficult to express in the cut-off theory as we already
commented. Similarly we meet a similar difficulty to express the BRST
invariance condition $\tilde{Q}\Psi +b_{0}c_{0}\Psi \star ^{W}\Psi =0$ in
the Siegel gauge in MSFT at this stage.

\subsection{Computing Feynman graphs including fermionic ghost sector}

We have defined the gauge fixed action Eq.(\ref{eq:action_in_MSFT0}) in MSFT
language. Based on it, we discuss the Feynman rules in MSFT and show simple
examples explicitly. Computations in the matter sector have already been
presented in \cite{BKM1}.

\paragraph{Vertex}

In MSFT the $n$-string interaction vertex is represented by $n$-th Moyal $%
\star $ product and its trace (\ref{eq:total_star}). In Fourier basis, $e^{i%
\bar{\xi}\eta }e^{-\bar{\xi}^{gh}\eta ^{gh}}$, this amounts to a phase
factor to represent the vertex as follows: 
\begin{eqnarray}
&&\mathrm{Tr}\left( (e^{i\bar{\xi}\eta _{1}}e^{-\bar{\xi}^{gh}\eta
_{1}^{gh}})\star \cdots \star (e^{i\bar{\xi}\eta _{n}}e^{-\bar{\xi}^{gh}\eta
_{n}^{gh}})\right)  \notag \\
&=&{\frac{(-)^{N}(\theta ^{\prime })^{2N}}{(2\pi \theta )^{Nd}}}\,\exp
\left( -{\frac{1}{2}}\sum_{i<j}\bar{\eta}_{i}\sigma \eta _{j}-{\frac{1}{2}}%
\sum_{i<j}\bar{\eta}_{i}^{gh}\Sigma \eta _{j}^{gh}\right)  \notag \\
&&\times (2\pi )^{2Nd}\delta ^{2Nd}(\eta _{1}+\cdots +\eta _{n})\,\delta
^{4N}(\eta _{1}^{gh}+\cdots +\eta _{n}^{gh})\,.
\end{eqnarray}
The constant factor comes from $|\det (2\pi \sigma )|^{-d/2}\det \sigma'
 $ in the definition of the trace.

\paragraph{Propagator}

It is convenient to introduce the propagator $\Delta (\eta ,\eta ^{\prime
},\tau ,p)$ in Fourier basis. This was computed in the matter sector in \cite
{BKM1}. Here we give the complete form, including the fermionic ghost sector 
\begin{eqnarray}
\Delta (\eta ,\eta ^{\prime },\tau ,p)&:= &\int {\frac{d^{2Nd}\xi }{(2\pi
)^{2Nd}}}\,d^{4N}\xi ^{gh}\,(e^{-i\bar{\xi}\eta }e^{\bar{\xi}^{gh}\eta
^{gh}})\,e^{-\tau L_{0}(p)}(e^{i\bar{\xi}\eta ^{\prime }}e^{-\bar{\xi}%
^{gh}{\eta'}^{gh}})  \notag \\
&=&e^{{\frac{d-2}{2}}\tau \sum_{n>0}\kappa _{n}+{\frac{1}{2}}(1+\bar{w}%
w)l_{s}^{2}p^{2}}\int {\frac{d^{2Nd}\xi }{(2\pi )^{2Nd}}}\,d^{4N}\xi
^{gh}\,e^{-i\bar{\xi}\eta }e^{\bar{\xi}^{gh}\eta ^{gh}}  \notag \\
&&\times \,e^{-{\frac{il_{s}\tau }{2}}pw_{e}\eta _{e}^{x}+{\frac{\tau }{4}}%
\bar{\eta}^{\prime }M_{0}^{-1}\tilde{\kappa}\eta ^{\prime }+\tau \bar{\frac{%
\partial }{\partial \eta ^{\prime }}}\tilde{\kappa}M_{0}{\frac{\partial }{%
\partial \eta ^{\prime }}}}e^{i\bar{\xi}\eta^{\prime }}  \notag \\
&&\times \,e^{{\frac{\tau }{4}}\bar{\eta}'{}^{gh}\varepsilon
M_{0}^{gh-1}\tilde{\kappa}^{gh}{\eta'}^{gh}-\tau \bar{\frac{\partial 
}{\partial {\eta'}^{gh}}}\varepsilon \tilde{\kappa}^{gh}M_{0}^{gh}{%
\frac{\partial }{\partial {\eta'}^{gh}}}}e^{-\bar{\xi}^{gh}{\eta'}
^{gh}}\,.  \label{eq:propagator_def}
\end{eqnarray}
Here $L_{0}(p)$ is given by setting $\beta _{0}=l_{s}p$ in 
Eq.(\ref{eq:total_L0}). 
Using Eqs.(\ref{eq:formula_matter_heat},\ref{eq:formula_heat}
), we obtain the propagator in the $bc$ ghost sector 
\begin{eqnarray}
\Delta (\eta ,\eta ^{\prime },\tau ,p)&=&g(\tau ,p)\,e^{-\bar{\eta}F(\tau
)\eta -\bar{\eta}^{\prime }F(\tau )\eta ^{\prime }+2\bar{\eta}G(\tau )\eta
^{\prime }+(\bar{\eta}+\bar{\eta}^{\prime })H(\tau ,p)}\nonumber\\
&&\times \,e^{\bar{\eta}^{gh}
F^{gh}(\tau )\eta^{gh} +\bar{\eta}'{}^{gh}F^{gh}(\tau ){\eta'}^{gh}-2\bar{
\eta}'{}^{gh}G^{gh}(\tau )\eta^{gh}}\label{eq:propagator_total}
\end{eqnarray}
where 
\begin{eqnarray}
g(\tau ,p) &=&\left( \frac{\theta }{2\pi }\right) ^{Nd}\,{\frac{(-1)^{N}}{{%
\theta ^{\prime }}^{2N}}}(1+\bar{w}w)^{\frac{d+2}{4}}  \notag\\
&&\times \,\left( \prod_{e>0}(1-e^{-2\tau \kappa
_{e}})\prod_{o>0}(1-e^{-2\tau \kappa _{o}})\right) ^{-\frac{d-2}{2}%
}e^{-\left( {\frac{\tau }{2}}+\bar{w}{\ \frac{\tanh (\frac{\tau \kappa _{e}}{%
2})}{\kappa _{e}}}w\right) l_{s}^{2}p^{2}}\,,  \notag \\
F(\tau ) &=&{\frac{1}{4}}M_{0}^{-1}(\tanh (\tau \tilde{\kappa}))^{-1}=\left( 
\begin{array}{cc}
{\frac{l_{s}^{2}}{2\kappa _{e}}}(\tanh (\tau \kappa _{e}))^{-1} & 0 \\ 
0 & \frac{\theta ^{2}}{8l_{s}^{2}}\bar{R}\kappa _{o}(\tanh (\tau \kappa
_{o}))^{-1}R
\end{array}
\right) ,  \notag \\
G(\tau ) &=&{\frac{1}{4}}M_{0}^{-1}(\sinh (\tau \tilde{\kappa}))^{-1}=\left( 
\begin{array}{cc}
{\frac{l_{s}^{2}}{2\kappa _{e}}}(\sinh (\tau \kappa _{e}))^{-1} & 0 \\ 
0 & \frac{\theta ^{2}}{8l_{s}^{2}}\bar{R}\kappa _{o}(\sinh (\tau \kappa
_{o}))^{-1}R
\end{array}
\right) ,   \notag\\
H(\tau ,p) &=&\frac{\tanh (\tau \kappa _{e}/2)}{\kappa _{e}}wl_{s}^{2}p\,, 
 \\
F^{gh}(\tau ) &=&{\frac{1}{4}}\varepsilon M_{0}^{gh-1}(\tanh \tau {\tilde{%
\kappa}}^{gh})^{-1}=\varepsilon \left( 
\begin{array}{cc}
-{\frac{i}{2}}T\kappa _{o}^{-1}(\tanh \tau \kappa _{o})^{-1}\bar{T} &  \\ 
& -{\frac{i\theta ^{^{\prime }2}}{8}}\kappa _{e}(\tanh \tau \kappa _{e})^{-1}
\end{array}
\right) , \notag  \\
G^{gh}(\tau ) &=&{\frac{1}{4}}\varepsilon M_{0}^{gh-1}(\sinh \tau \tilde{%
\kappa}^{gh})^{-1}=\varepsilon \left( 
\begin{array}{cc}
-{\frac{i}{2}}T\kappa _{o}^{-1}(\sinh \tau \kappa _{o})^{-1}\bar{T} &  \\ 
& -{\frac{i\theta ^{^{\prime }2}}{8}}\kappa _{e}(\sinh \tau \kappa _{e})^{-1}
\end{array}
\right) .\notag
\end{eqnarray}
The ghost structure of the quadratic term in the exponent is similar to the
matter one.

\paragraph{1-loop vacuum amplitude}

By taking the trace of the propagator Eq.(\ref{eq:propagator_total}), we
have 
\begin{eqnarray}
\int d^{d}p\,\mathrm{Tr}\,e^{-\tau (L_{0}-1)} &=&\int d^{d}p\int d^{2Nd}\eta
\,d^{4N}\eta ^{gh}\,e^{\tau}\,\Delta (\eta ,\eta ,p,\tau )  \notag \\
&=&(2\pi )^{\frac{d}{2}}l_{s}^{-d}\tau ^{-{\frac{d}{2}}}\,e^{\tau
}\prod_{e>0}(1-e^{-\tau \kappa _{e}})^{-(d-2)}\prod_{o>0}(1-e^{-\tau \kappa
_{o}})^{-(d-2)}\,.~~~~
\end{eqnarray}
This reproduces the expected partition function, with the correct spectrum,
including the ghost contribution. If we take the open string limit $\kappa
_{e}=e,\kappa _{o}=o,N=\infty $ naively in the formula of $L_{0}$ (\ref
{eq:L0_matter},\ref{eq:L0_bc_even}), namely at the Lagrangian level, we lose
the information on odd spectrum. This is one of the indications that the $%
\gamma $ term plays a nontrivial role. Of course, we obtain the correct
limiting partition function by taking the limit at the last stage of the
computation, which is given above.

\paragraph{External state}

As external states in Feynman graphs it is enough to consider monoid
elements such as 
\begin{eqnarray}
A_{\mathcal{N},M,\lambda ,M^{gh},\lambda ^{gh}}=\mathcal{N}\,e^{ip\bar{x}%
}e^{-\bar{\xi}M\xi -\bar{\xi}\lambda }e^{-\bar{\xi}^{gh}M^{gh}\xi ^{gh}-\bar{%
\xi}^{gh}\lambda ^{gh}}\,.  \label{eq:monoid_total}
\end{eqnarray}
In fact, we can compute various perturbative diagrams by preparing a
particular class of gaussian external states given by $M=M_{0},M^{gh}=%
\varepsilon M_{0}^{gh}$, where these matrices were given in Eqs.(\ref
{eq:total_L0}). If we also take $\bar{\lambda}=(-iw_{e}p,0),\lambda ^{gh}=0$%
, this external field represents perturbative vacuum ${\hat{c}}_{1}|p,\Omega
\rangle $ with momentum $p,$ which represents the perturbative tachyon with
proper normalization 
\begin{eqnarray}
&&A_{p}(\xi ) =2^{N(d-2)}(1+\bar{w}w)^{-{\frac{d+2}{8}}}e^{ip\bar{x}}e^{-%
\bar{\xi}M_{0}\xi +ipw_{e}x_{e}}e^{-\bar{\xi}^{gh}\varepsilon M_{0}^{gh}\xi
^{gh}}\,,~~~\mathrm{Tr}(A_{p}(\xi )^{\dagger }\star A_{p}(\xi ))=1\,.~~~~~~
\label{eq:perturbative_vacuum_even}
\end{eqnarray}
We omitted an overall $\xi _{0}$ because it drops out in the Siegel gauge
action (\ref{eq:action_in_MSFT0}). Excited states (that correspond to
polynomials multiplying this $A_{p}(\xi )$) can be obtained by
differentiating $A_{p}(\xi )e^{-\bar{\xi}\lambda -\bar{\xi}^{gh}\lambda
^{gh}}$ with respect to general $\lambda ,\lambda ^{gh}$ appropriately, and
then setting $\lambda =(-iw_{e}p,0),\lambda ^{gh}=0$. Therefore an explicit
computation of Feynman graphs with general $\lambda ,\lambda ^{gh},M,M^{gh}$
has many physical applications.

\paragraph{$\protect\tau$-evolved monoid element}

It is convenient to have a $\tau $-evolved formula for a gaussian Eq.(\ref
{eq:monoid_total}) to compute Feynman graphs in $\xi $-basis \cite{BKM1}. We
can derive it explicitly by evaluating: 
\begin{eqnarray}
&&e^{-\tau L_{0}}A_{\mathcal{N},M,\lambda ,M^{gh},\lambda ^{gh}}(\xi ) 
\notag \\
&=&\int d^{2Nd}\eta \,d^{4N}\eta ^{gh}\,e^{-i\bar{\xi}\eta }e^{\bar{\xi}%
^{gh}\eta ^{gh}}\left( \int d^{2Nd}\eta^{\prime }\,d^{4N}{\eta'}^{gh}
\Delta (\eta ,\eta ^{\prime },\tau ,p)\,\tilde{A}_{\mathcal{N},M,\lambda
,M^{gh},\lambda ^{gh}}(\eta ^{\prime })\right) \,,  \notag \\
&&\tilde{A}_{\mathcal{N},M,\lambda ,M^{gh},\lambda ^{gh}}(\eta )=\int {\frac{%
d^{2Nd}\xi }{(2\pi )^{2Nd}}}\,d^{4N}\xi ^{gh}\,e^{-i\bar{\xi}\eta }e^{\bar{%
\xi}^{gh}\eta ^{gh}}A_{\mathcal{N},M,\lambda ,M^{gh},\lambda ^{gh}}(\xi )\,.
\end{eqnarray}
By gaussian integration we have the following formula 
\begin{equation}
e^{-\tau L_{0}}A_{\mathcal{N},M,\lambda ,M^{gh},\lambda ^{gh}}(\xi )=%
\mathcal{N}\mathcal{N}^{m}(\tau )\mathcal{N}^{gh}(\tau )\,e^{ip\bar{x}}e^{-%
\bar{\xi}M(\tau )\xi -\bar{\xi}\lambda (\tau )}e^{-\bar{\xi}^{gh}M^{gh}(\tau
)\xi ^{gh}-\bar{\xi}^{gh}\lambda ^{gh}(\tau )}~~
\end{equation}
where 
\begin{align}
M\!\left( \tau \right) & =\left[ \sinh \tau \tilde{\kappa}+\left( \sinh \tau 
\tilde{\kappa}+M_{0}M^{-1}\cosh \tau \tilde{\kappa}\right) ^{-1}\right]
\left( \cosh \tau \tilde{\kappa}\right) ^{-1}M_{0}\,,   \\
\lambda \!\left( \tau \right) & =\left[ \left( \cosh \tau \tilde{\kappa}%
+MM_{0}^{-1}\sinh \tau \tilde{\kappa}\right) ^{-1}\left( \lambda +iwp\right) %
\right] -iwp\,,   \\
\mathcal{N}^{m}\!\left( \tau \right) & =\frac{e^{-\frac{1}{2}%
l_{s}^{2}p^{2}\tau }~\exp \left[ \frac{1}{4}\left( \bar{\lambda}+ip\bar{w}%
\right) \left( M+\coth \tau \tilde{\kappa}~M_{0}\right) ^{-1}\left( \lambda
+iwp\right) \right] }{\det \left( \frac{1}{2}\left( 1+MM_{0}^{-1}\right) +%
\frac{1}{2}\left( 1-MM_{0}^{-1}\right) e^{-2\tau \tilde{\kappa}}\right)
^{d/2}}
\end{align}
for the matter sector \cite{BKM1}\footnote{%
We supposed that $M$ is Lorentz symmetric.} and 
\begin{align}
M^{gh}(\tau )=& \left[ \sinh \tau {\tilde{\kappa}^{gh}+\left( \sinh \tau 
\tilde{\kappa}^{gh}+\varepsilon M_{0}^{gh}M^{gh-1}\cosh \tau \tilde{\kappa}%
^{gh}\right) }^{-1}\right] (\cosh \tau \tilde{\kappa}^{gh})^{-1}\varepsilon
M_{0}^{gh}\,,  \label{eq:tau-evolved} \\
\lambda ^{gh}(\tau )=& \left[ \cosh \tau \tilde{\kappa}^{gh}-M^{gh}%
\varepsilon M_{0}^{gh-1}\sinh \tau \tilde{\kappa}^{gh}\right] ^{-1}\lambda
^{gh}\,,   \\
\mathcal{N}^{gh}(\tau )=& \,e^{-{\frac{1}{4}}\bar{\lambda}^{gh}(M^{gh}+\coth
(\tau \tilde{\kappa}^{gh})\varepsilon M_{0}^{gh})^{-1}\lambda ^{gh}}  \notag
\\
& \times \,\left[ \det \left( {\frac{1}{2}}(1-M^{gh}\varepsilon
M_{0}^{gh-1})+{\frac{1}{2}}(1+M^{gh}\varepsilon M_{0}^{gh-1})e^{-2\tau 
\tilde{\kappa}^{gh}}\right) \right] ^{\frac{1}{2}}  
\end{align}
for ghost sector. When we consider a class of monoid such that $\bar{\xi}%
^{gh}M^{gh}\xi ^{gh}$ is $SU(1,1)$-symmetric and twist even\footnote{%
In \S \ref{sec:SU11}, we defined $SU(1,1)$-symmetric and twist even monoid
element which has $\lambda ^{gh}=0$ (\ref{eq:singlet}). Here we permit $%
\lambda ^{gh}\neq 0$ case.} then the evolved $M(\tau )$ also has this
symmetry. We can see this explicitly by noting that $f(\tilde{\kappa}%
)M_{0}^{gh}$ is a block diagonal and symmetric matrix (where $f(x)$ is an
arbitrary function). In this case with $M^{gh}=\varepsilon M^{gh^{\prime }}$%
, where $M^{gh^{\prime }}$ is a $2N\times 2N$ matrix, the above formula
becomes 
\begin{align}
M^{gh}(\tau )=& \varepsilon \left[ \sinh \tau {\tilde{\kappa}^{gh}+\left(
\sinh \tau \tilde{\kappa}^{gh}+M_{0}^{gh}M^{gh^{\prime }-1}\cosh \tau \tilde{%
\kappa}^{gh}\right) }^{-1}\right] (\cosh \tau \tilde{\kappa}%
^{gh})^{-1}M_{0}^{gh}\,,  \label{eq:tau-evolved2} \\
\lambda ^{gh}(\tau )=& \left[ \cosh \tau \tilde{\kappa}^{gh}+M^{gh^{\prime
}}M_{0}^{gh-1}\sinh \tau \tilde{\kappa}^{gh}\right] ^{-1}\lambda ^{gh}\,, 
 \\
\mathcal{N}^{gh}(\tau )=& \,e^{-{\frac{1}{4}}\bar{\lambda}^{gh}\varepsilon
(M^{gh^{\prime }}+\coth (\tau \tilde{\kappa}^{gh})M_{0}^{gh})^{-1}\lambda
^{gh}}\nonumber\\
& \times \det \left( {\frac{1}{2}}(1+M^{gh^{\prime }}M_{0}^{gh-1})+{\frac{1}{2}}
(1-M^{gh^{\prime }}M_{0}^{gh-1})e^{-2\tau \tilde{\kappa}^{gh}}\right) \,. 
\end{align}
We can use this reduced formula to compute the $\tau $-evolved monoid of $n$%
-th product of the perturbative vacuum: $e^{-\tau L_{0}}\left( A_{0}(\xi
)e^{-\bar{\xi}\lambda _{1}}\star \cdots \star A_{0}(\xi )e^{-\bar{\xi}%
\lambda _{n}}\right) $ because the coefficient matrix $M_{0}^{(n)}$
(\ref{eq:star_a2}) in the quadratic term in the exponent is proportional
to $\varepsilon $.

With the above preparation, we have all that is needed to compute the ghost
contribution in various amplitudes, by following the methods that were
developed in the matter sector in \cite{BKM1}.

\paragraph{4-tachyon amplitude}

The 4 point amplitude for tachyons is computed by putting together several
diagrams that are related to each other by permutations of the external
legs. For a typical 4-pt diagram $_{1}^{2}{>}-{<}_{4}^{3}$ the MSFT
expression is 
\begin{equation}
{}_{12}A_{34}=\int d^{d}\bar{x}\,\mathrm{Tr}\left( e^{-\tau L_{0}}\left(
A_{1}(\xi )\star A_{2}(\xi )\right) \star \left( A_{3}(\xi )\star A_{4}(\xi
)\right) \right)   \label{four}
\end{equation}
where $\tau $ is the length of the propagator. When $A_{i}(\xi
),~(i=1,\cdots,4)$ are gaussians, we can compute this quantity easily by
taking the $\star $ product between the pairs of gaussians 
(\ref{e_gh_structure_a}--\ref{e_gh_structure}), 
evolving by $\tau $ (\ref{eq:tau-evolved2}), and computing
the trace (\ref{eq:gaussian-trace}). At each step we only use the properties
of the monoid.

In the case of tachyons, the matter contribution was already computed in 
\cite{BKM1}. For the ghost contribution, we set the external field to 
$A_{i}^{gh}(\xi)=2^{-2N}(1+\bar{w}w)^{-{\frac{1}{4}}}\,e^{-\bar{\xi}
^{gh}\varepsilon M_{0}^{gh}\xi ^{gh}}$ and obtain 
\begin{eqnarray}
{}_{12}A_{34}^{gh} &=&(\det (2m_{0}^{gh}))^{-1}(\det
(1-(m_{0}^{gh})^{2}))^{2}  \notag \\
&&\times \left[ \det \left( 4\sinh \tau \tilde{\kappa}^{gh}\left( \left(
\cosh \tau \tilde{\kappa}^{gh}+{\frac{2}{1+(m_{0}^{gh})^{2}}}\sinh \tau 
\tilde{\kappa}^{gh}\right) ^{2}-1\right) ^{-1}e^{\tau \tilde{\kappa}%
^{gh}}\right) \right] ^{-1}  \notag \\
&=&2^{-8N}(1+\bar{w}w)^{\frac{3}{2}}\left( \det (1+3t\bar{t})\right) ^{4} 
\notag \\
&&\times \det \left( 1-\left( {\frac{t\bar{t}-1}{1+3t\bar{t}}}\,e^{-\kappa
_{e}\tau }\right) ^{2}\right) \det \left( 1-\left( {\frac{\bar{t}t-1}{1+3%
\bar{t}t}}\,e^{-\kappa _{o}\tau }\right) ^{2}\right)
\end{eqnarray}
where $m_{0}^{gh}=M_{0}^{gh}\sigma',\,t=\kappa _{e}^{1/2}T\kappa
_{o}^{-1/2}$. Including the matter sector \cite{BKM1} we obtain 
\begin{eqnarray}
{}_{12}A_{34} &=&2^{4(d-2)N}(1+\bar{w}w)^{-{\frac{d}{4}}+{\frac{3}{2}}%
}\left( \det (1+3t\bar{t})\right) ^{4}\left( \det (3+t\bar{t})\right)
^{-2d}(2\pi )^{d}\delta ^{d}(p_{1}+p_{2}+p_{3}+p_{4})  \notag \\
&&\times {\frac{\det \left( 1-\left( {\frac{t\bar{t}-1}{1+3t\bar{t}}}%
\,e^{-\kappa _{e}\tau }\right) ^{2}\right) \det \left( 1-\left( {\frac{\bar{t%
}t-1}{1+3\bar{t}t}}\,e^{-\kappa _{o}\tau }\right) ^{2}\right) }{\left[ \det
\left( 1-\left( {\frac{t\bar{t}-1}{3+t\bar{t}}}\,e^{-\kappa _{e}\tau
}\right) ^{2}\right) \det \left( 1-\left( {\frac{\bar{t}t-1}{3+\bar{t}t}}%
\,e^{-\kappa _{o}\tau }\right) ^{2}\right) \right] ^{\frac{d}{2}}}}  \notag
\\
&&\times \exp \left( -\frac{1}{2}l_{s}^{2}(p_{1}+p_{2})^{2}(\tau +\alpha
(\tau ))+l_{s}^{2}(p_{1}+p_{3})^{2}\beta (\tau )+\frac{1}{2}%
l_{s}^{2}\sum_{i=1}^{4}p_{i}^{2}\gamma (\tau )\right)
\label{eq:4-tachyon_amp}
\end{eqnarray}
where 
\begin{eqnarray}
\alpha \left( \tau \right) &=&\bar{v}\kappa _{o}^{-{\frac{1}{2}}}\biggl[\bar{%
t}\left( 1+t\bar{t}+{\frac{1}{2}}(1+t\bar{t})\,\mathrm{coth}{\frac{\tau
\kappa _{e}}{2}}\,(1+t\bar{t})\right) ^{-1}t \\
&&~~~~~+\left( 1+\bar{t}t+{\frac{1}{2}}(1+\bar{t}t)\,\mathrm{coth}{\frac{%
\tau \kappa _{o}}{2}}\,(1+\bar{t}t)\right) ^{-1}\biggr]\kappa _{o}^{-{\frac{1%
}{2}}}v\,,  \notag\\
\beta \left( \tau \right) &=&2\bar{v}\kappa _{o}^{-{\frac{1}{2}}}\biggl(%
4\sinh \tau \kappa _{o}+(1+\bar{t}t)\,\sinh \tau \kappa _{o}\,(1+\bar{t}t) \\
&&~~~~~~~~~~~~~+2(1+\bar{t}t)\cosh \tau \kappa _{o}+2\cosh \tau \kappa
_{o}\,(1+\bar{t}t)\biggr)^{-1}\kappa _{o}^{-{\frac{1}{2}}}v\,,  \notag\\
\gamma \left( \tau \right) &=&-\bar{v}\kappa _{o}^{-{\frac{1}{2}}}\,\mathrm{%
coth}{\frac{\tau \kappa _{o}}{2}}\left( 2+(1+\bar{t}t)\mathrm{coth}{\frac{%
\tau \kappa _{o}}{2}}\right) ^{-1}\kappa _{o}^{-{\frac{1}{2}}}v\,.
\end{eqnarray}
We note that the matrices $t\bar{t},\bar{t}t$ in the determinant factors in
the matter sector come out inverted in the ghost sector. By integrating with
the measure $\int_{o}^{\infty }d\tau e^{\tau }$ and adding permutations of
the diagram, we should reproduce the Veneziano amplitude\footnote{%
The expressions here are slightly further simplified than the ones in \cite
{BKM1}.} when all the tachyons are on-shell, $l_{s}^{2}p_{i}^{2}=2$ in the
open string limit $\kappa _{e}=e,\kappa _{o}=o,N=\infty $.

Our formula (\ref{eq:4-tachyon_amp}) has a counterpart in the operator
formalism. Although our expressions are simpler it is not easy to compare
results analytically because of the different formalisms. We have managed to
compare and agree with the determinant factor available in the computations
in \cite{Wati} in the operator formalism, by inserting the MSFT ghost
Neumann coefficients given in (\ref{eq:Neumann_X}) and matter Neumann
coefficients taken from \cite{BM2} 
\begin{equation}
\mathcal{M}^{(0)}={\frac{\tilde{m}_{0}^{\ast 2}-1}{{\tilde{m}}_{0}^{\ast 2}+3%
}}\,,~~~~CX^{(0)}=\sqrt{\kappa }{\frac{\tilde{m}_{0}^{\ast 2}-1}{1+3{\tilde{m%
}}_{0}^{\ast 2}}}{\frac{1}{\sqrt{\kappa }}}\,,~~~~\tilde{m}_{0}^{\ast
2}=\left( 
\begin{array}{cc}
t\bar{t} & 0 \\ 
0 & \bar{t}t
\end{array}
\right) \,.
\end{equation}
The difference of normalization compared to \cite{Wati} comes from that of
cubic term of the action (\ref{normalization}). We note that so far it has not
been demonstrated yet that either the operator formalism or the MSFT
approach reproduce the Veneziano amplitude analytically, although this is
expected to be true.

\section{Discussion}

In this paper we provided the details of the Moyal star formulation for
fermionic ghosts. Following the similar construction in the matter sector,
the split string formalism was used as an intermediate step. However, as in
the matter sector, the midpoint needed additional considerations to insure
that MSFT is in agreement with the operator formulation of string field
theory. MSFT then provides an alternative method of computation in string
field theory which is in many ways simpler and more efficient.

The regularization of the fermionic ghost sector, which is needed to avoid
the associativity anomaly, is made consistently with the matter sector. The
correctness of the formulation, including the regularization, was tested by
computing the Neumann coefficients by using MSFT methods and comparing them
to an independent computation that relies on conformal field theory. The
MSFT result generalizes the Neumann coefficients by computing them for any
set of oscillator frequencies $\kappa _{e},\kappa _{o}$ for any finite
number of oscillators $2N.$ These agree with conformal theory results in the
open string limit $\kappa _{e}=e,$ $\kappa _{o}=o,$ $N=\infty .$ The
agreement was established both analytically as well as numerically.

In numerical study of string field theory one necessarily deals with a
finite number of modes. One may debate which version of Neumann coefficients
is more consistent for such numerical study: the finite $N$ version of the
Neumann coefficients given in MSFT, or the level truncation of the infinite
Neumann matrices practiced in previous literature? The numerical analysis
that we provided could be helpful in understanding the issue and developing
the most appropriate numerical approximation scheme. We hope to address this
point, together with numerical studies of certain quantities in the near
future.

The regularized MSFT formulation is now complete in the Siegel gauge. It has
already been applied to the computation of perturbative Feynman graphs \cite
{BKM1} as well as to the analytic study of nonperturbative classical
solutions of string theory, including the nonperturbative vacuum of open
string theory \cite{BKM2}.

An open problem is the construction of a \textit{regularized} version of the
BRST operator. The regularization is indispensable to tame the associativity
anomaly and to have a well defined theory. Along with the successful
regularization in MSFT, the BRST operator is also needed to insure gauge
invariance in the general formalism, and to be able to work outside of the
Siegel gauge. In particular, the BRST operator can be used to impose the
additional gauge invariance conditions in the Siegel gauge on the
nonperturbative solutions we have obtained in \cite{BKM2}. Some of the
issues surrounding this problem are outlined following Eq.(\ref
{eq:action_bra_ket}). These remarks apply not only to MSFT, but also to any
version of string theory that uses a cutoff of the string modes (including
level truncation), since the Virasoro algebra does not close with a finite
number of modes. A substitute for the Virasoro algebra at finite $N,$ which
tends to the Virasoro algebra at infinite $N,$ is the key to solving this
problem.

\begin{center}
\noindent{\large \textbf{Acknowledgments}}
\end{center}

I.K. would like to thank H.~Hata, T.~Kawano and K.~Ohmori for valuable
discussions and comments. I.B. is supported in part by a DOE grant
DE-FG03-84ER40168. I.K. is supported in part by JSPS Research Fellowships
for Young Scientists. Y.M. is supported in part by Grant-in-Aid (\#
13640267) from the Ministry of Education, Science, Sports and Culture of
Japan.

\appendix

\section{Brief review of MSFT in matter sector\label{sec:review_app}}

\subsection{Half-string for cosine modes\label{sec:cosine}}

Here we review the split string formulation and its regularization which was
developed in \cite{BM1} and fix notation in this paper. Although it was
constructed to formulate the matter and the bosonized ghost sector, we can
apply the same formalism to fermionic functions which have a Fourier
expansion in terms of cosine mode in the full string basis.

A full string function $\phi (\sigma )$ which satisfies Neumann boundary
conditions at the end points 
\begin{equation}
\left. {\frac{d}{d\sigma }}\phi (\sigma )\right| _{\sigma =0}=\left. {\frac{d%
}{d\sigma }}\phi (\sigma )\right| _{\sigma =\pi }=0\,,
\end{equation}
has a Fourier expansion in terms of cosine modes 
\begin{equation}
\phi (\sigma )=\phi _{0}+\sqrt{2}\sum_{n=1}^{\infty }\phi _{n}\cos n\sigma
\,,~~~~~\phi _{0}={\frac{1}{\pi }}\int_{0}^{\pi }d\sigma \phi (\sigma
)\,,\quad \phi _{n}={\frac{\sqrt{2}}{\pi }}\int_{0}^{\pi }d\sigma \phi
(\sigma )\cos n\sigma \,.~~~~~~~
\end{equation}
Then split string functions $l(\sigma ),r(\sigma )$ for $\phi (\sigma )$ are
defined as 
\begin{eqnarray}
&&\phi (\sigma )=\left\{ 
\begin{array}{lc}
l(\sigma ) & \quad (0\leq \sigma \leq {\frac{\pi }{2}}) \\ 
r(\pi -\sigma ) & \quad ({\frac{\pi }{2}}\leq \sigma \leq \pi )
\end{array}
\right. \,,   \\
&&\phi _{0}={\frac{1}{\pi }}\int_{0}^{\frac{\pi }{2}}d\sigma (l(\sigma
)+r(\sigma ))\,,\qquad \phi _{n}={\frac{\sqrt{2}}{\pi }}\int_{0}^{\frac{\pi 
}{2}}d\sigma (l(\sigma )+(-1)^{n}r(\sigma ))\cos n\sigma \,.\quad
\label{eq:philrExp}
\end{eqnarray}

\subsubsection{Dirichlet at midpoint \label{sec:cosND}}

Split string functions with Neumann boundary conditions at the end points
and Dirichlet boundary conditions at the midpoint 
\begin{equation}
l^{\prime }(0)=r^{\prime }(0)=0\,,\qquad l(\pi /2)=r(\pi /2)=\bar{\phi}%
\,(:=\phi (\pi /2))\,,
\end{equation}
have a Fourier expansion in terms of odd cosine modes $o=1,3,5,\cdots $ 
\begin{equation}
l(\sigma )=\bar{\phi}+\sqrt{2}\sum_{o=1}^{\infty }l_{o}\cos o\sigma \,,\quad
r(\sigma )=\bar{\phi}+\sqrt{2}\sum_{o=1}^{\infty }r_{o}\cos o\sigma \,.
\end{equation}
The correspondence between $\{\bar{\phi},l_{o},r_{o}\}$ and $\{\phi
_{e},\phi _{o}\}$ is 
\begin{eqnarray}
&&\bar{\phi}=\phi _{0}-\bar{w}\phi _{e}\,,~~~~~~l_{o}=\phi _{o}+R\phi
_{e}\,,~~~~~~r_{o}=-\phi _{o}+R\phi _{e}\,,  \label{eq:oddcos} \\
&&\phi _{0}=\bar{\phi}+{\frac{1}{2}}\bar{v}(l_{o}+r_{o})\,,~~\phi _{e}={%
\frac{1}{2}}T(l_{o}+r_{o})\,,~~\phi _{o}={\frac{1}{2}}(l_{o}-r_{o})\,,
\end{eqnarray}
where we used matrix notation for simplicity, and denoted 
\begin{eqnarray}
&&R_{oe} ={\frac{4}{\pi }}\int_{0}^{\frac{\pi }{2}}d\sigma \cos o\sigma
\left( \cos e\sigma -\cos \frac{e\pi }{2}\right) =\frac{4e^{2}\,i^{o-e+1}}
{\pi o(e^{2}-o^{2})}\,,\qquad\notag   \\
&&T_{eo} ={\frac{4}{\pi }}\int_{0}^{\frac{\pi }{2}}d\sigma \cos e\sigma \cos
o\sigma =\frac{4o\,i^{o-e+1}}{\pi (e^{2}-o^{2})}\,, \label{eq:wRTvinfty} \\
&&v_{o} =\frac{2\sqrt{2}\,i^{o-1}}{\pi o}{=}\frac{1}{\sqrt{2}}T_{0o}\,,
~~~~~~~~~~~
w_{e} =\sqrt{2}\,i^{-e+2}\,.\notag
\label{eq:wRTvinfty}
\end{eqnarray}

\subsubsection{Neumann at midpoint \label{sec:cosNN}}

Split string functions with Neumann boundary conditions at the end points
and Neumann boundary conditions at the midpoint 
\begin{equation}
l^{\prime }(0)=r^{\prime }(0)=0\,,\qquad l^{\prime }(\pi /2)=r^{\prime }(\pi
/2)=0\,,
\end{equation}
are expanded in terms of even cosine modes $e=2,4,6,\cdots $ 
\begin{equation}
l(\sigma )=\bar{\phi}+\sqrt{2}\sum_{e=2}^{\infty }l_{e}\left( \cos e\sigma
-i^{e}\right) \,,\quad r(\sigma )=\bar{\phi}+\sqrt{2}\sum_{e=2}^{\infty
}r_{e}\left( \cos e\sigma -i^{e}\right) \,.~~~~
\end{equation}
The correspondence between split string and full string modes is 
\begin{eqnarray}
&&\bar{\phi}=\phi _{0}-\bar{w}\phi _{e}\,,~~~~~~~l_{e}=\phi _{e}+T\phi
_{o}\,,~~~~~~~r_{e}=\phi _{e}-T\phi _{o}\,,  \label{eq:evencos} \\
&&\phi _{0}=\bar{\phi}+{\frac{1}{2}}\bar{w}(l_{e}+r_{e})\,,~~\phi _{e}={%
\frac{1}{2}}(l_{e}+r_{e})\,,~~\phi _{o}={\frac{1}{2}}R(l_{e}-r_{e})\,.
\end{eqnarray}

\subsubsection{Regularization \label{sec:cos_reg}}

The infinite matrices $T,R$ and vectors $v,w$ defined in Eq.(\ref
{eq:wRTvinfty}) satisfy the following relations 
\begin{equation}
R_{oe}=o^{-2}T_{eo}e^{2},\quad R_{oe}=T_{eo}+v_{o}w_{e},\quad
v_{o}=\sum_{e>0}T_{eo}w_{e},\quad w_{e}=\sum_{o>0}R_{oe}v_{e}\,.
\label{eq:infinite_def}
\end{equation}
As noted in \cite{BM1}, there is an ambiguity in naive computation using
these matrices and vectors. For example, $T$ has an inverse matrix given by $%
R,$ and yet it has a zero eigenvalue $Tv=0$. This is possible only because
they are infinite dimensional matrices. It causes associativity anomalies.
To avoid the ambiguous results that come from the associativity anomaly, a
finite matrix regularization is proposed in \cite{BM1}. We define $N\times N$
matrices $T,R$ and $N$-vectors $v,w$ by 
\begin{equation}
R_{oe}=\left( \kappa _{o}\right) ^{-2}\bar{T}_{oe}\left( \kappa _{e}\right)
^{2},\quad R_{oe}=\bar{T}_{oe}+v_{o}\bar{w}_{e},\quad v_{o}=\bar{T}%
_{oe}w_{e},\quad w_{e}=\bar{R}_{eo}v_{o}\,,  \label{define}
\end{equation}
where a bar means transpose, and we introduced a set of $2N$ frequencies $%
\kappa _{e},\kappa _{o}$. These relations are identical to the ones
satisfied by the infinite matrices in Eq.(\ref{eq:infinite_def}), but we now
use them as defining relations for finite dimensional matrices and arbitrary
frequencies. We can solve the equations in Eq.(\ref{define}) explicitly in
term of the frequencies 
\begin{eqnarray}
&&T_{eo}=\frac{w_{e}v_{o}\kappa _{o}^{2}}{\kappa _{e}^{2}-\kappa _{o}^{2}}\,,
\quad R_{oe}=\frac{w_{e}v_{o}\kappa _{e}^{2}}{\kappa _{e}^{2}-\kappa
_{o}^{2}}\,,  \label{TR_exp} \\
&&w_{e}={i^{2-e}}\frac{\prod_{o^{\prime }}\left| \kappa _{e}^{2}/\kappa
_{o^{\prime }}^{2}-1\right| ^{\frac{1}{2}}}{\prod_{e^{\prime }\neq e}\left|
\kappa _{e}^{2}/\kappa _{e^{\prime }}^{2}-1\right| ^{\frac{1}{2}}}\,,\quad
v_{o}={i^{o-1}}\frac{\prod_{e^{\prime }}\left| 1-\kappa _{o}^{2}/\kappa
_{e^{\prime }}^{2}\right| ^{\frac{1}{2}}}{\prod_{o^{\prime }\neq o}\left|
1-\kappa _{o}^{2}/\kappa _{o^{\prime }}^{2}\right| ^{\frac{1}{2}}}\,.
\label{wv_exp}
\end{eqnarray}
By using only the defining relations we can show the following further
relations for the regularized version of $T,R,v,w$. 
\begin{eqnarray}
&&TR=1_{e},\quad RT=1_{o},\quad \bar{R}R=1+w\bar{w},\quad \bar{T}T=1-v\bar{v}%
,  \notag \\
&&T\bar{T}=1-\frac{w\bar{w}}{1+\bar{w}w},\quad Tv=\frac{w}{1+\bar{w}w},\quad 
\bar{v}v=\frac{\bar{w}w}{1+\bar{w}w},  \label{relations} \\
&&Rw=v(1+\bar{w}w),\quad R\bar{R}=1+v\bar{v}\left( 1+\bar{w}w\right) . 
\notag
\end{eqnarray}
The original $T,R,v,w$ in Eq.(\ref{eq:wRTvinfty}) are reproduced by setting
the open string limit: 
\begin{equation}
\kappa _{e}=e\,,~~\kappa _{o}=o\,,~~N\rightarrow \infty \,.
\label{eq:openstringlimit}
\end{equation}
We note that at this limit $\bar{w}w$ diverges 
\begin{equation}
1+\bar{w}w=\left( \prod_{n=1}^{N}{\frac{\kappa _{2n}}{\kappa _{2n-1}}}%
\right) ^{2}~\rightarrow \left( \frac{\sqrt{\pi }\Gamma \left( N+1\right) }{%
\Gamma \left( N+\frac{1}{2}\right) }\right) ^{2}\rightarrow ~\infty \,.
\end{equation}

\subsection{Some results in matter sector {\label{sec:review}}}

Here we summarize notation and conventions in the matter sector in MSFT.%
\footnote{%
In this subsection, we use the same symbols for matter as we did for the
ghosts in the main text for some quantities, such as positions and momenta.
We can avoid confusion from the context. We omit some definitions and
details because we can refer to Ref.\cite{BM2} for them.}

\subsubsection{Oscillators in MSFT}

\begin{itemize}
\item  Mode expansion in matter sector ($\mu =0,1,\cdots ,d-1$): 
\begin{eqnarray}
X^{\mu }(\sigma ) &=&\hat{x}_{0}^{\mu }+\sqrt{2}\sum_{n=1}^{\infty }\hat{x}%
_{n}^{\mu }\cos n\sigma =\hat{x}_{0}^{\mu }+i\sqrt{2\alpha ^{\prime }}%
\sum_{n=1}^{\infty }{\frac{1}{n}}(\alpha _{n}^{\mu }-\alpha _{-n}^{\mu
})\cos n\sigma \,,  \label{position} \\
P_{\mu }(\sigma ) &=&{\frac{1}{\pi }}\left( \hat{p}_{0\mu }+\sqrt{2}%
\sum_{n=1}^{\infty }\hat{p}_{n\mu }\cos n\sigma \right) ={\frac{1}{\pi }}%
\left( \hat{p}_{0\mu }+{\frac{1}{\sqrt{2\alpha ^{\prime }}}}%
\sum_{n=1}^{\infty }\eta _{\mu \nu }(\alpha _{n}^{\nu }+\alpha _{-n}^{\nu
})\cos n\sigma \right) \,.  \notag
\end{eqnarray}
Nonzero modes in matter sector 
\begin{eqnarray}
&&\alpha _{n}^{\mu }=\sqrt{\kappa _{n}}\hat{a}_{n}^{\mu }\,,\ \alpha
_{-n}^{\mu }=\sqrt{\kappa _{n}}\hat{a}_{n}^{\dagger \mu }\,,\ \ [\alpha
_{n}^{\mu },\alpha _{m}^{\nu }]=\epsilon (n)\kappa _{n}\delta _{n+m,0}\eta
^{\mu \nu }\,,\ \ [\hat{a}_{n}^{\mu },\hat{a}_{m}^{\dagger \nu }]=\delta
_{n,m}\eta ^{\mu \nu }\,,  \notag \\
&&x_{n}^{\mu }={\frac{i}{\sqrt{2\kappa _{n}}}}l_{s}(\hat{a}_{n}^{\mu }-\hat{a%
}_{n}^{\dagger \mu })\,,\ p_{n\mu }=\sqrt{\frac{\kappa _{n}}{2}}{\frac{\eta
_{\mu \nu }}{l_{s}}}(\hat{a}_{n}^{\nu }+\hat{a}_{n}^{\dagger \nu })\,,\ \ [%
\hat{x}_{n}^{\mu },\hat{p}_{m\nu }]=i\delta _{n,m}\delta _{\nu }^{\mu }\,, 
\notag \\
&&\alpha _{n}={\frac{1}{\sqrt{2}}}\left( l_{s}\hat{p}_{|n|}-i\epsilon (n){%
\frac{\kappa _{|n|}}{l_{s}}}\hat{x}_{|n|}\right) \,,  \label{eq:xmu_alpha}
\end{eqnarray}
where we define the symbols $l_{s}^{2}=2\alpha ^{\prime },\kappa _{n}=n$.

Zero mode in matter sector, with $a_{0}^{\mu }:=l_{s}\hat{p}_{0}^{\mu }\,$ 
\begin{eqnarray}
\hat{x}_{0}^{\mu }={\frac{i}{2}}l_{s}\sqrt{b}(\hat{a}_{0}^{\mu }-\hat{a}%
_{0}^{\dagger \mu })\,,\ \hat{p}_{0\mu }=\frac{1}{l_{s}\sqrt{b}}\eta _{\mu
\nu }(\hat{a}_{0}^{\nu }+\hat{a}_{0}^{\dagger \nu })\,,\ \ [\hat{x}_{0}^{\mu
},\hat{p}_{0\nu }]=i\delta _{\nu }^{\mu }\,,\ [\hat{a}_{0}^{\mu },\hat{a}%
_{0}^{\dagger \nu }]=\eta ^{\mu \nu }\quad  \label{eq:xmu0_a}
\end{eqnarray}
where $b$ is some positive constant.

\item  Position eigenstates 
\begin{eqnarray}
&&\langle x_{0},x_{n}|\hat{x}_{n}=\langle x_{0},x_{n}|x_{n}\,,\qquad \langle
x_{0},x_{n}|\hat{x}_{0}=\langle x_{0},x_{n}|x_{0}\,,  \notag \\
&&\hat{x}_{n}|x_{0},x_{n}\rangle =x_{n}|x_{0},x_{n}\rangle \,,\qquad \hat{x}%
_{0}|x_{0},x_{n}\rangle =x_{0}|x_{0},x_{n}\rangle ,
\end{eqnarray}
are given as squeezed states in Fock space 
\begin{eqnarray}
\langle x_{0},x_{n}| &=&\langle x_{0}|e^{\sum_{n>0}\left( {\frac{1}{2\kappa
_{n}}}\alpha _{n}^{2}+{\frac{i\sqrt{2}}{l_{s}}}x_{n}\alpha _{n}-{\frac{%
\kappa _{n}}{2l_{s}^{2}}}x_{n}^{2}\right) }\prod_{n>0}\left( \frac{\kappa
_{n}}{\pi l_{s}^{2}}\right) ^{\frac{d}{4}}  \notag \\
&=&\langle x_{0}|e^{\sum_{n>0}\left( {\frac{1}{2}}\hat{a}_{n}^{2}+{\frac{i%
\sqrt{2\kappa _{n}}}{l_{s}}}x_{n}\hat{a}_{n}-{\frac{\kappa _{n}}{2l_{s}^{2}}}%
x_{n}^{2}\right) }\prod_{n>0}\left( \frac{\kappa _{n}}{\pi l_{s}^{2}}\right)
^{\frac{d}{4}}\,,  \notag \\
\langle x_{0}| &=&\langle 0|e^{{\frac{1}{2}}\hat{a}_{0}^{2}+i{\frac{2}{l_{s}%
\sqrt{b}}}x_{0}\hat{a}_{0}-{\frac{1}{l_{s}^{2}b}}x_{0}^{2}}\left( \frac{2}{%
\pi l_{s}^{2}b}\right) ^{\frac{d}{4}}\,,  \notag \\
|x_{0},x_{n}\rangle &=&\prod_{n>0}\left( \frac{\kappa _{n}}{\pi l_{s}^{2}}%
\right) ^{\frac{d}{4}}e^{\sum_{n>0}\left( {\frac{1}{2\kappa _{n}}}\alpha
_{-n}^{2}-{\frac{i\sqrt{2}}{l_{s}}}x_{n}\alpha _{-n}-{\frac{\kappa _{n}}{%
2l_{s}^{2}}}x_{n}^{2}\right) }|x_{0}\rangle  \notag \\
&=&\prod_{n>0}\left( \frac{\kappa _{n}}{\pi l_{s}^{2}}\right) ^{\frac{d}{4}%
}e^{\sum_{n>0}\left( {\frac{1}{2}}\hat{a}_{n}^{\dagger 2}-{\frac{i\sqrt{%
2\kappa _{n}}}{l_{s}}}x_{n}\hat{a}_{n}^{\dagger }-{\frac{\kappa _{n}}{%
2l_{s}^{2}}}x_{n}^{2}\right) }|x_{0}\rangle \,,  \notag \\
|x_{0}\rangle &=&\left( \frac{2}{\pi l_{s}^{2}b}\right) ^{\frac{d}{4}}e^{{%
\frac{1}{2}}\hat{a}_{0}^{\dagger 2}-i{\frac{2}{l_{s}\sqrt{b}}}x_{0}\hat{a}%
_{0}^{\dagger }-{\frac{1}{l_{s}^{2}b}}x_{0}^{2}}|0\rangle \,.
\label{eq:matter_basis}
\end{eqnarray}
They satisfy normalization and completeness conditions 
\begin{eqnarray}
&&\langle x_{0},x_{n}|x_{0}^{\prime },x_{n}^{\prime }\rangle =\delta
^{d}(x_{0}-x_{0}^{\prime })\prod_{n>0}\delta ^{d}(x_{n}-x_{n}^{\prime })\,, 
\notag \\
&&\int d^{d}x_{0}\prod_{n>0}d^{d}x_{n}|x_{0},x_{n}\rangle \langle
x_{0},x_{n}|=1\,.
\end{eqnarray}

\item  Oscillators as differential operators in position space 
\begin{eqnarray}
&&\langle x_{0},x_{n}|\alpha _{m}|\Psi \rangle =-{\frac{i}{\sqrt{2}}}\left(
\epsilon (m){\frac{\kappa _{|m|}}{l_{s}}}x_{|m|}+l_{s}{\frac{\partial }{%
\partial x_{|m|}}}\right) \langle x_{0},x_{n}|\Psi \rangle \,,  \notag \\
&&\langle x_{0},x_{n}|\alpha _{0}|\Psi \rangle =-il_{s}{\frac{\partial }{%
\partial x_{0}}}\langle x_{0},x_{n}|\Psi \rangle \,.
\end{eqnarray}

\item  Transformation from position space to Moyal space 
\begin{eqnarray}
A(\bar{x},x_{e},p_{e}) &=&(\det (2T))^{d/2}\int dx_{o}e^{-{\frac{2i}{\theta }%
}p_{e}Tx_{o}}\Psi (x_{0},x_{n})  \notag \\
&=&2^{{\frac{Nd}{2}}}(1+\bar{w}w)^{-{\frac{d}{4}}}\int dx_{o}e^{-{\frac{2i}{%
\theta }}p_{e}Tx_{o}}\langle x_{0},x_{n}|\Psi \rangle =:\langle \bar{x}%
,x_{e},p_{e}|\Psi \rangle \,,  \label{eq:MSFTfield_matter}
\end{eqnarray}
where the midpoint $\bar{x}^{\mu }:=X^{\mu }\left( \pi /2\right) $ is
related to the center of mass $x_{0}$ through Eq.(\ref{position}) 
\begin{equation}
\bar{x}^{\mu }:=X^{\mu }\left( \pi /2\right) =x_{0}^{\mu
}-\sum_{e}x_{e}^{\mu }w_{e},\quad \langle \bar{x}|=\langle x_{0}|\exp
\left( i\hat{p}\cdot \sum_{e}x_{e}w_{e}\right)\,.  \label{midpoint}
\end{equation}
The Moyal space $\langle \bar{x},x_{e},p_{e}|$ is given by a squeezed state 
\begin{eqnarray}
&&\langle \bar{x},x_{e},p_{e}|=\langle \bar{x}|e^{{\frac{\alpha _{e}^{2}}{%
2\kappa _{e}}}-{\frac{\alpha _{o}^{2}}{2\kappa _{o}}}-\bar{\xi}M_{0}\xi -%
\bar{\xi}\lambda }{\det }(4{\kappa _{e}}^{1/2}T\kappa _{o}^{-1/2})^{\frac{d}{%
2}}  \notag \\
&&\qquad \qquad \quad =\langle \bar{x}|e^{{\frac{\alpha _{e}^{2}}{2\kappa
_{e}}}-{\frac{\alpha _{o}^{2}}{2\kappa _{o}}}-\bar{\xi}M_{0}\xi -\bar{\xi}%
\lambda }\,2^{Nd}(1+\bar{w}w)^{-{\frac{d}{8}}}\,,  \notag \\
&&M_{0}=\left( 
\begin{array}{cc}
{\frac{\kappa _{e}}{2l_{s}^{2}}} & 0 \\ 
0 & {\frac{2l_{s}^{2}}{\theta ^{2}}}T\kappa _{o}^{-1}\bar{T}
\end{array}
\right) \,,~~~~\lambda =\left( 
\begin{array}{c}
-i{\frac{\sqrt{2}}{l_{s}}}\alpha _{e}-i\hat{p}w_{e} \\ 
-{\frac{2\sqrt{2}l_{s}}{\theta }}T\kappa _{o}^{-1}\alpha _{o}
\end{array}
\right) \,.~
\end{eqnarray}

\item  Moyal $\star $ product and trace: 
\begin{eqnarray}
&&\star =\exp \left( {\frac{1}{2}}\overleftarrow{\partial _{\xi }}\sigma 
\overrightarrow{\partial _{\xi }}\right) =\exp \left( {\frac{i\theta }{2}}%
\left( \overleftarrow{\partial _{x_{e}}}\overrightarrow{\partial _{p_{e}}}-%
\overleftarrow{\partial _{p_{e}}}\overrightarrow{\partial _{x_{e}}}\right)
\right) \,,\quad \sigma =i\theta \left( 
\begin{array}{cc}
0 & 1 \\ 
-1 & 0
\end{array}
\right) \,,  \label{eq:Moyal_star_matter} \\
&&\mathrm{Tr}\,A(\bar{x},\xi )=|\det (2\pi \sigma )|^{-{\frac{d}{2}}}\int
dx_{e}dp_{e}A(\bar{x},\xi )=(2\pi \theta )^{-Nd}\int dx_{e}dp_{e}A(\bar{x}%
,\xi )\,.
\end{eqnarray}
The normalization of a field $A$ in Moyal space coincides with the
normalization of its image in Fock space (\ref{eq:MSFTfield_matter}) 
\begin{equation}
\langle \Psi |\Psi \rangle =\int d^{d}\bar{x}\,\mathrm{Tr}\left( A^{\dagger
}(\bar{x},\xi )\star A(\bar{x},\xi )\right) \,.
\end{equation}

\item  Oscillators as differential operators in Moyal space 
\begin{eqnarray}
&&\langle \bar{x},x_{e},p_{e}|\alpha _{0}|\Psi \rangle =-il_{s}\frac{%
\partial }{\partial \bar{x}}\langle \bar{x},x_{e},p_{e}|\Psi \rangle =:\beta
_{0}\langle \bar{x},x_{e},p_{e}|\Psi \rangle \,,  \notag \\
&&\langle \bar{x},x_{e},p_{e}|\alpha _{e}|\Psi \rangle =\left( \bar{\beta}%
_{e}^{x}-\frac{w_{|e|}}{\sqrt{2}}\beta _{0}\right) \langle \bar{x}%
,x_{e},p_{e}|\Psi \rangle =\left( \bar{\beta}_{e}^{x}-w_{e}^{\prime }\beta
_{0}\right) \langle \bar{x},x_{e},p_{e}|\Psi \rangle  \notag \\
&&\qquad \qquad =\beta _{e}^{x}\langle \bar{x},x_{e},p_{e}|\Psi \rangle \,, 
\notag \\
&&\langle \bar{x},x_{e},p_{e}|\alpha _{o}|\Psi \rangle =:\beta
_{o}^{p}\langle \bar{x},x_{e},p_{e}|\Psi \rangle =\sum_{e\neq 0}\bar{\beta}%
_{e}^{p}\langle \bar{x},x_{e},p_{e}|\Psi \rangle U_{-e,o}\,,  \notag \\
&&\beta _{o}^{p}=\sum_{e>0}\frac{1}{\sqrt{2}}\left( \epsilon (o)\frac{\theta
\kappa _{|o|}}{2l_{s}}R_{|o|e}\frac{\partial }{\partial p_{e}}+\frac{2l_{s}}{%
\theta }p_{e}T_{e|o|}\right) =\sum_{e\neq 0}\bar{\beta}_{e}^{p}U_{-e,o}\,, 
\notag \\
&&\bar{\beta}_{e}^{x}=-\frac{i}{\sqrt{2}}\left( \epsilon (e)\frac{\kappa
_{|e|}}{l_{s}}x_{|e|}+l_{s}\frac{\partial }{\partial x_{|e|}}\right)
\,,\qquad \bar{\beta}_{e}^{p}=\frac{1}{\sqrt{2}}\left( \frac{\theta \kappa
_{|e|}}{2l_{s}}\epsilon (e)\frac{\partial }{\partial p_{|e|}}+\frac{2l_{s}}{%
\theta }p_{|e|}\right) \,.~~~~
\end{eqnarray}
These satisfy ordinary commutation relation: 
\begin{eqnarray}
&&[\bar{\beta}_{e}^{x},\bar{\beta}_{e^{\prime }}^{x}]=\epsilon (e)\kappa
_{|e|}\delta _{e+e^{\prime }}\,,\quad \lbrack \bar{\beta}_{e}^{p},\bar{\beta}%
_{e^{\prime }}^{p}]=\epsilon (e)\kappa _{|e|}\delta _{e+e^{\prime }}\,,\quad
\lbrack \bar{\beta}_{e}^{x},\bar{\beta}_{e^{\prime }}^{p}]=0\,,  \notag \\
&&[\beta _{e}^{x},\beta _{e^{\prime }}^{x}]=\epsilon (e)\kappa _{|e|}\delta
_{e+e^{\prime }}\,,\quad \lbrack \beta _{o}^{p},\beta _{o^{\prime
}}^{p}]=\epsilon (o)\kappa _{|o|}\delta _{o+o^{\prime }}\,,\quad \lbrack
\beta _{e}^{x},\beta _{o^{\prime }}^{p}]=0\,.
\end{eqnarray}

\item  Oscillators as fields in Moyal space\footnote{%
The convention for $\beta _{n}$ here is the same as \cite{BKM2} $\beta _{n}^{%
\mathrm{BKM}}=\beta _{n},$ but differs by a factor from the convention in 
\cite{BM2} $\beta _{n}^{\mathrm{BM}}=\sqrt{\frac{\kappa _{|n|}}{2}}\beta
_{n} $.} 
\begin{eqnarray}
&&\bar{\beta}_{e}^{x}A=\sqrt{\frac{\kappa _{|e|}}{2}}\left( \beta _{e}\star
A-A\star \beta _{-e}\right) \,,\quad \bar{\beta}_{e}^{p}A=\sqrt{\frac{\kappa
_{|e|}}{2}}\left( \beta _{e}\star A+A\star \beta _{-e}\right) \,,  \notag \\
&&\beta _{e}:=\frac{1}{\sqrt{\kappa _{|e|}}}\left( -\frac{i}{2l_{s}}\epsilon
(e)\kappa _{|e|}x_{|e|}+\frac{l_{s}}{\theta }p_{|e|}\right) \,,  \notag \\
&&[\beta _{e},\beta _{e^{\prime }}]_{\star }=\epsilon (e)\delta
_{e+e^{\prime }}\,.
\end{eqnarray}
We can also define odd mode fields through a Bogoliubov transformation 
\begin{equation}
\sqrt{\kappa_{|o|}}\beta _{o}:=\sum_{e\neq 0}\sqrt{\kappa_{|e|}}
\beta _{e}U_{-e,o}\,.
\end{equation}
The following relations hold 
\begin{eqnarray}
&&\beta _{o}^{p}A=\sqrt{\frac{\kappa _{|o|}}{2}}\left( \beta _{o}\star
A+A\star \beta _{-o}\right) \,,  \notag \\
&&[\beta _{o},\beta _{o^{\prime }}]_{\star }=\epsilon (o)\delta
_{o+o^{\prime }}\,,\qquad \lbrack \beta _{-e},\beta _{o}]_{\star }=-\epsilon
(e)\kappa_{|e|}^{1\over 2}U_{-e,o}\kappa_{|o|}^{-{1\over 2}}\,.
\end{eqnarray}
\end{itemize}

\subsubsection{Butterfly projector}

The momentum independent butterfly state $A_{B}(\xi )$ satisfies 
\begin{equation}
\beta _{e}\star A_{B}=A_{B}\star \beta _{-e}=0\,,\qquad \forall e>0\,.
\label{eq:matter_butterfly}
\end{equation}
There is a unique solution in monoid 
\begin{equation}
A_{B}(\xi )=2^{dN}\exp \left( -\sum_{e>0}\left( \frac{1}{2l_{s}^{2}}\bar{x}%
_{e}\kappa _{e}x_{e}+\frac{2l_{s}^{2}}{\theta ^{2}}\bar{p}_{e}{\frac{1}{%
\kappa _{e}}}p_{e}\right) \right)\,.
\end{equation}
It satisfies 
\begin{equation}
A_{B}\star A_{B}=A_{B},\quad Tr\left( A_{B}\right) =1.
\end{equation}

In ordinary oscillator language, Eq.(\ref{eq:matter_butterfly}) means 
\begin{equation}
\alpha _{e}|\Psi _{B}\rangle =0\,,\quad \sum_{o>0}\left( \alpha
_{o}U_{-o,e}^{-1}+\alpha _{-o}U_{o,e}^{-1}\right) |\Psi _{B}\rangle =0\,,
\qquad \forall e>0
\end{equation}
for zero momentum state. Now we take the ansatz 
\begin{equation}
|\Psi _{B}\rangle =\mathcal{N}\exp \left( -{\frac{1}{2}}\sum_{m,n\geq
1}a_{m}^{\dagger }V_{mn}^{B}a_{n}^{\dagger }\right) |\Omega \rangle
\end{equation}
which corresponds to a monoid element in MSFT. 
Then we have constraints for $V_{mn}^{B}$ : 
\begin{equation}
\sum_{o^{\prime }>0}V_{oo^{\prime }}^{B}\sqrt{\kappa _{o^{\prime }}}%
U_{-o^{\prime },e}^{-1}=\sqrt{\kappa _{o}}U_{o,e}^{-1}\,,\quad
o,e>0\,,\qquad V_{eo}^{B}=V_{oe}^{B}=V_{ee^{\prime }}^{B}=0\,.
\label{eq:butterfly_VB}
\end{equation}
At the open string limit $\kappa _{e}=e,\kappa _{o}=o,N=\infty $, we can
show that the matrix $V_{mn}^{B}$ which was obtained in \cite{Gaiotto:2002kf}
: 
\begin{equation}
V_{mn}^{B}=\left\{ 
\begin{array}{cl}
-(-1)^{\frac{m+n}{2}}\frac{\sqrt{mn}}{m+n}\frac{\Gamma \lbrack \frac{m}{2}%
]\Gamma \lbrack \frac{n}{2}]}{\pi \Gamma \lbrack \frac{m+1}{2}]\Gamma
\lbrack \frac{n+1}{2}]} & \mathrm{for}\;m\;\mathrm{and}\;n\;\mathrm{odd} \\ 
0 & \mathrm{for}\;m\;\mathrm{or}\;n\;\mathrm{even}
\end{array}
\right.
\end{equation}
satisfies Eq.(\ref{eq:butterfly_VB}). Namely, we have obtained the
correspondence: 
\begin{equation}
A_{B}\quad \leftrightarrow \quad |\Psi _{B}\rangle =\exp \left( -{\frac{1}{2}%
}L_{-2}\right) |\Omega \rangle \,,\quad \mathrm{for}\quad \kappa
_{e}=e,\kappa _{o}=o,N=\infty \,.
\end{equation}

\subsubsection{$L_0$ and $\mathcal{L}_0$ \label{sec:L0L0matter}}

In MSFT $L_{0}$ in matter sector is defined as 
\begin{eqnarray}
L_{0} &=&{\frac{1}{2}}\beta _{0}^{2}+\sum_{e>0}\beta _{-e}^{x}\beta
_{e}^{x}+\sum_{o>0}\beta _{-o}^{p}\beta _{o}^{p}  \notag \\
&=&{\frac{1}{2}}(1+\bar{w}w)\beta _{0}^{2}+\beta _{0}\sum_{e>0}il_{s}w_{e}{%
\frac{\partial }{\partial x_{e}}}-\frac{d}{2}\sum_{n>0}\kappa _{n}  \notag \\
&&+\sum_{e>0}\left( -\frac{l_{s}^{2}}{2}{\frac{\partial ^{2}}{\partial
x_{e}^{2}}}-\frac{\theta ^{2}}{8l_{s}^{2}}\kappa _{e}^{2}\frac{\partial ^{2}%
}{\partial p_{e}^{2}}+\frac{1}{2l_{s}^{2}}\kappa _{e}^{2}x_{e}^{2}+\frac{%
2l_{s}^{2}}{\theta ^{2}}p_{e}^{2}\right) -\frac{1}{1+\bar{w}w}\frac{%
2l_{s}^{2}}{\theta ^{2}}\left( \sum_{e>0}w_{e}p_{e}\right) ^{2}\,.~~
\label{eq:L0_matter}
\end{eqnarray}
The operator $L_{0}$ can be rewritten in terms of a field $\mathcal{L}_{0}$
using Moyal star product, plus a remnant $\gamma $ called the ``midpoint
correction term'' \cite{BKM2}, which is multiplied with an ordinary product 
\begin{eqnarray}
&&L_{0}A_{\beta _{0}}=\mathcal{L}_{0}(\beta _{0})\star A_{\beta
_{0}}+A_{\beta _{0}}\star \mathcal{L}_{0}(-\beta _{0})+\gamma A_{\beta
_{0}}\,,   \\
&&\mathcal{L}_{0}(\beta _{0}):=\sum_{e>0}\left( \frac{l_{s}^{2}}{\theta ^{2}}%
p_{e}^{2}+\frac{\kappa _{e}^{2}}{4l_{s}^{2}}x_{e}^{2}-\frac{l_{s}}{\theta }%
w_{e}p_{e}\beta _{0}\right) +\frac{1}{4}(1+\bar{w}w)\beta _{0}^{2}-\frac{d}{4%
}\sum_{n>0}\kappa _{n}\,,   \\
&&\gamma =-\frac{1}{1+\bar{w}w}\frac{2l_{s}^{2}}{\theta ^{2}}\left(
\sum_{e>0}w_{e}p_{e}\right) ^{2}\,.
\end{eqnarray}
The $\gamma $ term \textit{formally} goes to zero as $\kappa
_{n}=n,N\rightarrow \infty $, since $\bar{w}w\rightarrow \infty .$ However,
this is not true in computations due to contributions of the form $\infty
/\infty $ that are related to the associativity anomaly. In fact, $\gamma $
is indispensable to reproduce the correct spectrum of $L_{0}$ \cite{BM2}\cite
{BKM1}. The $\gamma $ term depends only on one special momentum mode $\hat{p}%
=\left( 1+\bar{w}w\right) ^{-1/2}\sum_{e>0}w_{e}p_{e}$ which we call the
anomalous midpoint momentum mode \cite{BM2}\cite{B2}. We can rewrite $%
\mathcal{L}_{0}$ in terms of oscillators 
\begin{equation}
\mathcal{L}_{0}(\beta _{0})=\sum_{e>0}\kappa _{e}\beta _{-e}\star \beta _{e}+%
\frac{d}{4}\left( \sum_{e>0}\kappa _{e}-\sum_{o>0}\kappa _{o}\right) +{\frac{%
1}{4}}(1+\bar{w}w)\beta _{0}^{2}-{\frac{l_{s}}{\theta }}\left( \bar{w}%
_{e}p_{e}\right) \beta _{0}\,,
\end{equation}
and then, acting on the butterfly projector (\ref{eq:matter_butterfly}), we
have 
\begin{equation}
L_{0}A_{B}=\gamma A_{B}\,.
\end{equation}

\subsubsection{$n$-string vertex and Neumann coefficients 
\label{sec:Neumann-matter}}

Here we give brief review of the correspondence of $n$-string vertex and
Neumann coefficients in MSFT. We note the properties of the momentum state
and coherent states in Fock space, together with the corresponding Moyal
images.

\begin{itemize}
\item  Momentum eigenstate (zero mode part): 
\begin{eqnarray}
&&|p_{0}\rangle =\left( 2\pi bl_{s}^{2}\right) ^{\frac{d}{4}}e^{-{\frac{1}{2}%
}\hat{a}_{0}^{\dagger 2}+\sqrt{b}l_{s}\hat{a}_{0}^{\dagger }p_{0}-{\frac{%
bl_{s}^{2}}{4}}p_{0}^{2}}|0\rangle \,,  \notag \\
&&\langle p_{0}|=\langle 0|e^{-{\frac{1}{2}}\hat{a}_{0}^{2}+\sqrt{b}l_{s}%
\hat{a}_{0}p_{0}-{\frac{bl_{s}^{2}}{4}}p_{0}^{2}}\left( 2\pi
bl_{s}^{2}\right) ^{\frac{d}{4}}\,,  \notag \\
&&\langle p_{0}|p_{0}^{\prime }\rangle =(2\pi )^{d}\delta
^{d}(p_{0}-p_{0}^{\prime })\,,  \notag \\
&&\langle p_{0}|x_{0}\rangle =e^{-ip_{0}x_{0}},\quad \langle
x_{0}|p_{0}\rangle =e^{ip_{0}x_{0}}.
\end{eqnarray}

\item  Coherent state :\footnote{%
Here we introduce the \textit{bra} coherent state. This is a different
convention from that in \cite{BM2}.} 
\begin{eqnarray}
&&\langle \Psi |\hat{a}^{\dagger }=\langle \Psi |\mu ^{\ast }\,,~~~\langle
\Psi |=\langle p|e^{\mu ^{\ast }\hat{a}}\,,  \notag \\
&&{}_{2}\langle \Psi |V_{2}\rangle _{12}=e^{-\mu ^{\ast }C\hat{a}^{\dagger
(1)}}|-p\rangle _{1}=:|\tilde{\Psi}\rangle _{1}\,,  \notag \\
&&\tilde{A}:=\langle \bar{x},x_{e},p_{e}|\tilde{\Psi}\rangle =2^{Nd}(1+\bar{w%
}w)^{-{\frac{d}{8}}}e^{-i\bar{x}p}e^{{\frac{1}{2}}\mu _{e}^{\ast 2}-{\frac{1%
}{2}}\mu _{o}^{\ast 2}-\bar{\xi}M_{0}\xi -\bar{\xi}\lambda}\,,\qquad 
\notag \\
&&\lambda =\left( 
\begin{array}{c}
{\frac{i\sqrt{2}}{l_{s}}}\kappa _{e}^{\frac{1}{2}}\mu _{e}^{\ast }+ipw_{e}
\\ 
-{\frac{2\sqrt{2}l_{s}}{\theta }}T\kappa _{o}^{-{\frac{1}{2}}}\mu _{o}^{\ast
}
\end{array}
\right) =2K^{\ast }(\mu ^{\ast }+Wp)\,,  \notag \\
&&K^{\ast }=\left( 
\begin{array}{cc}
{\frac{i}{l_{s}}}\sqrt{\frac{\kappa _{e}}{2}} & 0 \\ 
0 & -{\frac{l_{s}}{\theta }}T\sqrt{\frac{2}{\kappa _{o}}}
\end{array}
\right) \,,\quad W=\left( 
\begin{array}{c}
{\frac{l_{s}}{\sqrt{2\kappa _{e}}}}w \\ 
0
\end{array}
\right) \,,
\end{eqnarray}
where we used the reflector 
\begin{eqnarray}
\langle V_{2}| &=&\int \frac{d^{d}p^{(1)}}{(2\pi )^{d}}\frac{d^{d}p^{(2)}}{%
(2\pi )^{d}}\,\langle 0,p^{(1)}|
\langle 0,p^{(2)}|e^{-\sum_{n\geq 1}(-1)^{n}
\hat{a}_{n}^{(1)}\hat{a}_{n}^{(2)}}(2\pi
)^{d}\delta ^{d}(p^{(1)}+p^{(2)})\,,  \notag \\
|V_{2}\rangle &=&\int \frac{d^{d}p^{(1)}}{(2\pi )^{d}}\frac{d^{d}p^{(2)}}{%
(2\pi )^{d}}(2\pi )^{d}\delta ^{d}(p^{(1)}+p^{(2)})e^{-\sum_{n\geq 1}(-1)^{n}%
\hat{a}_{n}^{\dagger (1)}\hat{a}_{n}^{\dagger (2)}}|0,p^{(1)}\rangle
|0,p^{(2)}\rangle\,.~~~~
\end{eqnarray}
In particular we have the bra-ket correspondence for eigenstates of $x_{n}$: 
\begin{equation}
{}_{12}\langle V_{2}|x_{0},x_{n}\rangle _{2}={}_{1}\langle
x_{0},(-1)^{n}x_{n}|\,.
\end{equation}

\item  Compute the $n$-string vertex for coherent states in terms of unknown
Neumann coefficients $V_{(n)}^{rs},V_{0(n)}^{rs},V_{00(n)}^{rs}$ 
\begin{eqnarray}
|V_{n}\rangle &=&\int \frac{d^{d}p^{(1)}}{(2\pi )^{d}}\cdots \frac{%
d^{d}p^{(n)}}{(2\pi )^{d}}(2\pi )^{d}\delta ^{d}(p^{(1)}+\cdots +p^{(n)}) 
\notag \\
&&\times\, e^{-{\frac{1}{2}}\hat{a}^{\dagger (r)}V_{(n)}^{rs}\hat{a}^{\dagger
(s)}-p^{(r)}V_{0(n)}^{rs}\hat{a}^{\dagger (s)}-{\frac{1}{2}}%
p^{(r)}V_{00(n)}^{rs}p^{(s)}}|p^{(i)}\rangle \,,   \\
\langle \Psi _{1}|\cdots \langle \Psi _{n}|V_{n}\rangle &=&(2\pi
)^{d}\delta ^{d}(p^{(1)}+\cdots +p^{(n)})  \notag \\
&&\times\, e^{-{\frac{1}{2}}\mu ^{(r)\ast }V_{(n)}^{rs}\mu ^{(s)\ast
}-p^{(r)}V_{0(n)}^{rs}\mu ^{(s)\ast }-{\frac{1}{2}}%
p^{(r)}V_{00(n)}^{rs}p^{(s)}}\,.
\end{eqnarray}

\item  Compute the trace of the Moyal images of $n$ coherent states in MSFT 
\begin{eqnarray}
&&\int d^{d}\bar{x}\mathrm{Tr}\left( \tilde{A}_{1}(\bar{x},\xi )\star 
\tilde{A}_{2}(\bar{x},\xi )\star \cdots \star \tilde{A}_{n}(\bar{x},\xi
)\right)  \notag \\
&&=(-1)^{\frac{Nd}{2}}\left( \det ((1+m_{0})^{n}-(1-m_{0})^{n})\right) ^{-{%
\frac{d}{2}}}2^{nNd}(1+\bar{w}w)^{-{\frac{nd}{8}}}  \notag \\
&&~~~\times (2\pi )^{d}\delta ^{d}(p^{(1)}+\cdots
+p^{(n)})\,e^{E^{(n)}}\,,\qquad   \\
&&E^{(n)}=-{\frac{1}{2}}\sum_{r,s}\mu ^{(r)\ast }C\left( 2K^{\ast -1}m_{0}%
\mathcal{O}_{(s-r)}(m_{0})K^{\ast }-\delta _{r,s}\right) \mu ^{(s)\ast } 
\notag \\
&&-2\sum_{r,s}p^{(r)}\bar{W}K^{\ast -1}m_{0}\mathcal{O}_{(s-r)}(m_{0})K^{%
\ast }\mu ^{(s)\ast }-{\frac{1}{2}}\sum_{r,s}p^{(r)}\left( 2\bar{W}K^{\ast
-1}m_{0}\mathcal{O}_{(s-r)}(m_{0})K^{\ast }W\right) p^{(s)}\,,  \notag \\
&&m_{0}:=M_{0}\sigma =\left( 
\begin{array}{cc}
0 & {\frac{i\theta }{2l_{s}^{2}}}\kappa _{e} \\ 
-{\frac{2il_{s}^{2}}{\theta }}T\kappa _{o}^{-1}\bar{T} & 0
\end{array}
\right)  \label{eq:m0matterdef}
\end{eqnarray}
where we used 
\begin{equation}
CK^{\ast -1}m_{0}=-\bar{K}^{\ast }\sigma \,,\quad CK^{\ast -1}m_{0}K^{\ast
}=-K^{\ast -1}m_{0}K^{\ast }C\,,\quad \mathcal{O}_{(s-r)}(m_{0})=-\mathcal{O}%
_{(r-s)}(-m_{0})\,.  \label{eq:rel_matterCK}
\end{equation}
We note 
\begin{equation}
\tilde{m}_{0}^{\ast }:=K^{\ast -1}m_{0}K^{\ast }=\left( 
\begin{array}{cc}
0 & -\kappa _{e}^{\frac{1}{2}}T\kappa _{o}^{-{\frac{1}{2}}} \\ 
-\kappa _{o}^{-{\frac{1}{2}}}\bar{T}\kappa _{e}^{\frac{1}{2}} & 0
\end{array}
\right) =-\tilde{m}_{0}\,.  \label{eq:tilm0}
\end{equation}
The sign is changed compared to that in \cite{BM2} because we used 
\textit{bra} coherent state $\langle \Psi _{c}|$ to define the Moyal 
field $\tilde{A}$.

\item  The Neumann coefficients in the matter sector are obtained by
identifying the Fock space and MSFT expressions and comparing the exponents%
\footnote{%
Here we defined the Witten's $\ast $ product using the \textit{ket} $%
|V_{3}\rangle $ as Eq.(\ref{eq:Witten's*}) which is different convention
from that in \cite{BM2}. Also, here we have included the overall
normalization $\rho $ which does not play a role in the computation of the
Neumann coefficients.} 
\begin{eqnarray}
&&\int d^{d}\bar{x}\,\mathrm{Tr}\left( \tilde{A}_{1}(\bar{x},\xi )\star 
\tilde{A}_{2}(\bar{x},\xi )\star \cdots \star \tilde{A}_{n}(\bar{x},\xi
)\right) =\rho \langle \Psi _{1}|\langle\Psi _{2}|\cdots 
\langle \Psi _{n}|V_{n}\rangle
\,,   \\
&&~~~~~~~~~~~\rho =(-1)^{\frac{Nd}{2}}\left( \det
((1+m_{0})^{n}-(1-m_{0})^{n})\right) ^{-{\frac{d}{2}}}2^{nNd}
(1+\bar{w}w)^{-{\frac{nd}{8}}}
\end{eqnarray}
and using momentum conservation $\delta ^{d}(p^{(1)}+p^{(2)}+\cdots
+p^{(n)}) $. Then one has the Neumann coefficients 
\begin{eqnarray}
&&V_{(n)}^{rs}=C\left( 2K^{\ast -1}m_{0}\mathcal{O}_{(s-r)}(m_{0})K^{\ast
}-\delta _{r,s}\right) \,,  \notag \\
&&V^{r}{}_{0(n)}^{s}=-2K^{\ast -1}m_{0}\mathcal{O}_{(s-r)}(-m_{0})K^{\ast }W-%
\frac{2}{n}W\,,   \\
&&V_{00(n)}^{rs}=2\bar{W}K^{\ast -1}m_{0}\mathcal{O}_{(s-r)}(m_{0})K^{\ast
}W-\frac{2}{n}\bar{W}W\,.\notag
\end{eqnarray}
They satisfy Neumann matrix algebra as in Ref.\cite{BM2}. For the 3-string
vertex we write them explicitly 
\begin{eqnarray}
&&\mathcal{M}^{(0)}:=CV_{(3)}^{rr}=\frac{\tilde{m}_{0}^{\ast 2}-1}{\tilde{m}%
_{0}^{\ast 2}+3}\,,~~~\mathcal{M}^{(\pm )}:=CV_{(3)}^{r,r\pm 1}=\frac{2\pm 
\tilde{m}_{0}^{\ast }}{\tilde{m}_{0}^{\ast 2}+3}\,,  \notag \\
&&\mathcal{V}^{(0)}:=V^{r}{}_{0(3)}^{r}=\frac{4\tilde{m}_{0}^{\ast 2}}{3(%
\tilde{m}_{0}^{\ast 2}+3)}\kappa _{e}^{-{\frac{1}{2}}}\frac{wl_{s}}{\sqrt{2}}%
\,,~~~\mathcal{V}^{(\pm )}:=V^{r,}{}_{0(3)}^{r\pm 1}=\frac{-2\tilde{m}%
_{0}^{\ast 2}\mp 6\tilde{m}_{0}^{\ast }}{3(\tilde{m}_{0}^{\ast 2}+3)}\kappa
_{e}^{-{\frac{1}{2}}}\frac{wl_{s}}{\sqrt{2}}\,,  ~~~~~~~~~~~ 
\label{eq:Neumann-3-matter}\\
&&V_{00}:=V_{00(3)}^{rr}=l_{s}^{2}\,\bar{w}\kappa _{e}^{-{\frac{1}{2}}}\frac{%
t\bar{t}}{t\bar{t}+3}\kappa _{e}^{-{\frac{1}{2}}}w
\notag
\end{eqnarray}
where we redefined as $V_{00}^{r,s}=V_{00}\delta _{r,s}$ using momentum
conservation. We used the notation $\bar{w}=\left( w_{e},0\right) \,,$ and$%
~t=\kappa_{e}^{1/2}T\kappa _{o}^{-1/2}$.

\item  Fermionic ghost Neumann coefficients for the 3-vertex can be derived
from matter Neumann coefficients for the 6-vertex 
\begin{eqnarray}
&&X^{rs}:=(-1)^{r+s}\sqrt{\kappa _{n}}(V_{(6)}^{r,s}-V_{(6)}^{r,s+3}){\frac{1%
}{\sqrt{\kappa _{n}}}}\,,  \notag \\
&&X^{r}{}_{0}^{s}:=(-1)^{r+s}\sqrt{\kappa _{n}}%
(V^{r,}{}_{0(6)}^{s}-V^{r,}{}_{0(6)}^{s+3})l_{s}^{-1}\,.
\end{eqnarray}
This gives 
\begin{eqnarray}
&&X^{(0)}=C\frac{\hat{m}_{0}^{\ast 2}-1}{3\hat{m}_{0}^{\ast 2}+1}\,,\quad
X^{(+)}=-C\frac{2\hat{m}_{0}^{\ast }+2\hat{m}_{0}^{\ast 2}}{3\hat{m}%
_{0}^{\ast 2}+1}\,,\quad X^{(-)}=C\frac{2\hat{m}_{0}^{\ast }-2\hat{m}%
_{0}^{\ast 2}}{3\hat{m}_{0}^{\ast 2}+1}\,,  \notag \\
&&X_{~~0}^{(0)}=\frac{4\hat{m}_{0}^{\ast 2}}{3\hat{m}_{0}^{\ast 2}+1}\frac{w%
}{\sqrt{2}}\,,\quad X_{~~0}^{(+)}=\frac{2\hat{m}_{0}^{\ast }-2\hat{m}%
_{0}^{\ast 2}}{3\hat{m}_{0}^{\ast 2}+1}\frac{w}{\sqrt{2}}\,,\quad
X_{~~0}^{(-)}=-\frac{2\hat{m}_{0}^{\ast }+2\hat{m}_{0}^{\ast 2}}{3\hat{m}%
_{0}^{\ast 2}+1}\frac{w}{\sqrt{2}}\,,\qquad  \label{eq:Neumann-3-ghost}
\end{eqnarray}
where we defined 
\begin{equation}
\hat{m}_{0}^{\ast }:=\sqrt{\kappa _{n}}\tilde{m}_{0}^{\ast }{\frac{1}{\sqrt{%
\kappa _{n}}}}=\sqrt{\kappa _{n}}K^{\ast -1}m_{0}K^{\ast }{\frac{1}{\sqrt{%
\kappa _{n}}}}\,.
\end{equation}
\end{itemize}

\section{Derivation of regularized matrix formula \label{sec:are}}

Here we sketch a derivation of fundamental formulas for regularized matrices
presented in \S \ref{sec:regularization_ghost}. We begin from the defining
relations in Eq.(\ref{eq:Udef}). The first two equations imply 
Eq.(\ref{eq:UU^-1}) and then from the remaining equations we have 
\begin{equation}
\sum_{o}Q_{eo}(v_{o}^{\prime })^{2}=(\kappa _{e}^{\prime
})^{-1}\,,
~~~\sum_{e}Q_{eo}(w_{e}^{\prime })^{2}=(\kappa _{o}^{\prime })^{-1}\,,
\label{eq:eqvw}
\end{equation}
where 
\begin{equation}
Q_{eo}:={\frac{1}{\kappa _{e}^{\prime }-\kappa _{o}^{\prime }}}\,,
\end{equation}
with $e=\pm 2,\pm 4,\cdots \pm 2N,$ and $o=\pm 1,\pm 3,\cdots \pm (2N-1)$.
Now we regard $Q=(Q_{eo})$ as a $2N\times 2N$ matrix and compute its inverse 
\begin{equation}
(Q^{-1})_{oe}=(\kappa _{e}^{\prime }-\kappa _{o}^{\prime }){\frac{%
\prod_{o^{\prime }\neq o}(\kappa _{e}^{\prime }-\kappa _{o^{\prime
}}^{\prime })\prod_{e^{\prime }\neq e}(\kappa _{o}^{\prime }-\kappa
_{e^{\prime }}^{\prime })}{\prod_{e^{\prime }\neq e}(\kappa _{e}^{\prime
}-\kappa _{e^{\prime }}^{\prime })\prod_{o^{\prime }\neq o}(\kappa
_{o}^{\prime }-\kappa _{o^{\prime }}^{\prime })}}\,.
\end{equation}
To prove the above formula, it is convenient to define a rational function $%
f(z)$ which is determined by the setup $(N,\kappa _{e}^{\prime },\kappa
_{o}^{\prime })$ uniquely: 
\begin{equation}
f(z):=\frac{\prod_{o^{\prime }}(z-\kappa _{o^{\prime }}^{\prime })}{%
\prod_{e^{\prime }}(z-\kappa _{e^{\prime }}^{\prime })}\,.
\label{eq:rational}
\end{equation}
Next we compute $(Q^{-1}Q)_{oo^{\prime \prime }}$. We use contour
integration and residues, where we denote the residue of $f(z)$ at $z=z_{0}$
as $\mathrm{Res}_{z=z_{0}}f(z)$ and assume that the frequencies $\kappa
_{e}^{\prime },\kappa _{o}^{\prime }$ are nondegenerate and finite 
\begin{eqnarray}
&&\sum_{e}{\frac{\kappa _{e}^{\prime }-\kappa _{o}^{\prime }}{\kappa
_{e}^{\prime }-\kappa _{o^{\prime \prime }}^{\prime }}}{\frac{%
\prod_{o^{\prime }\neq o}(\kappa _{e}^{\prime }-\kappa _{o^{\prime
}}^{\prime })\prod_{e^{\prime }\neq e}(\kappa _{o}^{\prime }-\kappa
_{e^{\prime }}^{\prime })}{\prod_{e^{\prime }\neq e}(\kappa _{e}^{\prime
}-\kappa _{e^{\prime }}^{\prime })\prod_{o^{\prime }\neq o}(\kappa
_{o}^{\prime }-\kappa _{o^{\prime }}^{\prime })}}=\sum_{e}{\frac{-\mathrm{Res%
}_{z=\kappa _{e}^{\prime }}f(z)}{(\kappa _{e}^{\prime }-\kappa _{o}^{\prime
})(\kappa _{e}^{\prime }-\kappa _{o^{\prime \prime }}^{\prime })}}{\frac{%
\prod_{e^{\prime }}(\kappa _{o}^{\prime }-\kappa _{e^{\prime }}^{\prime })}{%
\prod_{o^{\prime }\neq o}(\kappa _{o}^{\prime }-\kappa _{o^{\prime
}}^{\prime })}}  \notag \\
&=&-{\frac{\prod_{e^{\prime }}(\kappa _{o}^{\prime }-\kappa _{e^{\prime
}}^{\prime })}{\prod_{o^{\prime }\neq o}(\kappa _{o}^{\prime }-\kappa
_{o^{\prime }}^{\prime })}}\sum_{e}\mathrm{Res}_{z=\kappa _{e}^{\prime }}%
\frac{f(z)}{(z-\kappa _{o}^{\prime })(z-\kappa _{o^{\prime \prime }}^{\prime
})}  \notag \\
&=&{\frac{\prod_{e^{\prime }}(\kappa _{o}^{\prime }-\kappa _{e^{\prime
}}^{\prime })}{\prod_{o^{\prime }\neq o}(\kappa _{o}^{\prime }-\kappa
_{o^{\prime }}^{\prime })}}\oint_{z=\kappa _{o}^{\prime },\kappa _{o^{\prime
\prime }}^{\prime }}\frac{dz}{2\pi i}\frac{f(z)}{(z-\kappa _{o}^{\prime
})(z-\kappa _{o^{\prime \prime }}^{\prime })}  \notag \\
&=&{\frac{\prod_{e^{\prime }}(\kappa _{o}^{\prime }-\kappa _{e^{\prime
}}^{\prime })}{\prod_{o^{\prime }\neq o}(\kappa _{o}^{\prime }-\kappa
_{o^{\prime }}^{\prime })}}\mathrm{Res}_{z=\kappa _{o}^{\prime }}\frac{f(z)}{%
(z-\kappa _{o}^{\prime })^{2}}\,\delta _{o,o^{\prime \prime }}=\delta
_{o,o^{\prime \prime }}\,.
\end{eqnarray}
This shows that we have the correct inverse matrix $Q^{-1}$.

Similarly, we can obtain $(v_{o}^{\prime })^{2},(w_{e}^{\prime })^{2}$ 
Eqs.(\ref{prime},\ref{we}) as follows:\footnote{
We choose the sign convention of $w_e',v_o'$ such that they are
consistent with Eq.(\ref{eq:UU^-1limit}).
} 
\begin{eqnarray}
(v_{o}^{\prime })^{2} &=&\sum_{e}(Q^{-1})_{oe}(\kappa _{e}^{\prime })^{-1}=-{%
\frac{\prod_{e^{\prime }}(\kappa _{o}^{\prime }-\kappa _{e^{\prime
}}^{\prime })}{\prod_{o^{\prime }\neq o}(\kappa _{o}^{\prime }-\kappa
_{o^{\prime }}^{\prime })}}\sum_{e}{\frac{1}{\kappa _{e}^{\prime }}}{\frac{%
\prod_{o^{\prime }\neq o}(\kappa _{e}^{\prime }-\kappa _{o^{\prime
}}^{\prime })}{\prod_{e^{\prime }\neq e}(\kappa _{e}^{\prime }-\kappa
_{e^{\prime }}^{\prime })}}  \notag \\
&=&-{\frac{\prod_{e^{\prime }}(\kappa _{o}^{\prime }-\kappa _{e^{\prime
}}^{\prime })}{\prod_{o^{\prime }\neq o}(\kappa _{o}^{\prime }-\kappa
_{o^{\prime }}^{\prime })}}\sum_{e}\mathrm{Res}_{z=\kappa _{e}^{\prime }}%
\frac{f(z)}{z(z-\kappa _{o}^{\prime })}={\frac{\prod_{e^{\prime }}(\kappa
_{o}^{\prime }-\kappa _{e^{\prime }}^{\prime })}{\prod_{o^{\prime }\neq
o}(\kappa _{o}^{\prime }-\kappa _{o^{\prime }}^{\prime })}}\mathrm{Res}_{z=0}%
\frac{f(z)}{z(z-\kappa _{o}^{\prime })}  \notag \\
&=&{\frac{\prod_{o^{\prime }\neq o}\kappa _{o^{\prime }}^{\prime
}\prod_{e^{\prime }}(\kappa _{e^{\prime }}^{\prime }-\kappa _{o}^{\prime })}{%
\prod_{e^{\prime }}\kappa _{e^{\prime }}^{\prime }\prod_{o^{\prime }\neq
o}(\kappa _{o^{\prime }}^{\prime }-\kappa _{o}^{\prime })}}
={\prod_{o'>0,o'\ne |o|}\kappa_{o'}^2\prod_{e'>0}
(\kappa_{e'}^2-\kappa_{|o|}^2)\over 2\prod_{e'>0}\kappa_{e'}^2
\prod_{o'>0,o'\ne |o|}(\kappa_{o'}^2-\kappa_{|o|}^2)}
\,,  \label{eq:der_vwp_a} \\
(w_{e}^{\prime })^{2} &=&\sum_{o}(Q^{-1})_{oe}(\kappa _{o}^{\prime })^{-1}={%
\frac{\prod_{o^{\prime }}(\kappa _{e}^{\prime }-\kappa _{o^{\prime
}}^{\prime })}{\prod_{e^{\prime }\neq e}(\kappa _{e}^{\prime }-\kappa
_{e^{\prime }}^{\prime })}}\sum_{o}{\frac{1}{\kappa _{o}^{\prime }}}{\frac{%
\prod_{e^{\prime }\neq e}(\kappa _{o}^{\prime }-\kappa _{e^{\prime
}}^{\prime })}{\prod_{o^{\prime }\neq o}(\kappa _{o}^{\prime }-\kappa
_{o^{\prime }}^{\prime })}}  \notag \\
&=&{\frac{\prod_{o^{\prime }}(\kappa _{e}^{\prime }-\kappa _{o^{\prime
}}^{\prime })}{\prod_{e^{\prime }\neq e}(\kappa _{e}^{\prime }-\kappa
_{e^{\prime }}^{\prime })}}\sum_{o}\mathrm{Res}_{z=\kappa _{o}^{\prime }}%
\frac{f(z)}{z(z-\kappa _{e}^{\prime })}=-{\frac{\prod_{o^{\prime }}(\kappa
_{e}^{\prime }-\kappa _{o^{\prime }}^{\prime })}{\prod_{e^{\prime }\neq
e}(\kappa _{e}^{\prime }-\kappa _{e^{\prime }}^{\prime })}}\mathrm{Res}_{z=0}%
\frac{f(z)}{z(z-\kappa _{e}^{\prime })}  \notag \\
&=&{\frac{\prod_{e^{\prime }\neq e}\kappa _{e^{\prime }}^{\prime
}\prod_{o^{\prime }}(\kappa _{e}^{\prime }-\kappa _{o^{\prime }}^{\prime })}{%
\prod_{o^{\prime }}\kappa _{o^{\prime }}^{\prime }\prod_{e^{\prime }\neq
e}(\kappa _{e}^{\prime }-\kappa _{e^{\prime }}^{\prime })}}
={\prod_{e'>0,e'\ne |e|}\kappa_{e'}^2\prod_{o'>0}
(\kappa_{|e|}^2-\kappa_{o'}^2)\over 2\prod_{o'>0}\kappa_{o'}^2
\prod_{e'>0,e'\ne |e|}(\kappa_{|e|}^2-\kappa_{e'}^2)}
\label{eq:der_vwp}
\end{eqnarray}
where we assumed $\kappa _{e}^{\prime },\kappa _{o}^{\prime }$ are
nonzero and Eqs.(\ref{eq:kekoprime}).
We note that the above formula (\ref{eq:der_vwp_a},\ref{eq:der_vwp})
 can also be rewritten as 
\begin{equation}
(v_{o}^{\prime })^{2}={\frac{1}{\kappa _{o}^{\prime }}}\mathrm{Res}%
_{z=\kappa _{o}^{\prime }}\frac{f(0)}{f(z)}\,,~~~~(w_{e}^{\prime })^{2}={%
\frac{1}{\kappa _{e}^{\prime }}}\mathrm{Res}_{z=\kappa _{e}^{\prime }}\frac{%
f(z)}{f(0)}\,.  \label{eq:res_vwp}
\end{equation}
Now we consider the open string limit (\ref{eq:openstringlimit}). By setting
the open string limit $\kappa _{e}^{\prime }=e,\kappa_{o}^{\prime }=o,
N\rightarrow \infty $, the rational function $%
f(z)$ (\ref{eq:rational}) becomes 
\begin{equation}
\frac{f(z)}{f(0)}=\frac{\prod_{n=1}^{\infty }(1-{\frac{z^{2}}{(2n-1)^{2}}})}{%
\prod_{n=1}^{\infty }(1-{\frac{z^{2}}{4n^{2}}})}=\left( \cos {\frac{\pi z}{2}%
}\right) \left( {\frac{2}{\pi z}}\sin {\frac{\pi z}{2}}\right) ^{-1}=\frac{%
\pi z}{2\tan {\frac{\pi z}{2}}}\,.
\end{equation}
With this formula and Eq.(\ref{eq:res_vwp}) we can show that the regularized
quantities reduce to the original ones in Eq.(\ref{eq:UU^-1limit}).

\section{$bc$-ghost sector in position space {\label{sec:functional_rep}}}

Here we consider the ghost position space representation of the $n$-point
string vertices for $n=1,2,3,$ starting with the Fock space formalism.

\paragraph{Identity state}

\begin{eqnarray}
&&|I\rangle =\left( \sum_{o>0}(-1)^{\frac{o-1}{2}}\hat{b}_{-o}\right) \left( 
\hat{b}_{0}+2\sum_{e>0}(-1)^{\frac{e}{2}}\hat{b}_{-e}\right)
e^{\sum_{n=1}^{\infty }(-1)^{n}\hat{c}_{-n}\hat{b}_{-n}}\hat{c}_{0}\hat{c}%
_{1}|\Omega \rangle \,,  \label{eq:identityket} \\
&&\langle c_{0},x_{n},y_{n}|I\rangle =\frac{i}{\sqrt{2}}\left(
\sum_{o>0}(-1)^{\frac{o-1}{2}}x_{o}\right) \prod_{e>0}\left( (i\sqrt{2}%
x_{e})(\sqrt{2}y_{e}-2(-1)^{\frac{e}{2}}c_{0})\right) \,.
\end{eqnarray}
This is BRST invariant\cite{GJ} although the form is rather complicated.

\paragraph{Reflector}

\begin{eqnarray}
&&{}_{12}\langle V_2|= {}_1\langle \Omega|\hat{c}_{-1}^{(1)}\,{}_2\langle
\Omega| \hat{c}_{-1}^{(2)}e^{-\sum_{n=1}^{\infty}(-1)^n(\hat{c}^{(1)}_n \hat{%
b}^{(2)}_n+\hat{c}^{(2)}_n \hat{b}^{(1)}_n)}(\hat{c}^{(1)}_0 +\hat{c}%
^{(2)}_0)\,,  \label{eq:V_2bra} \\
&&|V_2\rangle_{12}=(\hat{b}_0^{(1)}-\hat{b}_0^{(2)})
e^{\sum_{n=1}^{\infty}(-1)^n(\hat{c}_{-n}^{(1)}\hat{b}_{-n}^{(2)} +\hat{c}%
_{-n}^{(2)}\hat{b}_{-n}^{(1)})}\hat{c}_0^{(1)}\hat{c}_{1}^{(1)}
|\Omega\rangle_1\,\hat{c}_0^{(2)}\hat{c}_{1}^{(2)}|\Omega\rangle_2\,,
\label{eq:V_2ket} \\
&&{}_1\langle c_0^{(1)},x_n^{(1)},y_n^{(1)}|{}_2\langle c_0^{(2)},
x_n^{(2)},y_n^{(2)}|V_2\rangle_{12}=(c_0^{(1)}+c_0^{(2)})
\prod_{n=1}^{\infty}\left(-2i((-1)^nx_n^{(1)}+x_n^{(2)})
((-1)^ny_n^{(1)}+y_n^{(2)})\right)  \notag \\
&&\qquad\qquad =\langle c_0^{(1)},x_n^{(1)},y_n^{(1)}
|-c_0^{(2)},-(-1)^nx_n^{(2)},-(-1)^ny_n^{(2)}\rangle\,.
\label{eq:CV2pos}
\end{eqnarray}

\paragraph{3-string vertex}

\begin{eqnarray}
&&|V_{3}\rangle _{123}=e^{\sum_{r,s=1}^{3}\left( -\hat{c}^{\dagger (r)}X^{rs}%
\hat{b}^{\dagger (s)}-\hat{c}^{\dagger (r)}X^{r}{}_{0}^{s}\hat{b}%
_{0}^{(s)}\right) }\hat{c}_{0}^{(1)}\hat{c}_{1}^{(1)}|\Omega \rangle _{1}%
\hat{c}_{0}^{(2)}\hat{c}_{1}^{(2)}|\Omega \rangle _{2}\hat{c}_{0}^{(3)}\hat{c%
}_{1}^{(3)}|\Omega \rangle _{3}\,,  \label{eq:V_3ket} \\
&&{}_{1}\langle c_{0}^{(1)},x_{n}^{(1)},y_{n}^{(1)}|{}_{2}\langle
c_{0}^{(2)},x_{n}^{(2)},y_{n}^{(2)}|{}_{3}\langle
c_{0}^{(3)},x_{n}^{(3)},y_{n}^{(3)}|V_{3}\rangle _{123}  \notag \\
&=&-\det_{r,s,n,m}(\delta ^{r,s}\delta _{n,m}+X_{nm}^{rs})(c_{0}^{(1)}-\bar{w%
}y_{e}^{(1)})(c_{0}^{(2)}-\bar{w}y_{e}^{(2)})(c_{0}^{(3)}-\bar{w}%
y_{e}^{(3)})e^{i\sum_{r,s}y^{(r)}\left( {\frac{1-X}{1+X}}\right)
^{rs}x^{(s)}}\qquad\label{eq:CV3pos}
\end{eqnarray}
where we used the relation\cite{IK}\cite{Oku1}\cite{Oku2}, 
\begin{equation}
X^{r}{}_{0}^{s}=(\delta ^{rs}+X^{rs})\frac{w}{\sqrt{2}}\,.
\end{equation}
Witten's star product in ghost position space 
\begin{eqnarray}
&&\Psi _{1}\star^W \Psi _{2}(c_{0},x_{n},y_{n})=\int dc_{0}^{(2)}dc_{0}^{(3)}%
\frac{dx_{n}^{(2)}dy_{n}^{(2)}}{-2i}\frac{dx_{n}^{(3)}dy_{n}^{(3)}}{-2i} 
\notag \\
&&\quad \times {}_{1}\langle c_{0},x_{n},y_{n}|{}_{2}\langle \widetilde{%
c_{0}^{(2)},x_{n}^{(2)},y_{n}^{(2)}}|{}_{3}\langle \widetilde{%
c_{0}^{(3)},x_{n}^{(3)},y_{n}^{(3)}}|V_{3}\rangle _{123}\Psi
_{1}(c_{0}^{(2)},x_{n}^{(2)},y_{n}^{(2)})\Psi
_{1}(c_{0}^{(3)},x_{n}^{(3)},y_{n}^{(3)})  \notag \\
&=&\int dc_{0}^{(2)}dc_{0}^{(3)}\frac{dx_{n}^{(2)}dy_{n}^{(2)}}{-2i}\frac{%
dx_{n}^{(3)}dy_{n}^{(3)}}{-2i}{}_{1}\langle c_{0},x_{n},y_{n}|{}_{2}\langle
c_{0}^{(2)},x_{n}^{(2)},y_{n}^{(2)}|{}_{3}\langle
c_{0}^{(3)},x_{n}^{(3)},y_{n}^{(3)}|V_{3}\rangle _{123}  \notag \\
&&\qquad \times \Psi
_{1}(-c_{0}^{(2)},-(-1)^{n}x_{n}^{(2)},-(-1)^{n}y_{n}^{(2)})\Psi
_{2}(-c_{0}^{(3)},-(-1)^{n}x_{n}^{(3)},-(-1)^{n}y_{n}^{(3)})
\end{eqnarray}
where 
\begin{equation}
{}_{1}\langle \widetilde{c_{0},x_{n},y_{n}}|:={}_{12}\langle
V_{2}|c_{0},x_{n},y_{n}\rangle _{2}={}_{1}\langle
-c_{0},-(-1)^{n}x_{n},-(-1)^{n}y_{n}|\,.  \label{eq:real_ghost}
\end{equation}

\section{Moyal $\star $ for $\left( DD\right) _{b}$\thinspace $\left(
NN\right) _{c}$ split strings \label{DDNN}}

In this appendix, we examine the remaining choice of the midpoint boundary
condition for the split string variables as compared with our discussion in
section \ref{sec:Moyalbc}. Namely, we consider Dirichlet boundary
condition for $b(\sigma )$ and Neumann for $c(\sigma )$ at the midpoint $%
\sigma =\pi /2$. In this case the left and right half of $b(\sigma )$ : $%
l^{b}(\sigma ),r^{b}(\sigma )$ satisfy Dirichlet boundary condition at both $%
\sigma =0,\pi /2$, and those of $c(\sigma )$ : $l^{c}(\sigma ),r^{b}(\sigma )
$ satisfy Neumann at $\sigma =0,\pi /2$. The $l^{b}(\sigma ),r^{b}(\sigma )$
and $l^{c}(\sigma ),r^{c}(\sigma )$ are expanded in terms of \textit{even}
sine/cosine modes respectively 
\begin{align}
& l^{b}(\sigma )={\frac{2}{\pi }}\sigma \bar{b}+i\sqrt{2}\sum_{e=2}^{\infty
}l_{e}^{b}\sin e\sigma \,,\quad r^{b}(\sigma )={\frac{2}{\pi }}\sigma \bar{b%
}+i\sqrt{2}\sum_{e=2}^{\infty }r_{e}^{b}\sin e\sigma \,, \\
& l^{c}(\sigma )=\bar{c}+\sqrt{2}\sum_{e=2}^{\infty }l_{e}^{c}(\cos
e\sigma -i^e)\,,\quad r^{c}(\sigma )=\bar{c}+\sqrt{2}%
\sum_{e=2}^{\infty }r_e^{c}(\cos e\sigma -i^e)\,.
\end{align}
{} From Eqs.(\ref{eq:bsigma})(\ref{eq:csigma})(\ref{eq:sinDD_psi2lr}) (\ref
{eq:evencos}), we have relations between split and full string variables: 
\begin{eqnarray}
&&\bar{b}=\bar{\tilde{w}}x_{o}\,,\quad l_{e}^{b}=x_{e}+\tilde{T}%
x_{o}\,,\quad r_{e}^{b}=-x_{e}+\tilde{T}x_{o}\,,  \label{half_b_even} \\
\  &&\bar{c}=c_{0}-\bar{w}y_{e}\,,\quad l_{e}^{c}=y_{e}+Ty_{o}\,.\quad
r_{e}^{c}=y_{e}-Ty_{o}\,.  \label{half_c_even}
\end{eqnarray}
With this setup Witten type product in the split string formulation becomes: 
\begin{equation}
{\tilde{A}}^{\prime }\ast {\tilde{B}}^{\prime }(\bar{b},\bar{c}%
,l_{e}^{b},l_{e}^{c},r_{e}^{b},r_{e}^{c})=\int \prod_{e>0}\left( id\eta
_{e}^{b}d\eta _{e}^{c}\right) \,{\tilde{A}}^{\prime }(\bar{b},\bar{c}%
,l_{e}^{b},l_{e}^{c},\eta _{e}^{b},\eta _{e}^{c}){\tilde{B}}^{\prime }(\bar{b%
},\bar{c},\eta _{e}^{b},-\eta _{e}^{c},r_{e}^{b},r_{e}^{c})\,.
\label{bc_half_star_even}
\end{equation}
where the split string and full string fields in position space are the same 
\begin{equation}
\tilde{A}^{\prime }(\bar{b},\bar{c},l_{e}^{b},l_{e}^{c},r_{e}^{b},r_{e}^{c})%
\sim \Psi (c_{0},x_{n},y_{n})  \label{eq:corr_evenfield}
\end{equation}
by substituting on the right hand side the inverse maps obtained in Eqs.(\ref
{half_b_even})(\ref{half_c_even}) 
\begin{eqnarray}
&&x_{e}={\frac{1}{2}}(l_{e}^{b}-r_{e}^{b})\,,\quad x_{o}=\tilde{u}_{o}\bar{b}%
+{\frac{1}{2}}\bar{S}_{oe}(l_{e}^{b}+r_{e}^{b})\,, \\
&&c_{0}=\bar{c}+{\frac{1}{2}}\bar{w}_{e}(l_{e}^{c}+r_{e}^{c})\,,\quad y_{e}={%
\frac{1}{2}}(l_{e}^{c}+r_{e}^{c})\,,\quad y_{o}={\frac{1}{2}}%
R_{oe}(l_{e}^{c}-r_{e}^{c})\,.
\end{eqnarray}
These relations are valid only when $\bar{w}w=\infty $ in the limit: $\kappa
_{e}=e,\kappa _{o}=o,N=\infty $.

As the next step, we consider the Moyal formulation, including the
regularization with $(N,\kappa _{e},\kappa _{o})$. Comparing Eqs.(\ref
{half_b_even})(\ref{half_c_even})(\ref{bc_half_star_even}) with Eq.(\ref
{anti-from half}), we identify 
\begin{equation*}
x\sim \tilde{T}x_{o}\,,\quad y\sim 2x_{e}\,,\quad x^{\prime }\sim
-Ty_{o}\,,\quad y^{\prime }\sim 2y_{e}\,,
\end{equation*}
and \textit{define} new variables with even index $E$ by 
\begin{equation}
x_{E}:=\tilde{T}x_{o}\,,\quad y_{E}:=Ty_{o}\,.
\end{equation}
At the limit $\kappa _{e}=e,\kappa _{o}=o,N=\infty $, $\tilde{T}$ has a zero
mode $\tilde{u}$ (\ref{eq:til_mat_rel_infty}) and we meet as usual the
associativity anomaly, but at finite $N$ everything is well-defined. From
Eq.(\ref{anti-from half})(\ref{eq:corr_evenfield}) we obtain the Moyal image 
$A^{\prime }(\bar{b},\bar{c},x_{E},p_{E},y_{E},q_{E})$ of the position space
field $\Psi (c_{0},x_{n},y_{n})$ : 
\begin{eqnarray}
&&A^{\prime }(\bar{b},\bar{c},x_{E},p_{E},y_{E},q_{E})  \notag \\
&=&2^{-2N}\int \prod_{e>0}\left( i^{-1}dx_{e}dy_{e}\right)
e^{-2p_{E}x_{e}-2q_{E}y_{e}}\tilde{A}^{\prime }(\bar{b},\bar{c}%
,x_{e}+x_{E},y_{e}+y_{E},-x_{e}+x_{E},y_{e}-y_{E})  \notag \\
&=&2^{-2N}\int \prod_{e>0}\left( i^{-1}dx_{e}dy_{e}\right)
e^{-2p_{E}x_{e}-2q_{E}y_{e}}\Psi (\bar{c}+\bar{w}y_{e},x_{e},\tilde{u}\bar{b}%
+\bar{S}x_{E},y_{e},Ry_{E})\,,
\end{eqnarray}
and the corresponding Moyal $\star $ product becomes 
\begin{equation}
\star =e^{-{\frac{1}{2}}\left( {\frac{\overleftarrow{\partial }}{\partial
x_{E}}}{\frac{\overrightarrow{\partial }}{\partial p_{E}}}+{\frac{%
\overleftarrow{\partial }}{\partial y_{E}}}{\frac{\overrightarrow{\partial }%
}{\partial q_{E}}}+{\frac{\overleftarrow{\partial }}{\partial p_{E}}}{\frac{%
\overrightarrow{\partial }}{\partial x_{E}}}+{\frac{\overleftarrow{\partial }%
}{\partial q_{E}}}{\frac{\overrightarrow{\partial }}{\partial y_{E}}}\right)
}\,.
\end{equation}
In this case, the above formula is more complicated than our previous choice
due to the additional $b$-ghost midpoint mode $\bar{b}$.

\section{Ghost butterfly projector with even modes\label{sec:butterfly}}

The butterfly projector in Eq.(\ref{oddbuttterfly}) is based on the odd mode
oscillators in Eq.(\ref{oddmode_fields}). There is another choice using even
mode oscillators Eq.(\ref{evenmode_fields2}) 
\begin{equation}
\beta _{e}^{b}\star \hat{A}_{B}^{\prime }=\beta _{e}^{c}\star \hat{A}%
_{B}^{\prime }=\hat{A}_{B}^{\prime }\star \beta _{-e}^{b}=\hat{A}%
_{B}^{\prime }\star \beta _{-e}^{c}=0\,,\quad \forall e>0\,.
\label{evenbutterfly}
\end{equation}
Explicitly, we have the even butterfly state 
\begin{equation}
\hat{A}_{B}^{\prime }=\xi _{0}\,2^{-2N}\exp \left( -\sum_{e>0}\left(
ix_{e}^{b}\kappa _{e}x_{e}^{c}+{\frac{4i}{{\theta ^{\prime }}^{2}}}%
p_{e}^{b}\kappa _{e}^{-1}p_{e}^{c}\right) \right)  \label{eq:even_butterfly}
\end{equation}
in the Siegel gauge.

As we will see in the following, the even butterfly is the one defined by
Gaiotto-Rastelli-Sen-Zwiebach(GRSZ) using twisted ghosts \cite
{Gaiotto:2001ji}. The conditions in Eq.(\ref{evenbutterfly}) correspond to 
\begin{eqnarray}
&&\hat{b}_{e}|\Psi _{B}\rangle =0\,,\qquad 
\hat{c}_e|\Psi _{B}\rangle =0\,,\nonumber\\
&&\sum_{o>0}\left( \hat{b}_{o}U_{-o,e}^{-1}+\hat{b}_{-o}U_{o,e}^{-1}\right)
|\Psi _{B}\rangle =0\,,\quad \sum_{o>0}\left( U_{e,-o}\hat{c}_{o}+U_{e,o}%
\hat{c}_{-o}\right) |\Psi _{B}\rangle =0
\end{eqnarray}
for all $e>0$ in ordinary oscillator language. If we take a gaussian ansatz 
\begin{equation}
|\Psi _{B}\rangle =\mathcal{N}\exp \left( \sum_{n,m\geq 1}\hat{c}_{-m}{%
\tilde{V}}_{mn}^{B}\hat{b}_{-n}\right) \hat{c}_{1}|\Omega \rangle \,,
\end{equation}
the above conditions become 
\begin{eqnarray}
&&\tilde{V}_{eo}^{B}=\tilde{V}_{oe}^{B}=\tilde{V}_{ee^{\prime }}^{B}=0\,,
\qquad e,e',o>0\,,\label{eq:VVVB} \\
&&\sum_{o^{\prime }>0}\tilde{V}_{o^{\prime }o}^{B}U_{-o^{\prime
},e}^{-1}=-U_{o,e}^{-1}\,,\qquad e>0\,,  \label{eq:VUU-1} \\
&&\sum_{o^{\prime }>0}\tilde{V}_{oo^{\prime }}^{B}U_{e,-o^{\prime
}}=U_{e,o}\,,\qquad e>0\,.  \label{eq:VUU}
\end{eqnarray}
We can solve Eq.(\ref{eq:VUU-1}) by comparing it with the matter one (\ref
{eq:butterfly_VB}) as 
\begin{eqnarray}
\tilde{V}_{mn}^{B} &=&-\frac{1}{\sqrt{n}}V_{nm}^{B}\sqrt{m}=-\sqrt{m}%
V_{mn}^{B}{\frac{1}{\sqrt{n}}}  \notag \\
&=&\left\{ 
\begin{array}{cl}
(-1)^{\frac{m+n}{2}}\frac{m}{m+n}\frac{\Gamma \lbrack \frac{m}{2}]\Gamma
\lbrack \frac{n}{2}]}{\pi \Gamma \lbrack \frac{m+1}{2}]\Gamma \lbrack \frac{%
n+1}{2}]} & \mathrm{for}\;m\;\mathrm{and}\;n\;\mathrm{odd} \\ 
0 & \mathrm{for}\;m\;\mathrm{or}\;n\;\mathrm{even}
\end{array}
\right.\,.
\label{eq:VBtilVB}
\end{eqnarray}
We can check that this ${\tilde{V}}^{B}$ satisfies Eq.(\ref{eq:VUU}). The
ghost butterfly which we have obtained in MSFT as above can be identified
with GRSZ's (twisted) ones \cite{Gaiotto:2001ji}\footnote{%
There is a correspondence for the vacuum: $\hat{c}_{1}|\Omega \rangle \sim
|\Omega ^{\prime }\rangle $ .}: 
\begin{equation*}
\hat{A}_{B}^{\prime }\quad \leftrightarrow \quad |\Psi _{B}\rangle =\exp
\left( -{\frac{1}{2}}L_{-2}^{\prime }\right) |\Omega ^{\prime }\rangle \quad 
\mathrm{for}\quad \kappa _{e}=e,\kappa _{o}=o,N=\infty \,.
\end{equation*}
The relation between the matter and (twisted) ghost generating functions is
obtained by using \cite{Okuda:2002fj} 
\begin{equation*}
{\frac{\partial }{\partial z}}\tilde{S}(w,z)=S(z,w)
\end{equation*}
where we defined the generating functions 
\begin{eqnarray}
&&S(z,w):=\sum_{m,n=1}^{\infty }\sqrt{mn}(-z)^{m-1}(-w)^{n-1}S_{mn}=\frac{1}{%
(z-w)^{2}}-\frac{f^{\prime }(z)f^{\prime }(w)}{(f(z)-f(w))^{2}}\,, \\
&&{\tilde{S}}(z,w):=\sum_{m,n=1}^{\infty }(-z)^{m-1}(-w)^{n}{\tilde{S}}%
_{mn}=-\frac{w}{z(w-z)}+\frac{f(w)}{f(z)}\frac{f^{\prime }(z)}{f(w)-f(z)}\,,
\\
&&|S\rangle =\exp \left( -{\frac{1}{2}}\sum_{m,n=1}^{\infty }\hat{a}%
_{m}^{\dagger }S_{mn}\hat{a}_{n}^{\dagger }\right) |\Omega \rangle \,,\quad |%
\tilde{S}\rangle =\exp \left( \sum_{m,n=1}^{\infty }\hat{c}_{-m}\tilde{S}%
_{mn}\hat{b}_{-n}\right) |\Omega ^{\prime }\rangle
\end{eqnarray}
namely 
\begin{equation}
\tilde{S}_{mn}=-{\frac{1}{\sqrt{n}}}S_{nm}\sqrt{m}=-\sqrt{m}S_{mn}{\frac{1}{%
\sqrt{n}}}\,.
\end{equation}
Here we have $V_{mn}^{B}=S_{mn}$ for the conformal mapping 
$f(z)={\frac{z}{\sqrt{1+z^{2}}}}$ which represents (canonical) 
butterfly state $e^{-{\frac{1}{2}}L_{-2}}|\Omega \rangle $
\cite{Gaiotto:2002kf} and the relation Eq.(\ref{eq:VBtilVB}).

\section{Algebra of gaussian operators \label{sec:gaussian}}

We discuss algebraic relations for gaussians constructed from bosonic and
fermionic oscillators which we used to compute the propagator (\ref
{eq:propagator_def}) in MSFT.

For bosonic oscillators: $a,a^{\dagger },[a,a^{\dagger }]=1$, we can prove a
formula 
\begin{eqnarray}
e^{a^{\dagger }Aa^{\dagger }+aBa} &=&e^{-{\frac{1}{2}}\mathrm{Tr}\log (\cos
(2\sqrt{AB}))}e^{{\frac{1}{2}}a^{\dagger }\tan \left( 2\sqrt{AB}\right) 
\sqrt{AB}B^{-1}a^{\dagger }}  \notag \\
&&\times \,e^{-a^{\dagger }\log \left( \cos \left( 2\sqrt{AB}\right) \right)
a}e^{{\frac{1}{2}}aB{\frac{\tan \left( 2\sqrt{AB}\right) }{\sqrt{AB}}}a}
\end{eqnarray}
using a similar method to Appendix A in \cite{K-P}. Here $A,B$ are symmetric
matrices: $\bar{A}=A,\bar{B}=B$. Then we have 
\begin{eqnarray}
e^{\eta A\eta +{\frac{\partial }{\partial \eta }}B{\frac{\partial }{\partial
\eta }}}e^{i\xi \eta } &=&e^{-{\frac{1}{2}}\mathrm{Tr}\log (\cos (2\sqrt{AB}%
))}  \notag \\
&&\times \,e^{{\frac{1}{2}}\eta \tan \left( 2\sqrt{AB}\right) \sqrt{AB}%
B^{-1}\eta -{\frac{1}{2}}\xi B{\frac{\tan \left( 2\sqrt{AB}\right) }{\sqrt{AB%
}}}\xi +i\eta {\frac{1}{\cos \left( 2\sqrt{AB}\right) }}\xi }
\end{eqnarray}
where we used the relation 
\begin{equation}
\left[ {\frac{\partial }{\partial \eta }},\eta \right] =1,\quad e^{\eta C{%
\frac{\partial }{\partial \eta }}}e^{i\xi \eta }e^{-\eta C{\frac{\partial }{%
\partial \eta }}}=e^{\eta e^{C}\xi }.
\end{equation}
We obtain the formula for the propagator 
\begin{eqnarray}
&&\int d^{M}\xi \,e^{-i\xi \eta }e^{\eta ^{\prime }A\eta ^{\prime }+{\frac{%
\partial }{\partial \eta ^{\prime }}}B{\frac{\partial }{\partial \eta
^{\prime }}}}e^{i\xi \eta ^{\prime }}  \notag \\
&=&(2\pi )^{\frac{M}{2}}\,e^{-{\frac{1}{2}}\mathrm{Tr}\log \left( B{\frac{%
\sin (2\sqrt{AB})}{\sqrt{AB}}}\right) }e^{-{\frac{1}{2}}\eta {\frac{\sqrt{AB}%
}{\tan (2\sqrt{AB})}}B^{-1}\eta -{\frac{1}{2}}\eta ^{\prime }{\frac{\sqrt{AB}%
}{\tan (2\sqrt{AB})}}B^{-1}\eta ^{\prime }+\eta ^{\prime }{\frac{\sqrt{AB}}{%
\sin (2\sqrt{AB})}}B^{-1}\eta }\,.
\end{eqnarray}
When the momentum is nonzero in (\ref{eq:propagator_def}), we need the
modified version of the above formula: 
\begin{eqnarray}
&&\int d^{M}\xi \,e^{-i\xi \eta }e^{\eta ^{\prime }A\eta ^{\prime }+{\frac{%
\partial }{\partial \eta ^{\prime }}}B{\frac{\partial }{\partial \eta
^{\prime }}}+C\eta ^{\prime }}e^{i\xi \eta ^{\prime }}  \notag \\
&=&(2\pi )^{\frac{M}{2}}e^{-{\frac{1}{2}}\mathrm{Tr}\log \left( B{\frac{\sin
(2\sqrt{AB})}{\sqrt{AB}}}\right) }e^{-{\frac{1}{2}}\eta {\frac{\sqrt{AB}}{%
\tan (2\sqrt{AB})}}B^{-1}\eta -{\frac{1}{2}}\eta ^{\prime }{\frac{\sqrt{AB}}{%
\tan (2\sqrt{AB})}}B^{-1}\eta ^{\prime }+\eta ^{\prime }{\frac{\sqrt{AB}}{%
\sin (2\sqrt{AB})}}B^{-1}\eta }  \notag \\
&&\times \,e^{-{\frac{1}{4}}CA^{-1}\left( 1-{\frac{\tan \sqrt{AB}}{\sqrt{AB}}%
}\right) C+{\frac{1}{2}}(\eta +\eta ^{\prime }){\frac{\tan \sqrt{AB}}{\sqrt{%
AB}}}C}\,.  \label{eq:formula_matter_heat}
\end{eqnarray}

For fermionic oscillators: $a,a^{\dagger },\{a_{i},a_{j}^{\dagger }\}=\delta
_{ij}$, we have a similar formula 
\begin{eqnarray}
e^{\bar{a}^{\dagger }Aa^{\dagger }+\bar{a}Ba} &=&e^{{\frac{1}{2}}\mathrm{Tr}%
(\log \cosh (2\sqrt{AB}))}e^{{\frac{1}{2}}\bar{a}^{\dagger }\tanh (2\sqrt{AB}%
)\sqrt{AB}B^{-1}a^{\dagger }}  \notag \\
&&\times \,e^{-\bar{a}^{\dagger }\log (\cosh (2\sqrt{AB}))a}e^{{\frac{1}{2}}%
\bar{a}B{\frac{\tanh (2\sqrt{AB})}{\sqrt{AB}}}a}\,,
\end{eqnarray}
where $A,B$ are antisymmetric matrices $\bar{A}=-A,\bar{B}=-B$. Noting 
\begin{equation}
\left\{ \eta ,{\frac{\partial }{\partial \eta }}\right\} =1\,,\quad e^{\bar{%
\eta}C{\frac{\partial }{\partial \eta }}}e^{-\bar{\xi}\eta }e^{-\bar{\eta}C{%
\frac{\partial }{\partial \eta }}}=e^{\bar{\eta}e^{C}\xi }\,,
\end{equation}
we obtain 
\begin{equation*}
e^{\bar{\eta}A\eta +\bar{\frac{\partial }{\partial \eta }}B{\frac{\partial }{%
\partial \eta }}}e^{-\bar{\xi}\eta }=e^{{\frac{1}{2}}\mathrm{Tr}(\log \cosh
(2\sqrt{AB}))}e^{{\frac{1}{2}}\bar{\eta}\tanh (2\sqrt{AB})\sqrt{AB}%
B^{-1}\eta +{\frac{1}{2}}\bar{\xi}B{\frac{\tanh {2\sqrt{AB}}}{\sqrt{AB}}}\xi
+\bar{\eta}{\frac{1}{\cosh {2\sqrt{AB}}}}\xi }\,.
\end{equation*}
By integration we have 
\begin{eqnarray}
&&\int d\xi \,e^{\bar{\xi}\eta }e^{{\bar{\eta}^{\prime }A\eta ^{\prime }+%
\bar{\frac{\partial }{\partial \eta ^{\prime }}}}B{\frac{\partial }{\partial
\eta ^{\prime }}}}e^{-\bar{\xi}\eta ^{\prime }}  \notag \\
&=&{\det }^{\frac{1}{2}}\left( B{\frac{\sinh 2\sqrt{AB}}{\sqrt{AB}}}\right)
\,e^{{\frac{1}{2}}\bar{\eta}{\frac{\sqrt{AB}}{\tanh 2\sqrt{AB}}}B^{-1}\eta +{%
\frac{1}{2}}\bar{\eta}^{\prime }{\frac{\sqrt{AB}}{\tanh 2\sqrt{AB}}}%
B^{-1}\eta ^{\prime }-\bar{\eta}^{\prime }{\frac{\sqrt{AB}}{\sinh 2\sqrt{AB}}%
}B^{-1}\eta }\,.  \label{eq:formula_heat}
\end{eqnarray}

\section{Neumann coefficients}

\subsection{Neumann coefficients from CFT}

\label{sec:Neumann_GJ} In this subsection, we give a short summary of the
analytic expression of Neumann coefficients given in \cite{GJ} which are
obtained from conformal field theory. We introduce a set of numbers $%
A_{n},B_{n}$ ($n=0,1,2,\cdots $) which appear in the Taylor expansion, 
\begin{equation}
\left( \frac{1+ix}{1-ix}\right) ^{1/3}=\sum_{e\geq
0}A_{e}x^{e}+i\sum_{o>0}A_{o}x^{o}\,,~~~~\left( \frac{1+ix}{1-ix}\right)
^{2/3}=\sum_{e\geq 0}B_{e}x^{e}+i\sum_{o>0}B_{o}x^{o}\,.
\end{equation}
{}From these data, the Neumann coefficients ($N^{(0,\pm)}_{nm}$ for
matter sector and $\tilde{N}^{(0,\pm)}_{nm}$ for ghost sector)
 are written as follows. First when $n,m>0$ and $n\neq m$, 
\begin{eqnarray}
N_{nm}^{(0)} &=&\left\{ 
\begin{array}{ll}
\frac{(-1)^{n}}{3}\left( \frac{A_{n}B_{m}+B_{n}A_{m}}{(n+m)}+\frac{%
A_{n}B_{m}-B_{n}A_{m}}{(n-m)}\right) \quad  & n+m=\mbox{even} \\ 
0 & n+m=\mbox{odd}
\end{array}
\right. \,, \\
N_{nm}^{(\pm )} &=&\left\{ 
\begin{array}{ll}
\frac{-(-1)^{n}}{6}\left( \frac{A_{n}B_{m}+B_{n}A_{m}}{(n+m)}+\frac{%
A_{n}B_{m}-B_{n}A_{m}}{(n-m)}\right) \quad  & n+m=\mbox{even} \\ 
\frac{\pm \sqrt{3}}{6}\left( \frac{A_{n}B_{m}-B_{n}A_{m}}{(n+m)}+\frac{%
A_{n}B_{m}+B_{n}A_{m}}{(n-m)}\right) \quad  & n+m=\mbox{odd} \\ 
& 
\end{array}
\right. \,, \\
{\tilde{N}}_{nm}^{(0)} &=&\left\{ 
\begin{array}{ll}
\frac{-(-1)^{n}}{3}\left( \frac{A_{n}B_{m}+B_{n}A_{m}}{(n+m)}-\frac{%
A_{n}B_{m}-B_{n}A_{m}}{(n-m)}\right) \quad  & n+m=\mbox{even} \\ 
0 & n+m=\mbox{odd}
\end{array}
\right. \,, \\
\tilde{N}_{nm}^{(\pm )} &=&\left\{ 
\begin{array}{ll}
\frac{(-1)^{n}}{6}\left( \frac{A_{n}B_{m}+B_{n}A_{m}}{(n+m)}-\frac{%
A_{n}B_{m}-B_{n}A_{m}}{(n-m)}\right) \quad  & n+m=\mbox{even} \\ 
\frac{\mp \sqrt{3}}{6}\left( \frac{A_{n}B_{m}-B_{n}A_{m}}{(n+m)}-\frac{%
A_{n}B_{m}+B_{n}A_{m}}{(n-m)}\right) \quad  & n+m=\mbox{odd} \\ 
& 
\end{array}
\right. \,.
\end{eqnarray}
For the diagonal components ($n=m>0$), they are replaced by, 
\begin{eqnarray}
&&N_{nn}^{(0)}=\frac{1}{3n}\left(
2(-1)^{n}(1+\sum_{k=1}^{n}(-1)^{k}A_{k}^{2})-(-1)^{n}-A_{n}^{2}\right)
\,,~~~~~N_{nn}^{(\pm )}=-\frac{(-1)^{n}}{2n}-\frac{N_{nn}^{(0)}}{2}\,,~~ \\
&&{\tilde{N}}_{nn}^{(0)}=N_{nn}^{(0)}-\frac{2(-1)^{n}A_{n}B_{n}}{3n}%
\,,~~~~~~~~{\tilde{N}}_{nn}^{(\pm )}=-\frac{(-1)^{n}}{2n}-\frac{1}{2}{%
\tilde{N}}_{nn}^{(0)}\,.
\end{eqnarray}
For the zero mode, we use 
\begin{eqnarray}
&&N_{0m}^{(0)}=\left\{ 
\begin{array}{ll}
\frac{2}{3m}A_{m}\quad  & m=\mbox{even} \\ 
0 & m=\mbox{odd}
\end{array}
\right. \,,~~N_{0m}^{(\pm )}=\left\{ 
\begin{array}{ll}
\frac{-1}{3m}A_{m}\quad  & m=\mbox{even} \\ 
\frac{\mp \sqrt{3}}{3m}A_{m} & m=\mbox{odd}
\end{array}
\right. \,,~~N_{00}=-{\frac{1}{2}}\mathrm{ln}{\frac{3^{3}}{4^{2}}}\,,~~~~~~~
\\
&&{\tilde{N}}_{0m}^{(0)}=\left\{ 
\begin{array}{ll}
\frac{-2}{3m}B_{m}\quad  & m=\mbox{even} \\ 
0 & m=\mbox{odd}
\end{array}
\right. \,,~~~{\tilde{N}}_{0m}^{(\pm )}=\left\{ 
\begin{array}{ll}
\frac{1}{3m}B_{m}\quad  & m=\mbox{even} \\ 
\frac{\mp \sqrt{3}}{3m}B_{m} & m=\mbox{odd}
\end{array}
\right. \,.
\end{eqnarray}

There are some differences in the convention to make direct comparison of
these quantities with the corresponding ones obtained in Moyal language
which are given in (\ref{eq:Neumann-3-matter}, \ref{eq:Neumann-3-ghost}). We
summarize them as follows, 
\begin{eqnarray}
&&\mathcal{M}_{nm}^{(0,\pm )}(cft)\equiv -(-1)^{n}\sqrt{mn}N_{mn}^{(0,\pm )}
\,,~~
\mathcal{V}_{n}^{(0,\pm )}(cft)\equiv -l_{s}\sqrt{n}N_{0n}^{(0,\pm
)}\,,~~V_{00}(cft)\equiv -l_{s}^{2}N_{00}\,,~~~~~~~ \\
&&X_{nm}^{(0,\pm )}(cft)\equiv m{\tilde{N}}_{nm}^{(0,\pm
)}\,,~~~~~~~~~X_{m0}^{(0,\pm )}(cft)\equiv m{\tilde{N}}_{0m}^{(0,\pm )}\,.
\end{eqnarray}
The sign factor in $\mathcal{M}^{(0,\pm )}(cft)$ comes in because 
we include the
multiplication of $C_{nm}=(-1)^{n}\delta _{n,m}$ in the MSFT definition.

\subsection{Ratios of MSFT-regulated and CFT Neumann coefficients}

\label{sec:Neumann_ratio}

The MSFT-regulated Neumann coefficients are discussed in sections
\ref{sec:Neumann coef},\ref{sec:Neumann_numerical} and 
\ref{sec:Neumann-matter}. 
The numerical ratios $\mathcal{M}_{ee}^{(0)}\left( N\right)
/{\cal M}_{ee^{\prime }}^{\left( 0\right) }\left(cft\right) $ and 
$X_{ee}^{(0)}\left( N\right) /X_{ee^{\prime }}^{\left( 0\right) }\left(
cft\right) $ for $e,e^{\prime }=2,4,6,8$ at $N=5,20,100,400,$ shows the
convergence of these to the CFT values in the large $N$ limit.
\begin{center}
\begin{tabular}{|l|cccc||l||cccc|}
\hline
$\frac{{\cal M}_{ee^{\prime }}^{\left( 0\right) }\left( 5\right) }
{{\cal M}_{ee^{\prime
}}^{\left( 0\right) }\left( cft\right) }$ & {\small 2} & {\small 4} & {\small 6}
& {\small 8} & $\frac{{\cal M}_{ee^{\prime }}^{\left( 0\right) }
\left( 20\right) }{{\cal M}_{ee^{\prime }}^{\left( 0\right) }
\left( cft\right) }$ & {\small 2} & {\small 
4} & {\small 6} & {\small 8} \\ \hline
\multicolumn{1}{|c|}{\small 2} & \multicolumn{1}{|l}{\small 1.15355} & 
\multicolumn{1}{l}{\small 1.27359} & \multicolumn{1}{l}{\small 1.43938} & 
\multicolumn{1}{l||}{\small 1.71214} & \multicolumn{1}{|c||}{\small 2} & 
\multicolumn{1}{|l}{\small 1.02373} & \multicolumn{1}{l}{\small 1.04035} & 
\multicolumn{1}{l}{\small 1.05899} & \multicolumn{1}{l|}{\small 1.07957} \\ 
\multicolumn{1}{|c|}{\small 4} & \multicolumn{1}{|l}{\small 1.27359} & 
\multicolumn{1}{l}{\small 1.41879} & \multicolumn{1}{l}{\small 1.61307} & 
\multicolumn{1}{l||}{\small 1.92691} & \multicolumn{1}{|c||}{\small 4} & 
\multicolumn{1}{|l}{\small 1.04035} & \multicolumn{1}{l}{\small 1.05982} & 
\multicolumn{1}{l}{\small 1.08099} & \multicolumn{1}{l|}{\small 1.10391} \\ 
\multicolumn{1}{|c|}{\small 6} & \multicolumn{1}{|l}{\small 1.43938} & 
\multicolumn{1}{l}{\small 1.61307} & \multicolumn{1}{l}{\small 1.84084} & 
\multicolumn{1}{l||}{\small 2.20463} & \multicolumn{1}{|c||}{\small 6} & 
\multicolumn{1}{|l}{\small 1.05899} & \multicolumn{1}{l}{\small 1.08099} & 
\multicolumn{1}{l}{\small 1.10438} & \multicolumn{1}{l|}{\small 1.12937} \\ 
\multicolumn{1}{|c|}{\small 8} & \multicolumn{1}{|l}{\small 1.71214} & 
\multicolumn{1}{l}{\small 1.92691} & \multicolumn{1}{l}{\small 2.20463} & 
\multicolumn{1}{l||}{\small 2.64473} & \multicolumn{1}{|c||}{\small 8} & 
\multicolumn{1}{|l}{\small 1.07957} & \multicolumn{1}{l}{\small 1.10391} & 
\multicolumn{1}{l}{\small 1.12937} & \multicolumn{1}{l|}{\small 1.15628} \\ 
\hline
\end{tabular}
\vskip5mm

\begin{tabular}{|l|cccc||l||cccc|}
\hline
$\frac{M_{ee^{\prime }}^{\left( 0\right) }\left( 100\right) }{M_{ee^{\prime
}}^{\left( 0\right) }\left( cft\right) }$ & {\small 2} & {\small 4} &
 {\small 6}
& {\small 8} & $\frac{{\cal M}_{ee^{\prime }}^{\left( 0\right) }
\left( 400\right) }{{\cal M}_{ee^{\prime }}^{\left( 0\right) }
\left( cft\right) }$ & {\small 2} & {\small
4} & {\small 6} & {\small 8} \\ \hline
\multicolumn{1}{|c|}{\small 2} & \multicolumn{1}{|l}{\small 1.00272} & 
\multicolumn{1}{l}{\small 1.00459} & \multicolumn{1}{l}{\small 1.00664} & 
\multicolumn{1}{l||}{\small 1.00885} & \multicolumn{1}{|c||}{\small 2} & 
\multicolumn{1}{|l}{\small 1.00043} & \multicolumn{1}{l}{\small 1.00071} & 
\multicolumn{1}{l}{\small 1.00103} & \multicolumn{1}{l|}{\small 1.00137} \\ 
\multicolumn{1}{|c|}{\small 4} & \multicolumn{1}{|l}{\small 1.00459} & 
\multicolumn{1}{l}{\small 1.00675} & \multicolumn{1}{l}{\small 1.00907} & 
\multicolumn{1}{l||}{\small 1.01151} & \multicolumn{1}{|c||}{\small 4} & 
\multicolumn{1}{|l}{\small 1.00071} & \multicolumn{1}{l}{\small 1.00105} & 
\multicolumn{1}{l}{\small 1.0014} & \multicolumn{1}{l|}{\small 1.00178} \\ 
\multicolumn{1}{|c|}{\small 6} & \multicolumn{1}{|l}{\small 1.00664} & 
\multicolumn{1}{l}{\small 1.00907} & \multicolumn{1}{l}{\small 1.0116} & 
\multicolumn{1}{l||}{\small 1.01426} & \multicolumn{1}{|c||}{\small 6} & 
\multicolumn{1}{|l}{\small 1.00103} & \multicolumn{1}{l}{\small 1.0014} & 
\multicolumn{1}{l}{\small 1.00179} & \multicolumn{1}{l|}{\small 1.0022} \\ 
\multicolumn{1}{|c|}{\small 8} & \multicolumn{1}{|l}{\small 1.00885} & 
\multicolumn{1}{l}{\small 1.01151} & \multicolumn{1}{l}{\small 1.01426} & 
\multicolumn{1}{l||}{\small 1.01709} & \multicolumn{1}{|c||}{\small 8} & 
\multicolumn{1}{|l}{\small 1.00137} & \multicolumn{1}{l}{\small 1.00178} & 
\multicolumn{1}{l}{\small 1.0022} & \multicolumn{1}{l|}{\small 1.00263} \\ 
\hline
\end{tabular}
\vskip5mm

\begin{tabular}{|l|cccc||l||cccc|}
\hline
$\frac{X_{ee^{\prime }}^{\left( 0\right) }\left( 5\right) }{X_{ee^{\prime
}}^{\left( 0\right) }\left( cft\right) }$ & {\small 2} & {\small 4} &
 {\small 6}
& {\small 8} & $\frac{X_{ee^{\prime }}^{\left( 0\right) }\left( 20\right) }{
X_{ee^{\prime }}^{\left( 0\right) }\left( cft\right) }$ & {\small 2} &
 {\small 4} & {\small 6} & {\small 8} \\ \hline
\multicolumn{1}{|c|}{\small 2} & \multicolumn{1}{|l}{\small 1.28946} & 
\multicolumn{1}{l}{\small 1.42714} & \multicolumn{1}{l}{\small 1.60523} & 
\multicolumn{1}{l||}{\small 1.89334} & \multicolumn{1}{|c||}{\small 2} & 
\multicolumn{1}{|l}{\small 1.1113} & \multicolumn{1}{l}{\small 1.15228} & 
\multicolumn{1}{l}{\small 1.19097} & \multicolumn{1}{l|}{\small 1.22861} \\ 
\multicolumn{1}{|c|}{\small 4} & \multicolumn{1}{|l}{\small 1.42714} & 
\multicolumn{1}{l}{\small 1.56675} & \multicolumn{1}{l}{\small 1.75409} & 
\multicolumn{1}{l||}{\small 2.06284} & \multicolumn{1}{|c||}{\small 4} & 
\multicolumn{1}{|l}{\small 1.15228} & \multicolumn{1}{l}{\small 1.18898} & 
\multicolumn{1}{l}{\small 1.22491} & \multicolumn{1}{l|}{\small 1.26065} \\ 
\multicolumn{1}{|c|}{\small 6} & \multicolumn{1}{|l}{\small 1.60523} & 
\multicolumn{1}{l}{\small 1.75409} & \multicolumn{1}{l}{\small 1.9584} & 
\multicolumn{1}{l||}{\small 2.29906} & \multicolumn{1}{|c||}{\small 6} & 
\multicolumn{1}{|l}{\small 1.19097} & \multicolumn{1}{l}{\small 1.22491} & 
\multicolumn{1}{l}{\small 1.25898} & \multicolumn{1}{l|}{\small 1.29343} \\ 
\multicolumn{1}{|c|}{\small 8} & \multicolumn{1}{|l}{\small 1.89334} & 
\multicolumn{1}{l}{\small 2.06284} & \multicolumn{1}{l}{\small 2.29906} & 
\multicolumn{1}{l||}{\small 2.69597} & \multicolumn{1}{|c||}{\small 8} & 
\multicolumn{1}{|l}{\small 1.22861} & \multicolumn{1}{l}{\small 1.26065} & 
\multicolumn{1}{l}{\small 1.29343} & \multicolumn{1}{l|}{\small 1.32697} \\ 
\hline
\end{tabular}
\vskip5mm

\begin{tabular}{|l|cccc||l||cccc|}
\hline
$\frac{X_{ee^{\prime }}^{\left( 0\right) }\left( 100\right) }{X_{ee^{\prime
}}^{\left( 0\right) }\left( cft\right) }$ & {\small 2} & {\small 4} &
 {\small 6}
& {\small 8} & $\frac{X_{ee^{\prime }}^{\left( 0\right) }\left( 400\right) }{
X_{ee^{\prime }}^{\left( 0\right) }\left( cft\right) }$ & {\small 2} & {\small 
4} & {\small 6} & {\small 8} \\ \hline
\multicolumn{1}{|c|}{\small 2} & \multicolumn{1}{|l}{\small 1.03837} & 
\multicolumn{1}{l}{\small 1.052} & \multicolumn{1}{l}{\small 1.06441} & 
\multicolumn{1}{l||}{\small 1.07597} & \multicolumn{1}{|c||}{\small 2} & 
\multicolumn{1}{|l}{\small 1.01529} & \multicolumn{1}{l}{\small 1.02071} & 
\multicolumn{1}{l}{\small 1.02562} & \multicolumn{1}{l|}{\small 1.03019} \\ 
\multicolumn{1}{|c|}{\small 4} & \multicolumn{1}{|l}{\small 1.052} & 
\multicolumn{1}{l}{\small 1.06393} & \multicolumn{1}{l}{\small 1.07517} & 
\multicolumn{1}{l||}{\small 1.08587} & \multicolumn{1}{|c||}{\small 4} & 
\multicolumn{1}{|l}{\small 1.02071} & \multicolumn{1}{l}{\small 1.02544} & 
\multicolumn{1}{l}{\small 1.02988} & \multicolumn{1}{l|}{\small 1.03409} \\ 
\multicolumn{1}{|c|}{\small 6} & \multicolumn{1}{|l}{\small 1.06441} & 
\multicolumn{1}{l}{\small 1.07517} & \multicolumn{1}{l}{\small 1.08556} & 
\multicolumn{1}{l||}{\small 1.09557} & \multicolumn{1}{|c||}{\small 6} & 
\multicolumn{1}{|l}{\small 1.02562} & \multicolumn{1}{l}{\small 1.02988} & 
\multicolumn{1}{l}{\small 1.03397} & \multicolumn{1}{l|}{\small 1.0379} \\ 
\multicolumn{1}{|c|}{\small 8} & \multicolumn{1}{|l}{\small 1.07597} & 
\multicolumn{1}{l}{\small 1.08587} & \multicolumn{1}{l}{\small 1.09557} & 
\multicolumn{1}{l||}{\small 1.10504} & \multicolumn{1}{|c||}{\small 8} & 
\multicolumn{1}{|l}{\small 1.03019} & \multicolumn{1}{l}{\small 1.03409} & 
\multicolumn{1}{l}{\small 1.0379} & \multicolumn{1}{l|}{\small 1.0416} \\ 
\hline
\end{tabular}
\end{center}

\section{Twist and $SU(1,1)$ in the Siegel gauge \label{sec:SU11}}

The twist operator is usually given by $\hat{\Omega}=(-1)^{L_{0}}$. In MSFT
this becomes 
\begin{equation}
\hat{\beta}_{\Omega }\hat{A}(\bar{x},x_{e},p_{e};\xi
_{0},x_{o},p_{o},y_{o},q_{o})=\hat{A}(\bar{x},x_{e},-p_{e};\xi
_{0},-x_{o},p_{o},-y_{o},q_{o})\,.  \label{eq:twistMSFT}
\end{equation}
This follows from 
\begin{equation}
\left\langle
x_{0},x_{e},x_{o};c_{0},x_{e}^{gh},x_{o}^{gh},y_{e}^{gh},y_{o}^{gh}\right| 
\hat{\Omega}=\left\langle
x_{0},x_{e},-x_{o};c_{0},x_{e}^{gh},-x_{o}^{gh},y_{e}^{gh},-y_{o}^{gh}%
\right| \,.
\end{equation}
When we use the even variables $x_{e}^{b},p_{e}^{b},x_{e}^{c},p_{e}^{c}$, we
have the expression 
\begin{equation}
\hat{\beta}_{\Omega }\hat{A}(\bar{x},x_{e},p_{e};\xi
_{0},x_{e}^{b},p_{e}^{b},x_{e}^{c},p_{e}^{c})=\hat{A}(\bar{x}%
,x_{e},-p_{e};\xi _{0},-x_{e}^{b},p_{e}^{b},-x_{e}^{c},p_{e}^{c})\,.
\label{eq:twistMSFT_2}
\end{equation}
The $SU(1,1)$ generators are given by 
\begin{eqnarray}
&&\mathcal{\hat{G}}:=\sum_{n=1}^{\infty }(\hat{c}_{-n}\hat{b}_{n}-\hat{b}%
_{-n}\hat{c}_{n})\quad (=N_{\mathrm{gh}}-(\hat{c}_{0}\hat{b}_{0}+1))\,, 
\notag \\
&&\hat{X}:=-\sum_{n=1}^{\infty }n\hat{c}_{-n}\hat{c}_{n}\,,\qquad \hat{Y}%
:=\sum_{n=1}^{\infty }{\frac{1}{n}}\hat{b}_{-n}\hat{b}_{n}
\end{eqnarray}
by oscillator representation \cite{Gaiotto:2002wy}\cite{Zwiebach}. In MSFT
we translate them as 
\begin{eqnarray}
&&\hat{\beta}_{\mathcal{\hat{G}}}=\sum_{o>0}\left( y_{o}{\frac{\partial }{%
\partial y_{o}}}-x_{o}{\frac{\partial }{\partial x_{o}}}+p_{o}{\frac{%
\partial }{\partial p_{o}}}-q_{o}{\frac{\partial }{\partial q_{o}}}\right)
=\sum_{e>0}\left( x_{e}^{c}{\frac{\partial }{\partial x_{e}^{c}}}-x_{e}^{b}{%
\frac{\partial }{\partial x_{e}^{b}}}+p_{e}^{b}{\frac{\partial }{\partial
p_{e}^{b}}}-p_{e}^{c}{\frac{\partial }{\partial p_{e}^{b}}}\right) \,, 
\notag \\
&&\hat{\beta}_{\hat{X}}=i\sum_{o>0}\left( y_{o}\kappa _{o}{\frac{\partial }{%
\partial x_{o}}}-p_{o}\kappa _{o}{\frac{\partial }{\partial q_{o}}}\right)
=i\sum_{e>0}\left( x_{e}^{c}{\frac{\partial }{\partial x_{e}^{b}}}-p_{e}^{b}{%
\frac{\partial }{\partial p_{e}^{c}}}\right) \,,  \label{eq:su(1,1)}\\
&&\hat{\beta}_{\hat{Y}}=i\sum_{o>0}\left( x_{o}\kappa _{o}^{-1}{\frac{%
\partial }{\partial y_{o}}}-q_{o}\kappa _{o}^{-1}{\frac{\partial }{\partial
p_{o}}}\right) =i\sum_{e>0}\left( x_{e}^{b}{\frac{\partial }{\partial
x_{e}^{c}}}-p_{e}^{c}{\frac{\partial }{\partial p_{e}^{b}}}\right) 
\notag
\end{eqnarray}
on the fields in the Siegel gauge. These operators satisfy the $su(1,1)$
algebra: 
\begin{equation}
\lbrack \hat{\beta}_{\hat{X}},\hat{\beta}_{\hat{Y}}]=-\hat{\beta}_{\mathcal{%
\hat{G}}}\,,\quad \lbrack \hat{\beta}_{\mathcal{\hat{G}}},\hat{\beta}_{\hat{X%
}}]=2\hat{\beta}_{\hat{X}}\,,\quad \lbrack \hat{\beta}_{\mathcal{\hat{G}}},%
\hat{\beta}_{\hat{Y}}]=-2\hat{\beta}_{\hat{Y}}\,.
\end{equation}
They are derivations with respect to the Moyal $\star $ product: 
\begin{equation}
\hat{\beta}_{\mathcal{\hat{O}}}(A_{1}\star A_{2})=(\hat{\beta}_{\mathcal{%
\hat{O}}}A_{1})\star A_{2}+A_{1}\star (\hat{\beta}_{\mathcal{\hat{O}}%
}A_{2})\,,\qquad \mathcal{\hat{O}}=\mathcal{\hat{G}},\hat{X},\hat{Y}\,.
\label{eq:su11derivation}
\end{equation}
In fact, the above $su(1,1)$ generators (\ref{eq:su(1,1)}) are inner
derivations 
\begin{eqnarray}
&&\hat{\beta}_{\hat{\mathcal{O}}}A=[\beta _{\hat{\mathcal{O}}},A]_{\star
}\,,\qquad \mathcal{\hat{O}}=\mathcal{\hat{G}},\hat{X},\hat{Y}\,,  \notag \\
&&\beta _{\hat{\mathcal{G}}}={\frac{1}{\theta ^{\prime }}}%
\sum_{o>0}(y_{o}q_{o}-x_{o}p_{o})={\frac{1}{\theta ^{\prime }}}%
\sum_{e>0}\left( x_{e}^{c}p_{e}^{c}-x_{e}^{b}p_{e}^{b}\right) \,,   \\
&&\beta _{\hat{X}}={\frac{i}{\theta ^{\prime }}}\sum_{o>0}y_{o}\kappa
_{o}p_{o}={\frac{i}{\theta ^{\prime }}}\sum_{e>0}x_{e}^{c}p_{e}^{b}\,,~~~~~~%
\beta _{\hat{Y}}={\frac{i}{\theta ^{\prime }}}\sum_{o>0}x_{o}\kappa
_{o}^{-1}q_{o}={\frac{i}{\theta ^{\prime }}}\sum_{e>0}x_{e}^{b}p_{e}^{c}\,.
\notag
\end{eqnarray}
By Eqs.(\ref{eq:twistMSFT_2},\ref{eq:su11derivation}), we can restrict
solutions of the equations of motion (\ref{eq:EOM_MSFT}) to the twist even
and $SU(1,1)$ singlet sector: 
\begin{equation}
\hat{\beta}_{\hat{\Omega}}A(\xi )=A(\xi )\,,~~~~~\hat{\beta}_{\hat{\mathcal{O%
}}}A(\xi )=0\,,~~~\hat{\mathcal{O}}=\hat{\mathcal{G}},\hat{X},\hat{Y}
\label{eq:singlet}
\end{equation}
in the Siegel gauge consistently. In fact, we note that in the Siegel gauge 
\begin{eqnarray}
&&[\hat{\beta}_{\hat{\mathcal{O}}},L_{0}]=0\,,~~~~~~~~~~~\hat{\mathcal{O}}=%
\hat{\Omega},\hat{\mathcal{G}},\hat{X},\hat{Y}\,,  \notag \\
&&\hat{\beta}_{\hat{\Omega}}(A\star A)=(\hat{\beta}_{\hat{\Omega}}A){\tilde{%
\star}}(\hat{\beta}_{\hat{\Omega}}A)=(\hat{\beta}_{\hat{\Omega}}A)\star (%
\hat{\beta}_{\hat{\Omega}}A)\,,~~~~\tilde{\star}:=\star |_{-\theta ,-\theta
^{\prime }}
\end{eqnarray}
from Eqs.(\ref{eq:total_L0},\ref{eq:total_star}). 
This condition (\ref{eq:singlet}) can be used to search for the
nonperturbative tachyon vacuum.\cite{Gaiotto:2002wy}

In MSFT, it is convenient to note the monoid structure (\S \ref{sec:partII}%
). It consists of gaussian Moyal fields: $A_{\mathcal{N},M,\lambda }=%
\mathcal{N}e^{-\bar{\xi}M\xi -\bar{\xi}\lambda }$. We can consider the twist
and $SU(1,1)$ symmetric class within the monoid. From Eqs.(\ref
{eq:twistMSFT_2},\ref{eq:su(1,1)}) the restriction by this symmetry (\ref
{eq:singlet}) of a monoid element $A_{\mathcal{N},M,\lambda }(\xi )$ is
given by 
\begin{equation}
M=\varepsilon M^{\prime }:=\left( 
\begin{array}{cc}
0 & M^{\prime } \\ 
-M^{\prime } & 0
\end{array}
\right) \,,~~M^{\prime }=\left( 
\begin{array}{cc}
A & 0 \\ 
0 & B
\end{array}
\right) \,,~~~\bar{A}=A\,,~~~\bar{B}=B\,,~~~~~\lambda =0  \label{single}
\end{equation}
in the basis $\bar{\xi}=(x_{e}^{b},p_{e}^{b},x_{e}^{c},p_{e}^{c})$\footnote{%
If we use the odd basis $(x_{o},p_{o},y_{o},q_{o})$, this condition becomes $%
\bar{A}\kappa _{o}=\kappa _{o}A,\,\kappa _{o}\bar{B}=B\kappa _{o}$.}.
Namely, the coefficient matrix of the quadratic term in the exponent becomes
block diagonal and symmetric. For example, the perturbative vacuum and
butterfly states (\ref{eq:perturbative_vacuum_even},\ref{eq:even_butterfly})
are of the form of (\ref{single}). Their $\tau $-evolved gaussians are also
in this class (\ref{eq:tau-evolved2}). We note that this class of gaussian
is not closed within the monoid because of twist operator which changes the
sign of noncommutative parameters $\theta ,\theta ^{\prime }$ in the Moyal $%
\star $ product, while the $SU(1,1)$-symmetry is conserved in the Siegel
gauge by Eq.(\ref{eq:su11derivation}).

\end{document}